\documentclass[%
 reprint,
superscriptaddress,
 amsmath,amssymb,
 aps,
 pra,
]{revtex4-2}

\usepackage{amssymb,amsmath}
\usepackage{mathrsfs}
\usepackage{graphicx}%
\usepackage[tiny, center, uppercase]{titlesec}
\usepackage{dcolumn}%
\usepackage{bm}%
\usepackage{subcaption}
\usepackage{float}
\usepackage{caption}
\usepackage{lipsum}  
\usepackage{tabularx}

\usepackage{xcolor}

\newcommand{\trento}{T$\mathrel{\protect\raisebox{-2.1pt}{R}}$ENT$\mathrel{\protect\raisebox{0pt}{o}}$}

\newcommand{\nn}{{\nonumber}}

\newcommand{\Q}{\mathcal{Q}}

\newcommand{\D}{\mathcal{D}}

\newcommand{\la}{\langle}
\newcommand{\ra}{\rangle}

\newcommand{\bx}{\mathbf{x}}

\newcommand{\vx}{\vec{x}}

\newcommand{\tot}{\text{tot}}

\newcommand{\hyd}{\text{hyd}}

\newcommand{\pa}{\text{part}}
\newcommand{\mi}{\text{min}}
\newcommand{\fluc}{\text{fluc}}
\newcommand{\ch}{\text{ch}}
\newcommand{\OO}{\mathcal{O}}

\begin{document}

\title{Revealing initial state properties through ultra-central symmetric heavy-ion collisions}%

\author{S.~M.~A.~Tabatabaee Mehr}
\affiliation{School of Particles and Accelerators, Institute for Research in
	Fundamental Sciences (IPM), P.O. Box 19395-5531, Tehran, Iran}

\author{S.~F.~Taghavi}
 \affiliation{TUM School of Natural Sciences, Technische Universit\"at M\"unchen, Garching, Germany}%

\begin{abstract}

Heavy-ion experiments provide a new opportunity to gain a deeper understanding of the structure of nuclei. To achieve this, it is crucial to identify observables under circumstances that are minimally affected by the process that leads to the initial state of heavy-ion collisions from nuclear wavefunction. In this study, we demonstrate that when assuming scale-invariance, the effect of this stage on the initial energy or entropy density moments in ultra-central symmetric collisions is negligible for nucleon sizes of approximately 0.7 fm or larger for large nuclei. By borrowing cluster expansion method from statistical physics and using scale-invariance assumption, we calculate the average ellipticity of initial density at the presence of short-range correlation. We compare our calculations to Monte Carlo studies and assess the accuracy of various methods of short-range correlation sampling. Additionally, we find that the isobar ratio can constrain the initial state parameters, in addition to deformation. Our study indicates that the isobar ratios in ultra-central collisions are especially sensitive to the fluctuation in the weight of the nuclei constituents and the two-body correlation among nucleons. This insight is crucial for drawing conclusions about nuclear deformations based on isobar ratios.

\end{abstract}

\maketitle

\tableofcontents
\addtocontents{toc}{\protect\setcounter{tocdepth}{1}}
    \addcontentsline{file}{section_unit}{entry}
\section{Introduction}

The study of nuclear structure has become an intense area of research in recent years due to its close relation to our understanding of quantum chromodynamics (QCD) in a non-perturbative regime~\cite{Hergert:2020bxy}, as well as the link between the neutron skin of heavy nuclei and the equation of state in neutron-rich matter, similar to that of neutron stars~\cite{Brown:2000pd, Horowitz:2000xj}. The study of heavy-ion collisions has also gained popularity as a method of extracting information about the properties of the colliding nuclei~\cite{Giacalone:2021udy,Giacalone:2023cet}. To accomplish this, there is a need for a thorough comprehension of collectivity during the process. Alternatively, identifying observables that are minimally reliant on collective expansion can offer valuable insights into the initial state. For instance, the ratio between collision observables of isobar elements, such as ${}^{96}$Ru--${}^{96}$Ru and ${}^{96}$Zr--${}^{96}$Zr, can provide useful information as these observables are particularly sensitive to the deformation of these nuclei in central collisions~\cite{Jia:2021tzt,Zhang:2021kxj,Zhang:2022fou,Jia:2022qgl}.

It is essential to incorporate the specific characteristics of each nucleus, such as its overall geometry, and the correlations between nucleons, into the initial state of ion collisions. This is done through heavy-ion initial state models like  MC-Glauber~\cite{Miller:2007ri,Alver:2008aq}, MC-KLN~\cite{Kharzeev:2001yq,Drescher:2006pi}, \trento{}~\cite{Moreland:2014oya,Moreland:2018gsh}, and IP-Glasma~\cite{Schenke:2012wb,Schenke:2012hg}. The initial entropy of a system is a result of the interaction between constituents of colliding nuclei. However, the process of entropy production is quite complex and not yet fully understood, leading to substantial differences between initial state models.  \trento{} model takes a phenomenological approach by introducing reduced thickness function as a function of participant's initial thickness functions of the target and projectile,
\begin{equation}\label{reducedThickness}
	T_R(T_A,T_B)= \mathcal{N}\left(\frac{T_A^p+T_B^p}{2}\right)^{1/p}.   
\end{equation}
The reduced thickness function is proportional to the initial entropy density.
The participant thickness functions is obtained from participant nuclear matter density as $T_{A(B)}(x,y) = \int dz \rho^\text{part}(\vx)$ where $z$ direction is along the beam axis. In the most central collisions and in the case of large nucleon inelastic cross-section, $\rho^\text{part}(\vx)$ is equivalent to a one-body density of nuclei. In \trento{} model and in the present work, the one-body density is parameterized by Woods-Saxon (WS) distribution,
\begin{equation}\label{WSdist}
	\rho_\text{WS}(|\vec{x}|,\theta,\phi) = \frac{\rho_0}{1+\exp\left[(|\vec{x}|-R(\theta,\phi))/a_0\right]},
\end{equation}
where $ R(\theta,\phi) = R_0(1 + \beta_2 Y_2^0(\theta,\phi)+\beta_3 Y_3^0(\theta,\phi)+\cdots)$.
Here, $\rho_0$ is the nucleon density constant, $R_0$ is the radius, $a_0$ is the skin thickness, $\beta_2$ is quadruple, and $\beta_3$ is the octupole deformations. 

The definition of Eq.~\eqref{reducedThickness} has scale-invariance characteristic, meaning that when we scale $T_{A,B} \to c \,T_{A,B}$, the reduced thickness function scales the same way, $T_{R} \to c\, T_{R}$. This property is supported by elliptic flow measurements in ultra-central uranium-uranium collisions at RHIC~\cite{Pandit:2013uiv,Wang:2014qxa}. In our study, we demonstrate that in ultra-central symmetric collisions (UCSC), the actual form of $T_R$ has a negligible impact on the geometrical properties of the produced entropy and its fluctuation as long as the nucleons are large enough. Even more complicated functions for $T_R$ lead to the same conclusion. When the nucleons are large enough, the details of their positions become less important, and in UCSC, we can assume that $T_A(x,y) \approx T_B(x,y)$. In the specific example of the scale-invariant $T_R$ in Eq.~\eqref{reducedThickness}, this leads to independence on the parameter $p$. More generally, by only assuming the scale-invariance $T_R(c T_A, c T_B) = c T_R(T_A, T_B)$, we can express the reduced thickness function in UCSC as $T_R(T_A,T_A)\equiv g(T_A)$. Here, the scale-invariance is given by $g(c T_A) = c g(T_A)$, which implies that $g(x)$ is a one-dimensional, degree one, homogeneous function. From Euler's homogeneous function theorem, we conclude that a linear function is the only continuously differentiable function with this property, i.e., $g(T_A)\propto T_A$. The constant of proportionality is absorbed into the overall normalization. Therefore, the actual functionality of the reduced thickness function is irrelevant in this case. For large nuclei, such as Pb, Au, Ru, and Zr, the nucleon size $w \sim 0.7\,$[fm] is large enough to employ this approximation. Here, $w$ is the width of a Gaussian profile for nucleons.

In order to access nuclear structure information, assuming scale-invariance can help to reduce theoretical uncertainty related to entropy production in the evolution of ultra-central heavy-ion collisions. With this assumption, we can connect the initial density geometry and its event-by-event fluctuation to the parameters of the nuclei density, including WS parameters and the two-body correlations. In UCSC, the average elliptic shape of the initial entropy, represented by $\epsilon_2^2\{2\} = \la \epsilon_2^2 \ra$, relies solely on the fluctuation for two spherical nuclei. The fluctuation caused by point-like sources results in $\epsilon_2^2\{2\} = 5 / (7 A)$ (refer to section~\ref{EllipticityFluctuationSec}). The impact of the repulsive central core of nuclei with a radius of $d_\mi / 2$ can be described as follows: The volume of central core need to be subtracted from the available space for fluctuation. Therefore, for small values of $d_\mi / R_0$ one expects decreasing trend for  $\epsilon_2^2\{2\}$ proportional to $1-A(d_\mi / R_0)^3$. For larger $A$ and $d_\mi$, we have smaller space for fluctuation. By an analogy between initial state fluctuation and ensembles of a thermal system, we use the cluster expansion method developed in statistical mechanics to make the above estimation more rigorous in section~\ref{sec:analogyStat}. 

Another correction arises from the contribution of nucleon interactions in entropy production. It is assumed in Ref.~\cite{Bozek:2013uha} that the weight of each nucleon's contribution in the initial entropy density fluctuates based on a Gamma function with a unit mean value and width $\sigma_{\text{fluc}}$ to explain the large multiplicity fluctuation in proton-lead collisions. Since $\epsilon_2^2\{2\}$ is a second-order moment, one would expect that the source fluctuation $\epsilon_2^2\{2\}$ is proportional to the second moment of the Gamma function, which is $1+\sigma_{\text{fluc}}^2$ (see section~\ref{subsec:constituteWeight}).  For a fixed value of $\epsilon_2^2\{2\}$, the two competing contributions lead to the equation $\sigma_{\text{fluc}}^2 \sim A(d_\mi / R_0)^3 + \text{constant}$. A similar correlation between $d_\mi$ and $\sigma_{\text{fluc}}$ has been observed in Bayesian analysis studies~\cite{JETSCAPE:2020mzn,Nijs:2023yab}.

After investigating the impact of short-range correlation and constituent weight fluctuation on $\epsilon_2\{2\}$ in UCSC, we focus to the isobar ratio. This ratio, which compares $\epsilon_2\{2\}$ between two isobar nuclei, such as Ru and Zr,  helps to minimize the impact of the collective expansion stage. This is because of the linear hydrodynamic response approximation, where the elliptic flow $v_2$, the second Fourier coefficient of the final particle distribution in the azimuthal direction~\cite{Ollitrault:1992bk,E877:1996czs}, is proportional to ellipticity $\epsilon_2$~\cite{Niemi:2015qia}. This linear approximation is more accurate for large nuclei and central collisions. The ratio  $v_2^\text{Ru}\{2\} / v_2^\text{Zr}\{2\}$ has been measured by STAR collaboration~\cite{STAR:2021mii}. 
Utilizing the isobar ratio, we can minimize inaccuracies caused by assuming a finite value for nucleon width and analyze the contribution of initial state parameters to measurements. In particular, we find that isobar ratio is sensitive to $\sigma_{\text{fluc}}$ and $d_\mi$, although the effect on $\sigma_{\text{fluc}}$ is more pronounced.

The paper is structured as follows: first, in Section.~\ref{scaleInvSec}, we examine the moments of the initial entropy density for models with and without scale-invariance property for central and non-central and symmetric and asymmetric collisions. We discuss an  analytical approach to study event-by-event fluctuation in section~\ref{UCSCClusterExpansion}. In Section~\ref{EllipticityFluctuationSec}, we discuss the eccentricity average  and its isobar ratio. Using the isobar ratio, we study the constraints on the nuclear structure in Section~\ref{ConstrainigSec}. Finally, we summarize our study in Section ~\ref{SummarySec}.

\section{Ultra-central symmetric collisions and the impact of scale-invariance}\label{scaleInvSec}

In this section, we provide evidence to support the claim that the form of the reduced thickness function has a small impact on the geometry of the initial entropy density in UCSC, by assuming the scale-invariance.

The scale-invariance assumption in \trento{} model relies on elliptic flow observations in ultra-central U-U collisions~\cite{Moreland:2014oya}.  Uranium nuclei have a large quadruple deformation, which means that during a collision, the two nuclei may collide on their tips, bodies, or somewhere in between. It can be argued that in tip-tip collisions, there are more binary collisions compared to body-body collisions. This leads to an increase in entropy and, consequently, more particles. This concept is implemented in MC-Glauber, where the reduced thickness function is defined as
\begin{equation}\label{ReducedThicknessGlauber}
	T_R(T_A,T_B) \propto \frac{1}{2}\left(T_A+T_B\right) + \alpha T_A T_B,
\end{equation}
where $\alpha$ controls the contribution of binary collisions.
 In the presence of binary collisions, events with the highest multiplicities should be populated by tip-tip collisions with smaller ellipticity, which goes against the STAR data from the RHIC experiment~\cite{Pandit:2013uiv,Wang:2014qxa}. It is concluded in Ref.~\cite{Moreland:2014oya} that U-U collisions do not favor the entropy production with binary collisions.  On the other hand, by using the combination $(N_A + N_B)/2 + \alpha N_{col}$, it is possible to link the collision geometry to the final particle production. This approach has been studied in Refs.~\cite{ALICE:2010mlf,ALICE:2013hur}. The conclusion drawn is that the centrality dependence of the charged-particle multiplicity in Pb-Pb collisions can only be described when $\alpha$ is nonvanishing. This implies that one need to replace $(N_A + N_B)/2$ with an unspecified function $f(N_A , N_B)$ to open room for a function compatible with both U--U collision measurements and centrality dependence of the charged-particle multiplicity in Pb--Pb.
 
Since tip-tip and body-body U--U collisions lead to same multiplicity, it is possible to assume that all the collision configurations are equivalent to the collision of $N_A$ one-to-one nucleon. The latter configuration suggests that the number of produced particles should be directly proportional to the number of nucleons, i.e., $N_\ch \propto N_A$, where the constant of proportionality depends on the one-to-one nucleon interactions. Based on this argument, one can suggest that the scaling property holds true for the collision of any configuration of $N_A$ and $N_B$ nucleons:
\begin{equation}\label{ChargeMultiScaling}
	N_\ch \propto f(N_A,N_B)
\end{equation}
where $f$ has a scale-invariant property, $f(c N_A,c N_B) = c f( N_A, N_B)$, leading to linear scaling $N_\ch \propto N_A$ when $N_A = N_B$. It is important to note that the reverse logic is not valid, and we can have functions $f(x,y)$ that are not scale-invariant but collapse into a linear function at $f(x,x)$. Although microscopic theories based on CGC suggest reduced thickness function with scale-invariance property (which we will discuss shortly), there is still room for exploration regarding the scale-invariance in non-central collisions and the case where $N_A \neq N_B$ from an experimental point of view.

It is a known approximation that the final multiplicity is proportional to the initial entropy, $N_\ch \propto S$. Also, by definition, the participant thickness functions, 
\begin{equation}\label{participants}
	T_{A(B)}(\bx) = \kappa \sum_{i=1}^{N_{A(B)}}\int dz\,\rho_{\text{nucleon}}(\vec{x}-\vec{x}_i),
\end{equation}
scales similar to the number of participant $N_{A(B)}$ where $\bx$ is the coordinate in the transverse space. Here, we have assumed $\int d^3 x \rho_{\text{nucleon}}$ is dimensionless and normalized to unity, and $\kappa$ is a dimensionless constant for overall normalization. Therefore, by defining the reduced thickness function as the entropy density in the transverse direction,
$T_{R}(\bx) = \int dz\, s(\vec{x})$,
 one suggests a differential form of Eq.~\eqref{ChargeMultiScaling} as $T_R(T_A,T_B) \propto f(T_A,T_B)$. In Eq.~\eqref{reducedThickness}, an specific form of $f(T_A,T_B)$ is chosen.

It is important to note that the argument mentioned above assumes that the system has reached its local thermal equilibrium from its initial state. In order to extend the argument to the out-of-equilibrium period of the evolution, it is necessary to establish a connection between the entropy density, $s_\hyd$, at the start of local thermal equilibrium, $\tau_\hyd$, and the energy density, $e_0$, at an earlier time, $\tau_0 \gtrsim 0^+$. 
In Ref.~\cite{Giacalone:2019ldn}, the concept of hydrodynamic attractors was used to obtain, $s_\hyd \tau_\hyd \propto (e_0 \tau_0)^{2/3}$ where the right-hand side is in the out-of-equilibrium period. This formula can be used to calculate the final multiplicity, which is given by
\begin{equation}\label{EntropyEnergyConvert}
	\frac{dN_\ch}{d\eta} \propto \frac{dS}{d\eta} =A_\perp (s_\hyd \tau_\hyd) \propto A_\perp (e_0 \tau_0)^{2/3},
\end{equation}
where $A_\perp$ is the transverse size of the system at the time $\tau_\hyd$. At the pre-equilibrium stage, we anticipate a slight time dependence for $A_\perp$. To achieve scale-invariance in Eq.~\eqref{ChargeMultiScaling}, the energy reduced thickness function needs to scale through a homogeneous function with degree  3/2,
\begin{equation}\label{reducedEnergyScaleInv}
	T_R^{\text{energy}}(T_A,T_B) = \mathcal{N} \frac{\tau_\hyd^{3/2}}{\tau_0} f^{3/2}(T_A,T_B).
\end{equation}
Assuming $T_{A(B)}$ as the energy thickness function, the constant $\kappa$ in Eq.~\eqref{participants} has a dimension of energy too. 

In the context of the CGC framework, Ref.~\cite{Borghini:2022iym} obtains the initial energy density using the saturation scale $Q_{s,A(B)}^2$. Since $Q_{s,A(B)}^2 \propto T_{A(B)}$, the reduced thickness function can be defined as,
\begin{equation}\label{CGCTR}
\begin{split}
		T_R^{\text{energy}}&(T_A,T_B)  \propto   \\
		&\frac{T_A T_B}{(T_A + T_B)^{5/2}}\left[2T_A^2 +7T_A T_B +2 T_B^2\right].
\end{split}
\end{equation}
The scaling of $T_{A,B} \to c\,T_{A,B}$ leads to $T_R^\text{energy}\to c^{3/2} T_R^\text{energy}$, and consequently, $(dN_\ch/d\eta)\to c\,(dN_\ch/d\eta)$, indicating scale-invariance.

In Ref.~\cite{Nijs:2023yab}, a modified version of reduced thickness function is introduced,  
\begin{equation}\label{GeneralReducedThickness}
	T_R^{\text{energy}}(T_A,T_B) = \mathcal{N} E_{\text{ref}}^{2-2q} \left[(T_A^p+T_B^p)/2\right]^{q/p},
\end{equation}
which violates the scale-invariance for $q\neq 3/2$. This definition with $q=3/2$ and $E_{\text{ref}}= \tau_0 (\tau_\hyd\kappa)^{-3/2} $ is equivalent to Eq.~\eqref{reducedEnergyScaleInv}. 
The global Bayesian analysis in Ref.~\cite{Nijs:2023yab} suggests $q = 1.34\substack{+0.14 \\ -0.18}$ which is marginally compatible with $q=3/2$. In the mentioned study, the relation $e \propto s^{4/3}$ is used to convert entropy to energy density, which is different from our approach based on Eq.~\eqref{EntropyEnergyConvert}. Additionally, the study analyzes the scaling property of the IP-Glasma and shows that $e_0 \propto (T_A T_B)^{q/2}$, where the value of $q$ falls in the range of 0.54 to 1.71.

In summary, scale-invariance refers to the case that the initial state scaling results a same amount of scaling in the final multiplicity. If the initial entropy density is used, the reduced thickness function need to be a homogeneous function of degree one. If the energy density is used, the degree of homogeneity should be 3/2. For this study, we assume that the initial state is modeled using the entropy density.

\begin{figure*}
	\begin{center}
		\includegraphics[width=0.95\textwidth]{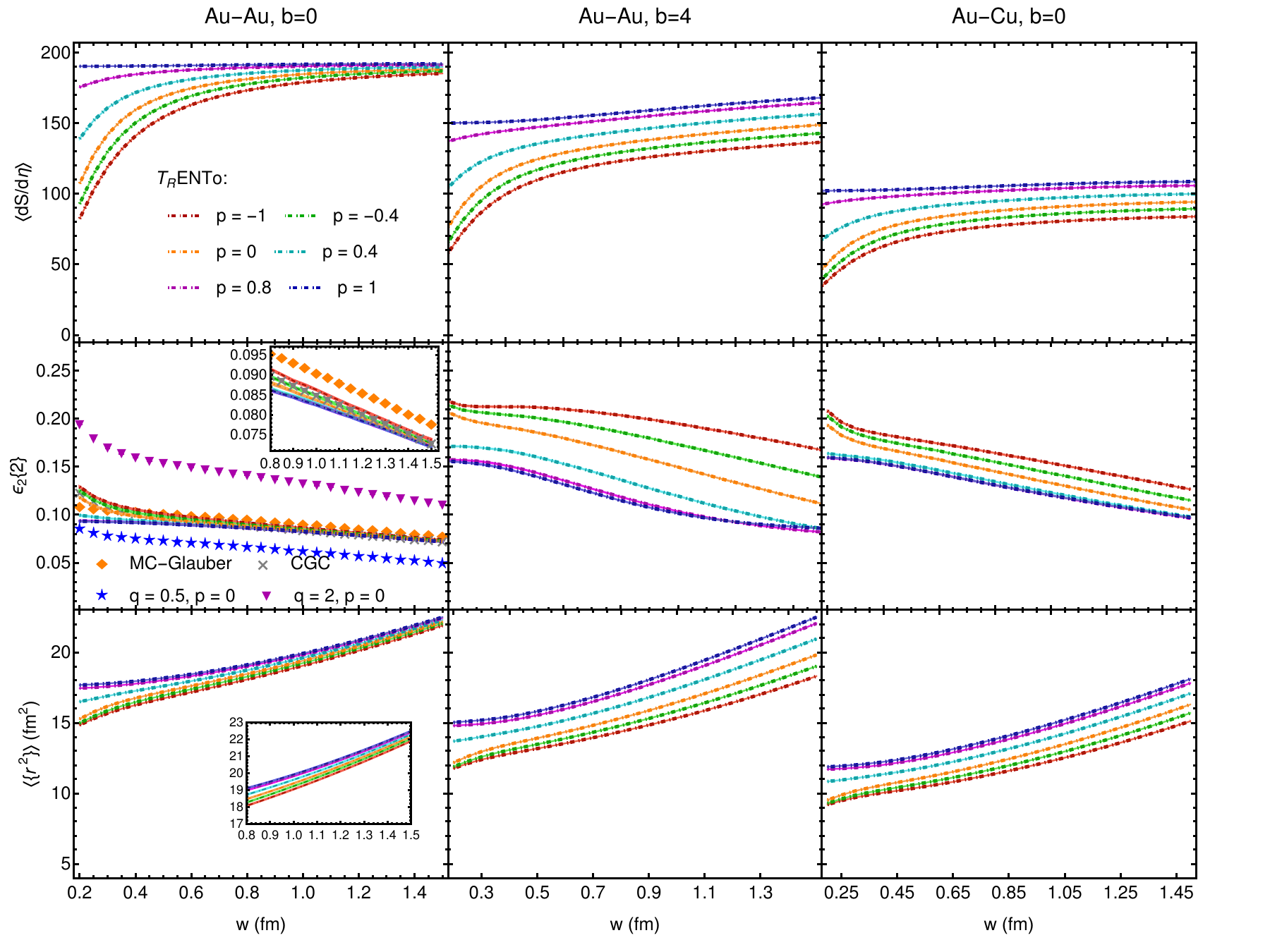}
		\caption{The convergence of initial entropy density moments is affected by impact parameter and collision system symmetry. When the impact parameter is zero, and the collision system is symmetric, the initial state moments converge to the same curve as $w$ increases. The moments do not converge to the same curve for events with non-zero impact parameters or asymmetric collision systems.}
		\label{ConvergencTestTrento}
	\end{center}
\end{figure*}

To investigate the relationship between the nucleon size, impact parameter, and asymmetric collisions, as well as the impact of scale-invariance on the moments of the initial entropy density in UCSC, we use \trento{} model and the model based on CGC, which have scale-invariance properties. We also use the MC-Glauber model and the generalized \trento{} model to examine scenarios with broken scale-invariance. The  MC-Glauber model, CGC model, and generalized \trento{} model are implemented in \trento{} event generator via Eqs.~\eqref{ReducedThicknessGlauber}, \eqref{CGCTR}, and \eqref{GeneralReducedThickness}. To quantitatively study $T_R$, we calculate the following moments,
\begin{equation}\label{ISMoments}
\{r^2\}, \qquad \epsilon_ne^{in\phi_n}=-\frac{ \{r^n e^{in\varphi}\} }{\{r^n\}},\quad n=2,3,
\end{equation}
where $(r,\varphi)$ is the polar coordinate in the transverse space $\bx$, and $\{\cdots\} = \int d\bx \cdots T_R(\bx)$. Using \trento{} event generator, we generate Au--Au collisions  with impact parameters $b=0\,$[fm] and $4\,$[fm], and Au--Cu with $b=0\,$[fm] at $\sqrt{s_{\text{NN}}}=200\,$GeV as an example of an asymmetric collision. Here, the fluctuation parameter and minimum distance between nucleons are set as follows: $\sigma_\fluc =1$, $d_\mi = 0.6\,$[fm]. We calculate the following many event average quantities, $\left\la dS/d\eta \right\ra$, $\epsilon_2\{2\}$, and $\la \{r^2\} \ra$
and plot them in terms of nucleon width $w$, shown in Fig.~\ref{ConvergencTestTrento}. As seen from the figure, when the collision system is symmetric, the moments for different values of $p$ tend to converge to the same values for large $w$ during most central collisions while it is not the case for non-central collisions and also for asymmetric collisions, meaning the functionality of reduced thickness function has a small impact on ultra-central symmetric collisions. In the appendix~\ref{moreMoments}, we examined $\epsilon_3\{2\}$,  $\la \epsilon_2^2\epsilon_3^2\ra - \la \epsilon_2^2\ra \la\epsilon_3^2\ra$, and $\la \epsilon_2^2 \epsilon_4 \cos \left(4\varphi_2-4\varphi_4\right)\ra$ which also shows similar behavior.

In appendix~\ref{SingleShotTR}, we study the effect of impact parameter for $p=0$ and $p=1$ in more detail for two limits $\pi w^2 \gg \pi R_0^2/A$ and $\pi w^2 \ll \pi R_0^2/A$. We summarized our result in table in the appendix~\ref{SingleShotTR}. For large $w$, the total entropy for \trento{} reduced thickness function in Eq.~\eqref{reducedThickness} with $p=1$ is proportional to $(N_A + N_B) / 2$ while for $p=0$ it is given by $\sqrt{N_A N_B}$. These two values approach the same value for the symmetric collisions $N_A \approx N_B$. Another example is the rms radius $\{r^2\}$. For $p=1$, we find
\begin{equation}\label{p1PointLike}
\begin{split}
	\left.\{r^2\} \right|_{p=1} = & \frac{N_A}{N_A+N_B}  \{x_{a}^2 + y_{a}^2\}_s \\
	& \hspace*{1cm} + \frac{N_B}{N_A+N_B}  \{x_{b}^2+y_b^2\}_s+2w^2,
\end{split}
\end{equation}
and for $p=0$,
\begin{equation}\label{p0PointLike}
\begin{split}
	& \left. \{r^2\} \right|_{p=0}  = 	\frac{1}{2}\Big( \{x_{a}^2 + y_{a}^2\}_s +  \{x_{b}^2 + y_{b}^2\}_s  +2   \{x_{a}\}_s   \{x_{b}\}_s   \\
	& \hspace*{3.5cm}  +2   \{y_{a}\}_s   \{y_{b}\}_s   \Big)  + 2w^2, 
\end{split}
\end{equation}
where $\bx_{a(b)}$ is the location of the target (projectile) sources and $\{ \cdots \}_s$ refers to the average over sources location. For the central collisions, we have $\{\bx_{a}\}_s \approx \{\bx_{b}\}_s  \approx 0$ and for symmetric collisions $N_A \approx N_B$, leading to $\left. \{r^2\} \right|_{p=1} \approx \left. \{r^2\} \right|_{p=0}$. We find similar behavior for $\{y^2-x^2\}$ and $\{2 x y\}$.

To examine whether this convergence is the result of the scale-invariance, we have plotted $\epsilon_2\{2\}$ for Au--Au collisions where the reduced thickness function are calculated from MC-Glauber model ($\alpha = 0.28$), CGC model, and (generalized) \trento{} model for $p=0$ with $q=1,1/2,2$. To translate the CGC model with initial energy density into scale-invariant entropy density, $T_R^{\text{energy}}(T_A,T_B)$ is raised to the power 2/3.
Also, Eq.~\eqref{GeneralReducedThickness} is assumed to be entropy density. In this scenario, $q=1$ represents the scale-invariant case, not $q=3/2$. We have also assumed $E_\text{ref} = 1\,$[fm$^{-1}$]. The parameter $\alpha$ in the MC-Glauber model is intentionally chosen to be larger than usual to emphasize the violation of scale invariance. As observed in Fig.~\ref{ConvergencTestTrento}, the CGC model and \trento{} model with $q=1$ converge to the same values, while the MC-Glauber and \trento{} model with $q=2$ and $1/2$ do not, indicating the violation of scale invariance in the latter models. The same behavior is observed for $\epsilon_3\{2\}$ as shown in appendix~\ref{moreMoments}. It is evident that a small violation of scale-invariance in the MC-Glauber model leads to a small convergence violation. This suggests that a small contribution of the reduced thickness function definition in UCSC is valid also in cases where scale-invariance is only approximately maintained.

\section{Many event average initial state in ultra-central symmetric collisions }\label{UCSCClusterExpansion}

In this section, we study the effect of short-range correlations and constituent weight fluctuation on the event-by-event source fluctuations. 
The technical points and details of this section can be found in the appendix~\ref{app:EbyEAndClusterExpandion}.

\subsection{Two-body correlations from a statistical mechanics analogy}\label{sec:analogyStat}

The position of the sources is determined by the wavefunction of the colliding nuclei. Specifically, the one-body density, denoted as $p_A(\vx)$, and the two-body correlation, denoted as $c_2(\vx_1,\vx_2)$, have a significant impact on the measurement of anisotropic flow. To establish a relationship between the anisotropic flow observables and $p_A(\vx)$ and $c_2(\vx_1,\vx_2)$ at UCSC, we make use of the approximation discussed in Section~\ref{scaleInvSec}. In this scenario, the reduced thickness function is assumed to be the sum of $T_A(\bx)$ and $T_B(\bx)$. It can be demonstrated that in this case, the moments of $T_R$ can be divided into two parts: one part is influenced by the location of the sources, and the other by the distribution of the constituents, such as nucleons or quarks (appendix~\ref{DecompositionSubsection}). This can be observed in Eqs.~\eqref{p1PointLike} and \eqref{p0PointLike}, where the contribution of $2w^2$ is separated from the average source location. In UCSC and assuming a sufficiently large nucleon-nucleon cross-section, it is assumed that all nucleons participate in the collision. This implies that the location of the initial state sources must be similar to those in each nucleus at the time of the collision. Consequently, the fluctuation of sources in a single nucleus can be translated into fluctuations of the entire initial state (appendix~\ref{sourceFlucObsSubSection}).

The process of sampling the nucleon sources becomes more complicated when considering correlations, especially in the context of deformed nuclei. In subsection~\ref{MCSamplingProblems}, we will explore this further. The approximation introduced in UCSC allows us to take a more direct approach to understanding the effect of short-range correlations by using classical techniques from statistical mechanics. We will explore this concept in the following.

The cornerstone concept of classical statistical mechanics is that to obtain thermodynamic quantities in an equilibrated system, the time average over physical quantities can be replaced by averaging over an ensemble of systems. The state of each system is distributed based on statistical distributions, mostly depending on variables such as temperature or chemical potential. In the study of fluctuation in heavy-ion physics, the ensemble of systems has a clear counterpart, which is the collection of independent events. The concept of an equilibrated system is translated to the fact that all events in heavy-ion collisions are equivalent to each other and are achieved under the same experimental conditions, such as the center of mass energy or the same species collisions. The ``state" of an event is the position of the sources, which are distributed based on the nucleus wavefunction. To make the analogy clearer, assume $P_A(\vx_1,\ldots,\vx_A)$ as the distribution of sources in a nucleus,  
\begin{equation}
	P_A(\vx) = \frac{1}{\mathcal{Z}_A} F_A(\vx),
\end{equation} 
where $\mathcal{Z}_A$ is $P_A$ normalization factor. If sources are distributed based on a Gaussian profile with width $\sigma$, one needs to define $F_A(\mathbf{x}) = e^{-\frac{1}{2\sigma^2} \sum_{i=1}^A |\mathbf{x}_i|^2   }$, analogous to a free non-relativistic gas with  $F_N(\vec{p}) = e^{-\beta E} = e^{-\frac{1}{2mkT} \sum_{i=1}^N |\vec{p}_i|^2 } $.

This analogy becomes particularly useful when attempting to account for short-range correlations. The cluster expansion method is utilized to determine thermodynamic quantities in the presence of short-range correlation (see Ref.~\cite{huang1987statistical}). To apply this method to our problem, we define 
\begin{equation}\label{correlationAndDistribution}
	F_A(\vx) = \left[\prod_{i=1}^{A}  p_A(\vx_i) e^{\frac{1}{A}\sum_a\lambda_a f_a(\vx_i)} \right] \left[\prod_{1\leq i<j \leq A} c_2(\vx_j,\vx_j)\right].
\end{equation}
In statistical mechanics, the partition function typically serves as the generating function. The parameter $\beta$ is used to extract the thermodynamic variables and their fluctuations. In our approach, we introduce an auxiliary term $e^{\frac{1}{A}\sum_a\lambda_a f_a(\vx_i)}$, where $\lambda_a$ plays a similar role to $\beta$, but it needs to be set to zero once the desired average is obtained by differentiating the partition function. In order to find the desired moment, it is necessary to select an appropriate function $f_a(\vx)$.

The cluster expansion provides a closed form for the grand-partition function (details in appendix~\ref{app:CESection}),
\begin{equation}
	\begin{split}
		\log \mathscr{Q}&(\la A \ra,\lambda_a) = \\ 
		b_1(\lambda_a)&\;\la A \ra + \left[b_2(\lambda_a)- 2b_2(0)b_1(\lambda_a) \right] \;\la A \ra^2 +\cdots.
	\end{split}
\end{equation}
The $\ell$-cluster integrals, $b_\ell(\lambda_a)$, connect partition function to $p_A(\vx)$ and $c_2(\vx_1,\vx_2)$,
\begin{subequations}
	\begin{align}
		b_1(\lambda_a) &=  \int  d^{3}x\; p_A(\vx)\; e^{\frac{1}{A}\sum_a\lambda_a f_a(\vx)},\\
		b_2(\lambda_a) &= \frac{1}{2} \int  d^{3}x_1d^{3}x_2\; p_A(\vx_1)p_A(\vx_2)\; \\
		& \hspace*{3cm} e^{\frac{1}{A}\sum_a\lambda_a (f_a(\vx_1) + f_a(\vx_2))} \;f_{12}.\nonumber
	\end{align}
\end{subequations}
where the so-called Mayer function is defined as $f_{12} = c_2(\vx_1,\vx_2) - 1$. Here, we assume that $\la A \ra \equiv A$ is large enough that we can disregard the fluctuation of $A$ and approximate $\mathscr{Q}(\la A \ra,\lambda_a) \propto Q_A(\lambda_a)$, where $Q_A(\lambda_a)$ is the canonical partition function at a fixed value of $A$. It's important to note that the calculated observables from canonical and grand-canonical partition functions might have differences in subleading orders in the $1/A$ expansion, which are ignored in the present study.

To find moments such as $\la \{ f_1  \}^{m_1} \{ f_2  \}^{m_2} \ra$, one should calculate $\partial_{\lambda_1}^{m_1} \partial_{\lambda_2}^{m_2} \mathscr{Q}(\la A \ra,\lambda_a) / \mathscr{Q}(\la A \ra,\lambda_a)$ and set $\lambda_a=0$. To obtain $\la \{r^2\}\ra$, we choose $f_1(\vx) = |\bx|^2$, $m_1=1$, and $m_2 = 0$. The functions $f_{1}(\vx) = f_{2}^*(\vx) =|\bx|^2e^{2i\varphi}$ with $m_1=m_2=1$ corresponds to the moment average $\left\la\left|\{r^2 e^{2i\varphi}\}\right|^2\right\ra$.

The method described above is general. For a specific choice of $p_A(\mathbf{x})$ as $\rho_{\text{WS}}(\mathbf{x})$ with radius $\tilde{R}_0$ and skin thickness $\tilde{a}_0$, and $c_2(\mathbf{x}_1,\mathbf{x}_2) = \Theta(|\mathbf{x}_1-\mathbf{x}_2|-d_\text{min})$, which corresponds to nucleons' hard inner core with a radius of $d_\text{min}/2$, we obtain an analytical estimate for the radius as
\begin{equation}\label{StepFubcCorrA}
	\begin{split}
		\left \la \{r^2\}_{s,A} \right\ra =  &\frac{2\tilde{R}_0^2}{5}\left[1 + \frac{7\pi^2}{3}\left(\frac{\tilde{a}_0}{\tilde{R}_0}\right)^2\right. \\& \left. \quad+ \frac{7\pi^2}{3}\left(\frac{\tilde{a}_0}{\tilde{R}_0}\right)^2\left(\frac{d_\mi}{\tilde{R}_0}\right)^3+\cdots\right].
	\end{split}
\end{equation}
Here, $\la \{\cdots\}_{s,A}\ra$ stands for the average over the position of nucleons in a single nucleus $A$. The elliptic shape of the initial density can also be found as 
\begin{equation}\label{EllipticShape}
	\begin{split}
		&\left \la \left|\{r^2 e^{2i\varphi}\}_{s,A}\right|^2 \right\ra =  \frac{1}{A} \frac{8 \tilde{R}_0^2}{35}\left[ 1+6\pi^2\left(\frac{\tilde{a}_0}{\tilde{R}_0}\right)^2 \right.  \\
		&\left. \hspace*{1cm}+\frac{31 \pi^4 }{3}\left(\frac{\tilde{a}_0}{\tilde{R}_0}\right)^4+ d'_\varphi  \left(\frac{d_\mi}{\tilde{R}_0}\right)^3 + \cdots\right],
	\end{split}
\end{equation}  
where:
\begin{equation}\label{dprimePhi}
	d'_\varphi = -A\left[1-6\pi^2\left(\frac{\tilde{a}_0}{\tilde{R}_0}\right)^2-\frac{31 \pi^4 }{3}\left(\frac{\tilde{a}_0}{\tilde{R}_0}\right)^4\right].
\end{equation}
When calculating the expression $d'_\varphi$, we replaced WS distribution with a step function $\Theta(\tilde{R_0} - |\vx|)$ to facilitate the analytical integration. We compared our analytical estimate for $d'_\varphi$ with the numerical result and found that the numerical value of $d'_\varphi$ is approximately half of the value obtained using the step function, indicating the significance of the skin thickness in determining this quantity. We confirmed our calculation using Monte Carlo sampling in sections~\ref{MCSamplingProblems} and \ref{EllipticityFluctuationSec}.

\subsection{Note on the one-body density and two-body correlation and Monte~Carlo event sampling}\label{MCSamplingProblems}

Short-range correlation changes the one-body distribution. For example,  nucleons with a hard central core will be pushed into the outer regions, altering the one-body density assumed by the WS distribution. The two-body density is given by
\begin{equation}\label{TwoBodyDensity}
	\begin{split}
		&p_A^{(2)}(\vx_1,\vx_2) = \\
		&\frac{\rho_{\text{WS}}(\vx_1) \rho_{\text{WS}}(\vx_2) \Theta(|\vx_1-\vx_2| - d_\mi) }{\int d\vx'_1 d\vx'_2 \rho_{\text{WS}}(\vx'_1) \rho_{\text{WS}}(\vx'_2) \Theta(|\vx'_1-\vx'_2| - d_\mi)},
	\end{split}
\end{equation}
where $\Theta(x)$ is the Heaviside step-function. It is important to note that $p_A(\vx_1) = \int d\vx_2 p_A^{(2)}(\vx_1,\vx_2)$ is not coincident with the WS distribution, although we might approximate it with the WS distribution with modified $R_0$ and $a_0$ at small values of $d_\mi$ (see section.~\ref{radiusRenorm}).  The two-body correlation also deviates from the step-function as discussed in  Ref.~\cite{Luzum:2023gwy}.

To overcome the problem of one-body density modification,  in \trento{} event generator, the radial direction of the sources is sampled first based on WS distribution to keep the radial size of the nucleus fixed. Afterward, the spherical angles of the sources are sampled iteratively. Sources are sorted in the radial direction and moving from smaller radii, the $i$th source is placed at $\vx_i$ if its distance from all other $i-1$ sources is greater than $d_\mi$, otherwise the $i$th source angel are rejected and resample the angles until it is placed somewhere in the space.  Although the pre-sampling of the radial direction is performed to keep the WS structure under control, it systematically leads to inaccuracy in placing the nucleons with minimum distance. Considering two nucleons are sampled at $r_1 $ and $ r_2$ radial direction, the maximum distance of these two nucleons is $|r_1-r_2|$, meaning if this distance is smaller than $d_\mi$, it is impossible to place the second source at any given spherical angles. In \trento{} event generator, the angles are sampled 1000 times and if it fails, the source leaves it at its last iteration. The failure percentage is negligible for small values of $d_\mi$, but this systematic inconsistency should be considered as a hint that this method may lead to an inaccurate two-body correlation. More importantly, for deformed nuclei, the nucleus radius depends on the orientation, and the method employed by \trento{} event generator is no longer applicable.

There is also a caveat in placing the sources iteratively.
One may assume that at each iteration, all three directions are generated based on the WS distribution, and the minimum distances with former generated sources are examined iteratively. However, in the iterative method starting from WS with no information from short-range behavior, the $i$th source generated in the $i$th iteration is only correlated with all $(i-1)$ sources generated in the previous iterations. The first source, in fact, contains no information from short-range correlation. On the other hand, if we start from $p_A(\vx)$ and $c_2(\vx_1,\vx_2)$, the information of the correlation is already encoded in the sampling from the first iteration. We have examined this claim in a 1D toy model with only two sources, where each source is generated based on a Gaussian distribution, and the correlation is imposed using minimum distance. We can compare our toy Monte Carlo with analytical calculations using this simplified toy model. The study mentioned in the appendix~\ref{1DToyModel} suggests that the iterative method works correctly only if we use the true one-body density that contains the short-range effect information. This indicates that in more realistic cases, the iterative method works well if we use the one-body density and two-body correlation extracted from Eq.~\eqref{TwoBodyDensity}. Short-range correlations can affect the radius of spherically symmetric nuclei. In the \trento{} model, the radius is kept fixed, so one considers that the short-range correlation is accounted for in this method, although pre-generating the radius can cause issues with source placement and not applicable for deformed nuclei.

\begin{figure}
	\begin{center}
		\includegraphics[width=0.40\textwidth]{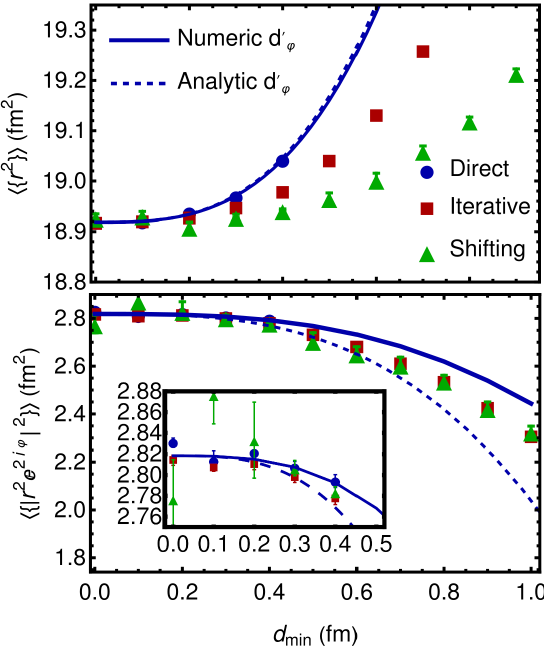}
		\caption{Comparing three different methods of source sampling of point-like sources in one Au nucleus at the presence of short-range correlation.}
		\label{fig:randomGenerator}
	\end{center}
\end{figure}

An alternative way of generating sources using WS distribution and minimum distance among nucleons is to generate $A$ random sources based on WS and then compare all the mutual distances. If there is any pair with condition $|\vx_1-\vx_2| < d_\mi$, the whole set of $A$ sources should be discarded, and a new set is generated. We call this method as direct method. In this way, we find consistent one-body density (which is different from the original WS) and two-body correlation. There is no problem with the position of the source when using this method, and it can be used for any deformed nuclei. However, it should be noted that this method takes longer to execute than the iterative method, especially for larger values of $d_\mi$. In  Ref.~\cite{Alvioli:2009ab}, the sources with appropriate two-body correlation are generated in a box using Metropolis random search, and then the WS probability distribution is imposed to choose those sources relevant to the one-body density. This method has been employed by SMASH initial state model in Ref.~\cite{Hammelmann:2019vwd}. We have examined a similar strategy in our 1D toy model by generating sources with some minimum distances and then imposing the Gaussian constraint. The result is coincident with direct method and leads to correct one-body density and two-body correlation. There is also a recent method (we call it as shifting method) in imposing the two-body correlations by generating independent sources first based on WS distribution and then shifting away the sources using a carefully constructed function based on the two-body correlation~\cite{Luzum:2023gwy}.

In Fig.~\ref{fig:randomGenerator}, we compare three different methods: direct method, iterative method, and the shifting method, to appreciate their differences. The iterative method generates all three coordinates of the sources, not just the angular part, which makes it different from the method implemented in \trento{}. To eliminate the complications arising from entropy production and nucleon participation, we consider only one Au nucleus to generate point-like sources and impose a two-body correlation using a step-function. We calculate $\la \{ r^2 \}_{s,A} \ra$ and $ \la \left|\{ r^2 e^{i2\varphi}  \}_{s,A} \right|^2  \ra $ for different values of $d_\mi$. In the direct method, we did not go beyond $d_\mi = 0.4\,$[fm] due to the inefficiency of the direct method. The figure shows that all three sets of events give the same values when there are zero short-range correlations. However, for $d_\mi > 0$, the main difference appears in $\la \{ r^2 \}_{s,A} \ra$, which quantifies the size of the one-body density. Short-range correlation modifies the one-body density, but the modification from the three methods is different. The dashed (solid) lines are analytical (numerical) estimations of $d'_\varphi$.  The cluster expansion calculation is compatible with direct method of source sampling. The difference in $ \la \left|\{ r^2 e^{i2\varphi}  \}_{s,A} \right|^2  \ra $ is less pronounced. The direct method and cluster expansion calculation rely on using consistent one-body density and two-body correlation. Iteratively sampling sources without considering the short-range correlation into the one-body density can lead to inconsistency. The authors still need to investigate in greater detail the observed difference between the direct method and the shifting method.

\subsection{Radius and skin thickness redefinition}\label{radiusRenorm}

As seen from Eq.~\eqref{StepFubcCorrA}, the short-range correlation modifies the size of the nucleus as expected. The physical values of the radius and skin thickness contain the effect of the short-range correlation. For that reason, we redefine ``bare'' values of $\tilde{R}_0$ and $\tilde{a}_0$ to absorb the term corresponding to the correlation in $\left \la \{r^2\}_{s,A} \right\ra$ and write the final result in terms of ``physical'' values $R_0$ and $a_0$. Using Eq.~\eqref{StepFubcCorrA}, we cannot fix both radius and skin thickness. We assume two cases: first, the skin thickness is fixed, $\tilde{a}_0 = a_0$, but the radius is changed, and second, the radius is fixed, $\tilde{R}_0 = R_0$, but the skin thickness is modified. For the former case, we find the value of $\tilde{R}_0$ in terms of physical quantities $R_0$ as follows:
\begin{equation}\label{R0Renormalization}
\begin{split}
		\tilde{R}_0^2 &= R_0^2 \left[1-A\frac{7\pi^2}{3}\left(\frac{a_0}{R_0}\right)^2\left(\frac{d_\mi}{R_0}\right)^3+\cdots\right],
\end{split}
\end{equation}
while $\tilde{a}_0$ for the latter reads as
\begin{equation}\label{a0Renormalization}
	\begin{split}
		\tilde{a}_0 &= a_0\left[ 1- \frac{A}{2} \left(\frac{d_\mi}{R_0}\right)^3 +\cdots \right].
	\end{split}
\end{equation}

\begin{figure}
	\centering
	\includegraphics[width=0.45\textwidth]{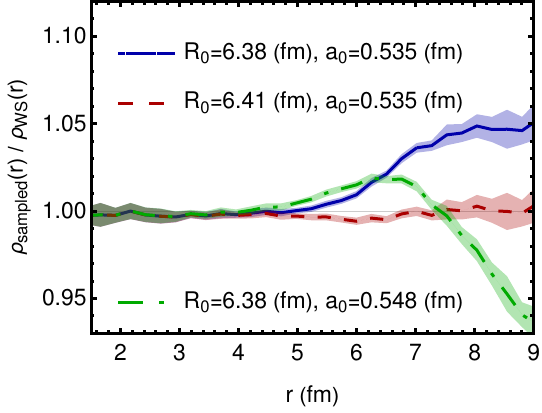}
	\caption{{The Au nucleus density sampled with $R_0 = 6.38\,$[fm], $a_0=0.535\,$[fm], and $d_\mi = 0.4$[fm] over WS distribution with the same radius and skin thickness (blue), larger radius fixed skin thickness (red, dashed), fixed radius larger skin thickness (green, dot-dashed). The sampled density is close to WS with a larger radius but fixed skin thickness. } }
	\label{fig:SampledDensityOverWS}
\end{figure}

In Fig.~\ref {fig:SampledDensityOverWS}, we show the modification of WS distribution in the presence of correlation. Here, we sampled a 200\,k Au nucleus with $\tilde{R}_0 = 6.38\,$[fm], $\tilde{a}_0=0.535\,$[fm], and $d_\mi = 0.4\,$[fm]. We divided the radial direction in the transverse direction into 50 bins and calculated the average number of sources in each bin. This average represents the density of the nucleus in the presence of the correlation. Then, we compared the density from the sampled sources with the original WS distribution. As seen from the blue curve in the figure, the density from the sampled sources differs from the WS distribution, which has the same radius and skin thickness. The density ratio goes above unity at the tail of the distribution, indicating that sources are pushed away due to the short-range correlation. We also calculated $R_0$ and $a_0$ from Eqs.~\eqref{R0Renormalization} and \eqref{a0Renormalization} and sampled distribution using redefined values. We observe that WS with a modified radius and fixed skin thickness can explain the new distribution while keeping $R_0$ fixed, but modifying $a_0$ does not match the sampling. In the present study, we follow the first strategy and modify the radius such that the final nucleus radius matches the physical radius.

\subsection{Constituent weight fluctuation}\label{subsec:constituteWeight}

In addition to the fluctuation in the position of the sources, the contribution of each source to the initial entropy density can fluctuate. Ref.~\cite{Bozek:2013uha} discusses that the final multiplicity fluctuation in p--Pb collision can be described by assuming that the participant contribution to the initial entropy density fluctuation follows a Gamma distribution,  
\begin{equation}\label{GammaDist}
	p_\fluc(\gamma)=\frac{k^k}{\Gamma(k)} \gamma^{k-1} e^{-k \gamma},\qquad \sigma_{\text{fluc}}=1/\sqrt{k}.
\end{equation}
To take the effect of fluctuation into account, we replace $p_A(\vx_i)$ with $p_\fluc(\gamma_i) p_A(\vx_i)$ and modify the auxiliary function to $e^{\frac{1}{A}\sum_a\lambda_a \gamma_i f_a(\vx_i)}$ in Eq.~\eqref{correlationAndDistribution}.  We have examined the contribution of $\sigma_\fluc^2 (d_\mi / R_0)^3$ in moment $\left \la \left|\{r^2 e^{2i\varphi}\}_{s,A}\right|^2 \right\ra$. This contribution is zero for the case WS distribution is replaced by step-function. Simulation based on \trento{} model also confirms that this contribution is small. Therefore, we  calculate the contribution of $\sigma_\fluc$ and $d_\mi$ separately. By ignoring the effect of two-body correlation, we find that the canonical partition function is written in terms of 1-cluster integral, $Q_A(\lambda) \propto b_1^A(\lambda)$. Employing $\partial_\lambda Q_A(\lambda) / Q_A(\lambda)|_{\lambda = 0} $ leading to      
\begin{equation}\label{SingleObsAvreave}
	\la \{ f \}_{s,A} \ra  =\int d^3 x \, f(\vx) \,  p_{A}(\vx).
\end{equation}
In the case of calculating second-order moments average,  from $\partial_{\lambda_1}\partial_{\lambda_2} Q_A(\lambda) / Q_A(\lambda)|_{\lambda_i = 0} $, we find 
\begin{equation}\label{TwoObsAvreave}
	\begin{split}
		\la \{ f_1  \}_{s,A} \{ f_2  \}_{s,A}\ra  &= \frac{1+\sigma_\fluc^2}{A}\int d^3 x \, f_1(\vx)f_2(\vx) \,  p_{A}(\vx)\\
		& + \frac{A-1}{A} \la \{ f_1  \}_{s,A}  \ra \la \{ f_2  \}_{s,A}  \ra,
	\end{split}
\end{equation}
where we have used the fact that $\int d\gamma\, \gamma \, p_\fluc(\gamma) = 1$ and $\int d\gamma\, \gamma^2 \, p_\fluc(\gamma) = 1 + \sigma_\fluc^2$. Approximating $Q_A(\lambda)$ by its grand-canonical counterpart, as discussed in section~\ref{sec:analogyStat}, would have led to replacing $A-1$ with $A$ in the second term of Eq.~\eqref{TwoObsAvreave}. This modification is negligible for large $A$ and ignored here.

To find the values of $\la \{r^2\} \ra$ and $\la \{r^2\}^2 \ra$, we need to substitute $f(\vx) = |\bx|^2$ and $f_1(\vx) = f_2(\vx) = |\bx|^2$ into Equations \eqref{SingleObsAvreave} and \eqref{TwoObsAvreave} respectively. It is evident from Equation \eqref{TwoObsAvreave} that the second term contributes the most when $A$ is large, which implies that $\la \{r^2\}^2 \ra \approx \la \{r^2\} \ra^2$. This shows that the value of $\la \{r^2\}^2 \ra$ is mainly influenced by the size and has a small contribution from fluctuation. On the other hand, to determine $\left \la \left|\{r^2 e^{2i\varphi}\}_{s,A}\right|^2 \right\ra$, we choose $f_{1}(\vx) = f_{2}^*(\vx) =|\bx|^2e^{2i\varphi}$. In this case, the second term in Equation \eqref{TwoObsAvreave} is zero, which means that the primary contribution is proportional to $(1+\sigma_{\text{fluc}}^2)/A$, indicating that it solely originates from the fluctuations.

In Equation \eqref{TwoObsAvreave}, the presence of $\sigma_\fluc^2$ arises from the calculation of second order moments. Therefore, the dependence on $\sigma_\fluc^2$ can be accurate for values greater than unity. When there is fluctuation present, we use the expression $\{f\}_s = \sum_i f(\vx_i) / N$ instead of $\{f\}_s = \sum_i \gamma_i f(\vx_i) / \sum_i \gamma_i$ to calculate many event averages. The approximation we make is that the numerator and denominator fluctuate independently event by event. We have provided a more detailed discussion in Appendix \ref{app:EbyEAndClusterExpandion}. We validated this approximation for $\sigma_{\text{fluc}}$ up to 2 using a toy Monte Carlo.

\subsection{From source fluctuation of a single nucleus to initial density fluctuation}

So far, we have calculated the impact of source fluctuation on a single nucleus, $A$. The first-order moment for combining sources from two nuclei is directly related to the single nucleus moment, $ \langle \{f\}_s \rangle = \langle \{f\}_{s,A} \rangle$. For a general second-order moment, we find 
\begin{equation}\label{ObservableDecompositionBI}
	\begin{split}	
		\la \{f_1\}_s \{f_2\}_s \ra &= \frac{1}{2}\left[ \left\la \{f_1\}_{s,A}\{f_2\}_{s,A} \right\ra + \right. \\
		& \hspace*{2cm} \left. \left\la \{f_1\}_{s,A}\right\ra \left\la \{f_2\}_{s,A} \right\ra \right]. 
	\end{split}
\end{equation}  
In addition, at a finite cross-section, there are some sources that do not participate in the interaction. This leads to a correction in the decomposition of fluctuations into independent target and projectile. Moreover, for the deformed nuclei in UCSC, an intrinsic value is introduced to the elliptic shape, and their random orientation leads to the fluctuation~\cite{Jia:2021tzt}. The details of these corrections, along with the effect of nucleon substructure with $n_c$ constituents and width $v$, are discussed in appendix \ref{appendix:extraParameters}. The final result for the initial state size is given as follows:
\begin{equation}\label{RsqFinalMoment}
	\begin{split}
		&\la \{r^2\}\ra^2 = \frac{4R_0^4\, \omega^2}{25}\Bigg[\omega^2+ \alpha_r\left(\frac{a_0}{R_0}\right)^2\\
		&\hspace*{0.0cm}+s_r\left(\frac{a_0}{R_0}\right)\left(\frac{ \pi R_0^2/A}{\sigma^{\text{NN}}_\text{inel}}\right)+b_{2,r} \beta_2^2+b_{3,r} \beta_3^2+\cdots\Bigg].
	\end{split}
\end{equation}
Accordingly, the elliptical shape is obtained as 
\begin{equation}\label{Rsq2PhiFinalMoment}
	\begin{split}
		&\left\la\left|\{r^2 e^{2i\varphi}\}\right|^2\right\ra = \frac{1}{A}\frac{4 R_0^4}{35}\left[1+\alpha_{\varphi} \left(\frac{a_0}{R_0}\right)^2\right.\\
		&\hspace*{1.7cm}\left.+s_{\varphi}\left(\frac{a_0}{R_0}\right)\left(\frac{\pi R_0^2/A}{\sigma^{\text{NN}}_\text{inel}}\right)+d_{\varphi}\left(\frac{d_\mi}{R_0}\right)^3\right.\\
		&\hspace*{1.7cm}  \left. +v_{\varphi}\frac{\sigma_\fluc^2}{n_c}+ b_{2,\varphi} \beta_2^2 +\cdots\right].
	\end{split}
\end{equation}
The coefficients are shown in table~\ref{tab:TableOfCoeff}. We note that $(w^2-v^2)/R_0^2$ is the expansion parameter, and it should be small, while $(w/R_0)^2$ is not. In fact, the estimation is more accurate for larger values of $w$.

\begin{table}[htb]
	\setlength{\extrarowheight}{2pt}
	\resizebox{0.95\columnwidth}{!}{%
		\begin{tabular}{c | c}
			coefficient & value            \\
			\hline
			$\omega^2$         &  $1+5\left( w / R_0 \right)^2$ \\
			$\alpha_r$         &  $14\pi^2/ 3$\\
			$\alpha_\varphi$   & $6\pi^2$ \\
			$s_r$              & $-6$   \\
			$s_\varphi$        & $51 / 4$  \\
			$ d_\varphi$       & $ d'_\varphi - \alpha_r A \left( a_0 / R_0 \right)^2$     \\
			$ v_\varphi$       & $ 1+10\left(w^2-v^2\right)/R_0^2+\alpha_{\varphi} \left( a_0 / R_0 \right)^2$   \\
			$  b_{2,r }$       & $ 29 / 8\pi$   \\
			$ b_{3,r }$        & $ 7 / 2\pi$    \\
			$ b_{2,\varphi }$  & $ 21\, A / 20 \pi$  \\
			$\alpha_\epsilon$  &  $\alpha_\varphi- \alpha_{r} / \omega^2 $\\ 
			$s_\epsilon$       &  $s_\varphi -  s_r / \omega^2 $\\
			$b_{2,\epsilon }$  &  $b_{2,\varphi } - b_{2,r} \omega^2 $\\
			$b_{3,\epsilon }$  &  $- b_{3,r} / \omega^2 $ 				
		\end{tabular}
	}
	\caption{Table of coefficients for $\la \{r^2\}\ra$, $\left\la\left|\{r^2 e^{2i\varphi}\}\right|^2\right\ra$, and $\epsilon_2\{2\}$ estimations. }\label{tab:TableOfCoeff}
\end{table}

\section{Ellipticity fluctuation and isobar ratio}\label{EllipticityFluctuationSec}

\begin{figure*}
	\centering
		\includegraphics[width=0.78\textwidth]{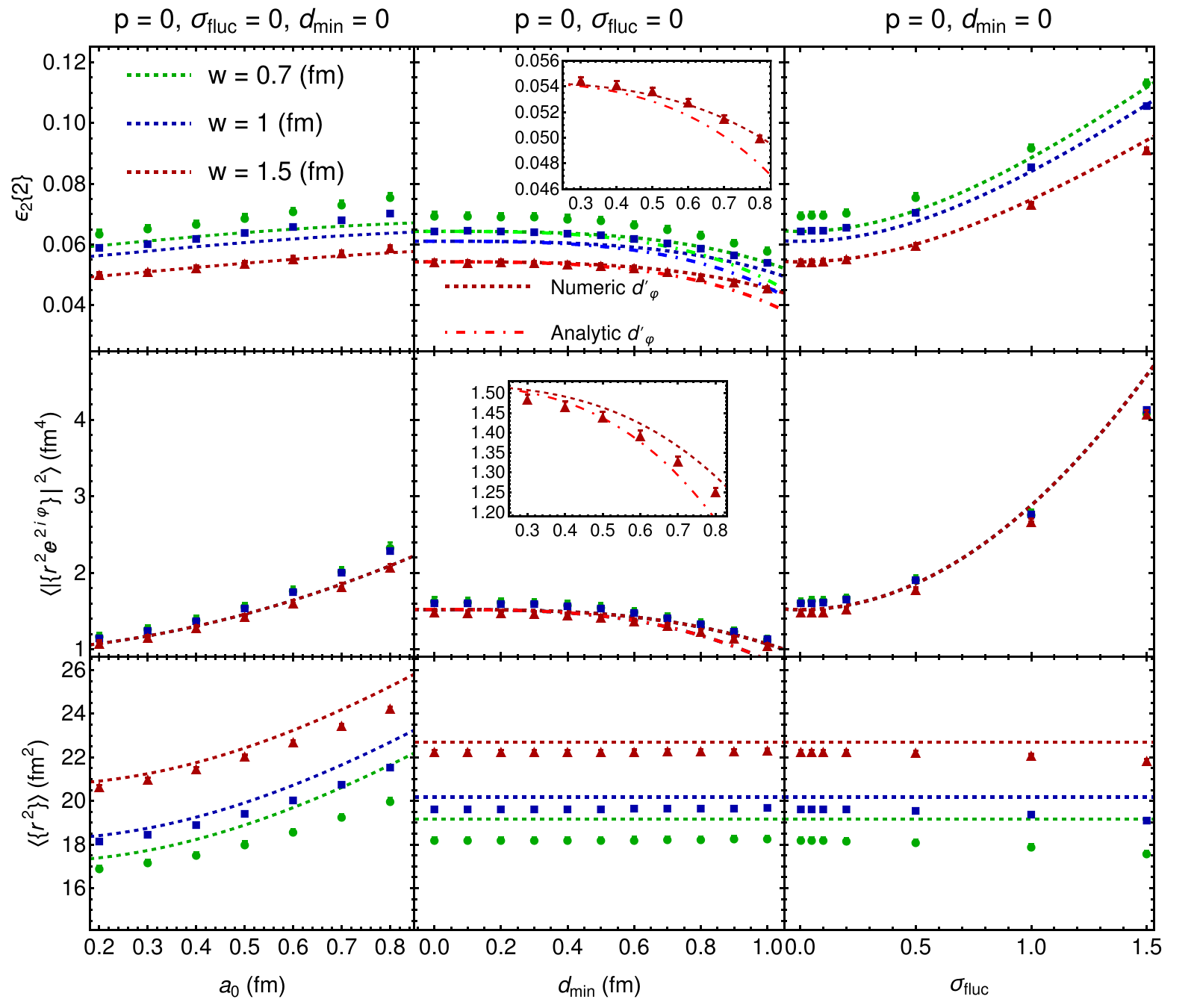} 
	\caption{ Comparing the analytical estimation (curves) for $\la \{r^2\}\ra $, $\left\la\left|\{r^2 e^{2i\varphi}\}\right|^2\right\ra$, and $\epsilon_2\{2\}$ with \trento{} model outcome (dots) for Au--Au collisions  with $\sigma_{\textbf{inel}}^{NN}=4.2\,$[fm$^2$]. The compatibility improves for larger nucleons size due to the improvement of approximation based on scale invariance. }
	\label{trentoValidation}
\end{figure*}

\textit{Ellipticity fluctuation.} In the previous section, we explicitly estimated two moments of the initial state entropy density. In this section, we will use these moments to derive an analytical estimate for the average ellipticity. Additionally, we will validate this estimation, along with the moments stated in Eqs.~\eqref{RsqFinalMoment} and \eqref{Rsq2PhiFinalMoment}, using the \trento{} model.

The quantity $\epsilon_2^2\{2\}$ can be obtained by dividing Eq.~\eqref{Rsq2PhiFinalMoment} by Eq.~\eqref{RsqFinalMoment},
\begin{equation}\label{eps2Approx}
\epsilon_2^2\{2\}=\la \epsilon_2^2\ra = \left\la\frac{\left|\{r^2 e^{2i\varphi}\}\right|^2}{ \{r^2\} ^2}\right\ra  \approx \frac{\left\la\left|\{r^2 e^{2i\varphi}\}\right|^2\right\ra}{\left\la \{r^2\}\right\ra^2 } .
\end{equation}
To justify the last approximation equality, we utilize the Power distribution, which can be expressed as $p(\epsilon_2) = (N-1) \epsilon_2(1-\epsilon_2^2)^{(N-3)/2}$ \cite{Yan:2013laa}. The Power distribution characterizes the distribution of $\epsilon_2$ for $N$ independent point-like sources that are distributed with a Gaussian profile. It is determined that $\langle \epsilon_2^2 \rangle = 2/(N+1)$ using the Power distribution. Furthermore, we can compute $\langle \{r^2\}^2 \rangle$ and $\left\langle\left|\{r^2 e^{2i\varphi}\}\right|^2\right\rangle$ for independent point-like sources distributed according to a Gaussian distribution, using the methods outlined in the preceding section. We obtain the same result for $\langle \left|\{r^2 e^{2i\varphi}\}\right|^2\rangle / \langle \{r^2\}^2\rangle$. Discrepancies arise for finite-size sources. The comparison between $\langle \epsilon_2^2 \rangle$ from the Power distribution with finite-size sources (not provided here) and $\langle \left|\{r^2 e^{2i\varphi}\}\right|^2\rangle / \langle \{r^2\}^2\rangle$ reveals a disparity on the order of $1/N^3$ which is neglected here. Hence, 
\begin{equation}\label{eps22FinalMoment}
	\begin{split}
		&\epsilon_2^2\{2\} = \frac{5}{7}\frac{1}{A\, \omega^4}\left[1+\alpha_{\epsilon} \left(\frac{a_0}{R_0}\right)^2\right.\\
		&\hspace*{1.7cm}\left.+s_{\epsilon}\left(\frac{a_0}{R_0}\right)\left(\frac{s_\perp}{\sigma^{\text{NN}}_\text{inel}}\right)+d_{\varphi}\left(\frac{d_\mi}{R_0}\right)^3\right.\\
		&\hspace*{1.7cm}  \left. +v_{\varphi}\frac{\sigma_\fluc^2}{n_c}+ b_{2,\epsilon} \beta_2^2 + b_{3,\epsilon} \beta_3^2 +\cdots\right].
	\end{split}
\end{equation}

To validate our estimate, we compare it with the realistic \trento{} model in the Au--Au collision with $\sigma_{\textbf{inel}}^{NN}=4.2\,$[fm$^2$]. The results are shown in Fig.~\ref{trentoValidation}. From the figure, we can see that our approximation improves as the nucleon width increases. However, it still provides reasonable agreement even for nucleon widths as small as $w=0.7\,$[fm]. The validation related to the substructure of nucleons and the finite value of the cross-section can be found in appendix~\ref{appendix:extraParameters}.

Based on the figure, we observe that the moment $ \la \{r^2\} \ra$ is not greatly impacted by the parameter $\sigma_{\text{fluc}}$, which aligns with our calculations. This parameter primarily affects the fluctuation, while the value of $ \la \{r^2\} \ra$ is mainly determined by the nucleus size. Additionally, the \trento{} outcome is not significantly influenced by $d_\mi$ because of the pre-generation of the radial direction of the nucleons' position, which prevents the modification of the WS radius. In our case, we accomplished this by redefining the radius. For each value of $d_\mi$, the radius of the WS distribution is adjusted to maintain a fixed $ \la \{r^2\} \ra$. Furthermore, increasing $a_0$ enlarges the total nucleus size, subsequently increasing $ \la \{r^2\} \ra$.

\begin{figure*}[t!]
	\centering
	\begin{tabular}{c}
		\includegraphics[width=0.9\textwidth]{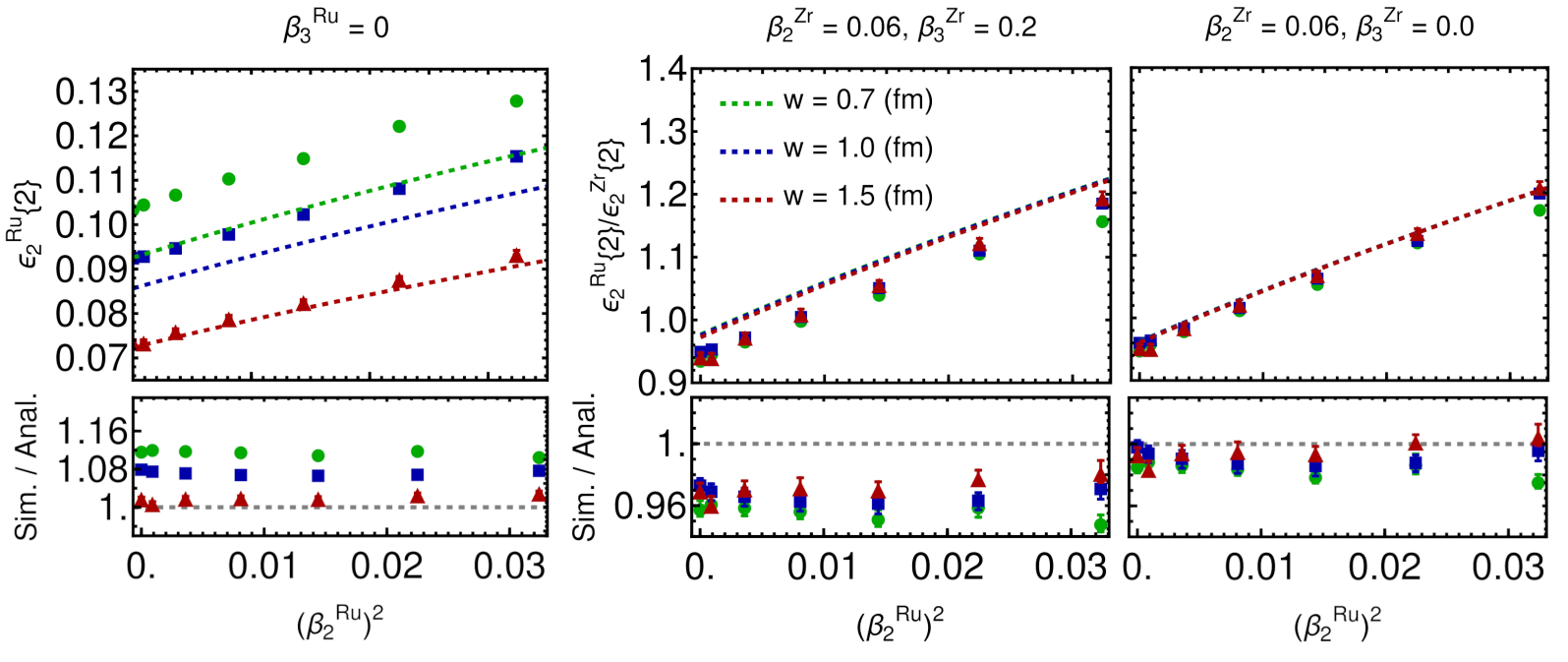}  \\ 
		\vspace*{0.2cm} 
	\end{tabular}
	\caption{Comparison of the isobar ratio between deformed nuclei Ru and Zr from \trento{} model (dots) with our analytical estimate (curves). The inaccuracy caused by the small nucleon size is eliminated when calculating the ratio. However, the impact of octupole deformation of Zr results in an inaccurate estimation of the ratio, as shown in the middle panels.}
	\label{trentoValidationRatio}
\end{figure*}

The value of the quantity $\left\la\left|\{r^2 e^{2i\varphi}\}\right|^2\right\ra$ is influenced only by the fluctuation in ultra-central Au-Au collisions. The figure illustrates that as $\sigma_\fluc$ increases, the value of the quantity also increases. The size of the system also affects this quantity by providing more room for fluctuation. Increasing $a_0$ leads to a larger system, while an opposite trend is observed as $d_\mi$ increases. This is because increasing $d_\mi$ reduces the available space for fluctuation by removing the volume of the hard inner core of the nucleons from the total volume in a nucleus.

We now discuss the behavior of $\epsilon_2\{2\}$ when we change the initial state parameters. The ellipticity average can be thought of as a normalized version of moment $\left\la\left|\{r^2 e^{2i\varphi}\}\right|^2\right\ra$. Therefore, the dependence of $\epsilon_2\{2\}$ on the parameters is mostly a competition between $\left\la\left|\{r^2 e^{2i\varphi}\}\right|^2\right\ra$ and $\left\la\{r^2\}\right\ra$. In this case, $\left\la\left|\{r^2 e^{2i\varphi}\}\right|^2\right\ra$ dominates. Our calculation shows a smaller difference compared to the \trento{} outcome since some errors are canceled out in calculating the ratio. However, inaccuracies due to small $w$ still affect $\epsilon_2\{2\}$ because $\left\la\left|\{r^2 e^{2i\varphi}\}\right|^2\right\ra$ has a smaller dependence on $w$ compared to $\left\la\{r^2\}\right\ra$. This dependence is reduced when we calculate the isobar ratio.

We have compared the dependence of $\epsilon_2\{2\}$ on $d_\mi$ between our numerical calculation of $d'_\varphi$ and the \trento{} model. We found that there is a very good agreement for $w=1.5\,$[fm]. We recall the discussion about the implementation of short-range correlation for Monte Carlo sampling in section~\ref{MCSamplingProblems}. We deduce that the pre-generation of radial direction is effectively similar to implementing the effect of short-range correlation into the WS distribution. This is because small values of $d_\mathrm{min}$ do not modify the functionality of WS much. In three dimensions, the issue for source placement is negligible, and the inaccuracy of the iterative method is small compared to what we observe in the 1D toy model shown in Fig.~\ref{1DCorrelationFig}~(right) in appendix~\ref{1DToyModel}. One may wonder if these issues become more critical in other moments.

\begin{figure*}
	\centering
	\includegraphics[width=0.7\textwidth]{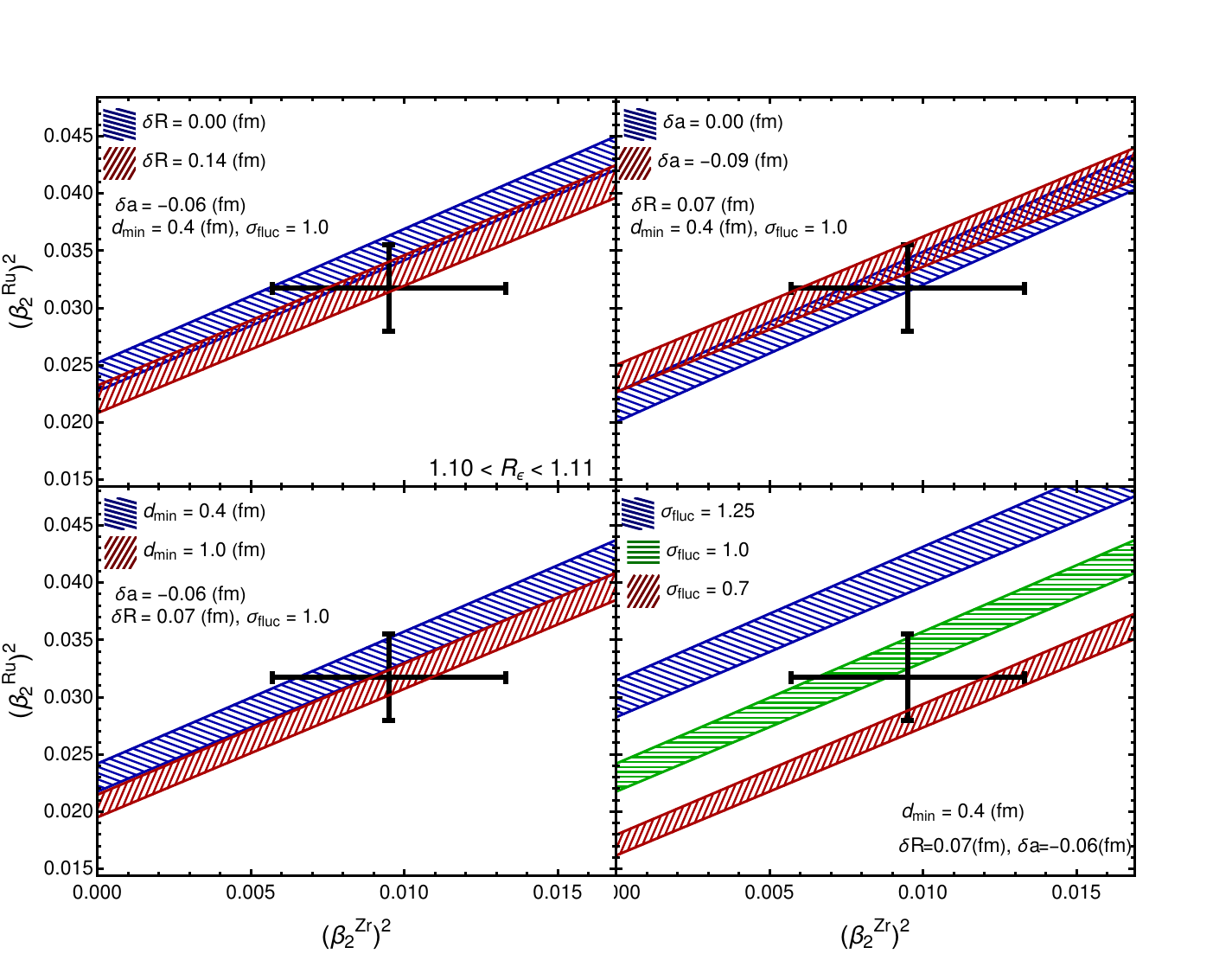}  
	\caption{ Constrainting on $(\beta_{2}^\text{Ru})^2$ and $(\beta_{2}^\text{Zr})^2$ for a given $R_\epsilon$, with different values of $d_\mi$, $\sigma_\fluc$, $\delta R$, and $\delta a$. Here, the WS parameters of Zr nucleus is chosen $R_0^{(\text{Zr})} = 5.02\,$[fm] and $a_0^{(\text{Zr})} = 0.52\,$[fm]. For Ru nucleus, parameters are given as $R_0^{(\text{Ru})} = R_0^{(\text{Zr})} +\delta R$ and $a_0^{(\text{Ru})} = a_0^{(\text{Zr})} +\delta a$. The other parameters are chosen as follows: $n_c = 1$, $w=0.7\,$[fm], and $\beta_3^{(\text{Zr})} = 0.2$. The data point shows the location of $\beta_2^\text{Zr}$ and $\beta_2^\text{Ru}$ inferred from electric transition rate~\cite{Pritychenko:2013gwa}. The results show that the most significant dependence comes form   $\sigma_{\text{fluc}}$ parameter and the least is the effect of $R_0$ and $a_0$. }
	\label{constrain2D}
\end{figure*}

\textit{Isobar ratio.}  We also investigate our approximation for the isobar ratio of deformed nuclei, where the WS parameters of two nuclei are different. Specifically, we have considered the isobar nuclei Ru and Zr. Nucleus Ru, with a radius of $R_0^{(\text{Ru})}$ and a skin thickness of $a_0^{(\text{Ru})}$, is considered to have only $\beta_{2}^{(Ru)}$. On the other hand, nucleus Zr, with $R_0^{(\text{Zr})}$ and $a_0^{(\text{Zr})}$, exhibits non-vanishing $\beta_{2}^{(\text{Zr})}$ and $\beta_3^{(\text{Zr})}$ deformations. All other parameters of the two nuclei are assumed to be the same.
The ratio between eccentricities of two nuclei is shown by $R_\epsilon$,
\begin{equation}
	R_\epsilon = \frac{\epsilon_2^{(\text{Ru})}\{2\}}{\epsilon_2^{(\text{Zr})}\{2\}}.
\end{equation}
We compare our analytical calculation with the results obtained from the \trento{} model. The comparison is shown in Fig.~\ref{trentoValidationRatio}. Our predictions of $\epsilon_2\{2\}$ underestimate the simulation for smaller values of nucleon width ($w$), but the agreement improves as we move to larger values of $w$. However, the $w$ dependence has a small contribution to the ratio calculation, both in the simulation and our analytical estimation. In the middle panel of Fig.~\ref{trentoValidationRatio}, we observe a systematic discrepancy between the simulation and the analytical estimation. In the right panel, we ignored the effect of octupole deformation of Zr and found excellent agreement. This observation suggests that our estimation of the contribution of deformation in the appendix~\ref{subsec:deform} requires improvement for nuclei that exhibit both quadrupole and octupole deformation simultaneously.

\section{Constraining the nuclear structure via isobar ratios}\label{ConstrainigSec}

It is known that in ultra-central heavy ion collisions, $\epsilon_2$ is linearly correlated with the final elliptic flow, $v_2$~\cite{Niemi:2015qia}. However, this relationship depends on the specific observable being studied and requires careful examination through hydrodynamic simulations. For example, in Ref.~\cite{Nijs:2021kvn}, the isobar ratio in both the initial and final states was calculated using the \trento{} model for the initial state and the \textit{Trajectum} hydrodynamic model for the collective expansion. For the 0-5$\%$ centrality class, it was found that
\begin{equation}
	\frac{v_2^{(\text{Ru})}\{2\}}{v_2^{(\text{Zr})}\{2\}} = k_\text{hyd} \frac{\epsilon_2^{(\text{Ru})}\{2\}}{\epsilon_2^{(\text{Zr})}\{2\}},
\end{equation}  
where $k_\text{hyd} \gtrsim 1$ is a constant. However, it should be noted that the 0-5$\%$ centrality class is not completely populated by ultra-central collisions. Therefore, it is important to examine the effect of hydrodynamic response in ultra-central collisions to estimate a more accurate value for $k_\text{hyd}$. In addition, we have found that our calculation at the presence of octupole deformation of Zr shows a systematic difference compared to simulation,  
\begin{equation}
	\frac{\epsilon_2^{(\text{Ru})}\{2\}}{\epsilon_2^{(\text{Zr})}\{2\}} = k_\text{corr} \left. \frac{\epsilon_2^{(\text{Ru})}\{2\}}{\epsilon_2^{(\text{Zr})}\{2\}}\right|_{\text{Analytic}},
\end{equation}   
where $ k_\text{corr} \approx 0.96$, in our particular example.  In Ref.~\cite{STAR:2021mii}, it was reported that the value of $v_2^{(\text{Ru})}\{2\} / v_2^{(\text{Zr})}\{2\}$ in the 0-5$\%$ centrality class is $1.0247\pm 0.0017$. Our calculation is more accurate for events in a smaller width of the centrality class (less than 1$\%$), in which a larger value for the ratio is expected~\cite{Zhang:2022fou}. Here, we assume $k_\text{corr} k_\text{hyd} \approx 0.97$

In the ultra-central events, the primary contribution to $R_\epsilon$ arises from the quadrupole deformation of Ru and Zr, denoted by $(\beta_{2}^\text{Ru})^2 - (\beta_{2}^\text{Zr})^2$, as referenced in Ref.~\cite{Zhang:2021kxj}. To assess the impact of various parameters in the UCSC (ultra-central collisions), we modify the parameters while keeping the value of $R_\epsilon$ constant. In Fig.~\ref{constrain2D}, we demonstrate $(\beta_{2}^\text{Ru})^2$ against $(\beta_{2}^\text{Zr})^2$ for a fixed $R_\epsilon$. The values of the parameters $d_\mi$, $\sigma_\fluc$, $\delta R = R_0^\text{Ru} - R_0^\text{Zr}$, and  $\delta a = a_0^\text{Ru} - a_0^\text{Zr}$ are presented in the panels of the figure. The following parameters are held constant in the plot: $R_0^{(\text{Zr})} = 5.02\,$[fm], $a_0^{(\text{Zr})} = 0.52\,$[fm], $\beta_3^{(\text{Zr})}  = 0.2$, $n_c  = 1$, and $w = 0.7$[fm]. The data point in the figure denotes the positions of $\beta_2^\text{Zr} $ and $\beta_2^\text{Ru}  $ inferred from electric transition rate~\cite{Pritychenko:2013gwa}. It is important to note that the inferred deformation values from transition rates are not directly comparable with those obtained from heavy-ion collisions. This depiction is presented to compare the model's sensitivity with the measurement accuracy. It is evident that the most influential parameters are $\sigma_{\text{fluc}}$ and $d_\mi$.  The effect of difference in the radius is in the order of $d_\mi$ only for large values of $\delta R$.

\section{Summary}\label{SummarySec}

In this paper, we demonstrated that the functionality of the reduced thickness function has minimal effect on the moments of initial state entropy density in UCSC when the nucleon width is large enough, and the initial state has scale-invariance property. We validated this argument by studying a few initial state moments using the \trento{} Monte Carlo event generator (see Figs.~\ref{ConvergencTestTrento} and \ref{ConvergencTestTrentoII}).

In our study on UCSC, we utilized the cluster expansion method to determine the impact of the one-body density and two-body correlation of the nucleus on moments  $\left\la\left|\{r^2 e^{2i\varphi}\}\right|^2\right\ra$, $\left\la \{r^2\}^2\right\ra $, and $\epsilon_2\{2\}$. To carry out these calculations, we employed the WS distribution for the distribution of nucleons and hard inner core for nucleons to enforce correlation among them.  Our findings indicate that even though the minimum distance between nucleons modifies the one-body density, it is still possible to describe this density with WS, with a larger radius up to at least $d_\mi = 0.4\,$[fm].  We also compared different methods of Monte Carlo sampling of sources in the presence of short-range correlations and demonstrated that our approach can be used to assess the accuracy of short-range correlations sampling. This revealed that incorporating information on short-range correlations in the one-body density is essential for obtaining accurate results.    We calculated the effect of cross-section and nucleon substructure. To validate, we compared results with  \trento{} model outcome. 

We investigated the impact of initial state parameters on the elliptic flow ratio for Ru--Ru and Zr--Zr in ultra-central collisions. Our findings confirmed that the quadrupole deformation of Ru and Zr has the most significant contribution. We also found that the effect of constituent weight fluctuation and two-body correlation must be taken into account when inferring the nuclei deformation. 

Although we compared our calculation with the \trento{} model in this study, our result should be valid for any other initial state models that satisfy the scale-invariance.
\section*{Acknowledgment}

The authors are thankful to Jiangyong Jia for discussion and comment, we also thank Matthew Luzum, and Jean-Yves Ollitrault for their helpful discussions. S.F.T. is supported by the Deutsche Forschungsgemeinschaft (DFG) through the grant TA 1975/1-1.

\appendix

\begin{center}
	\textbf{APPENDICES}
\end{center}

\section{Convergence of triangularity, symmetric cumulants and symmetry plane correlation}\label{moreMoments}

\begin{figure*}
	\begin{center}
		\includegraphics[width=0.89\textwidth]{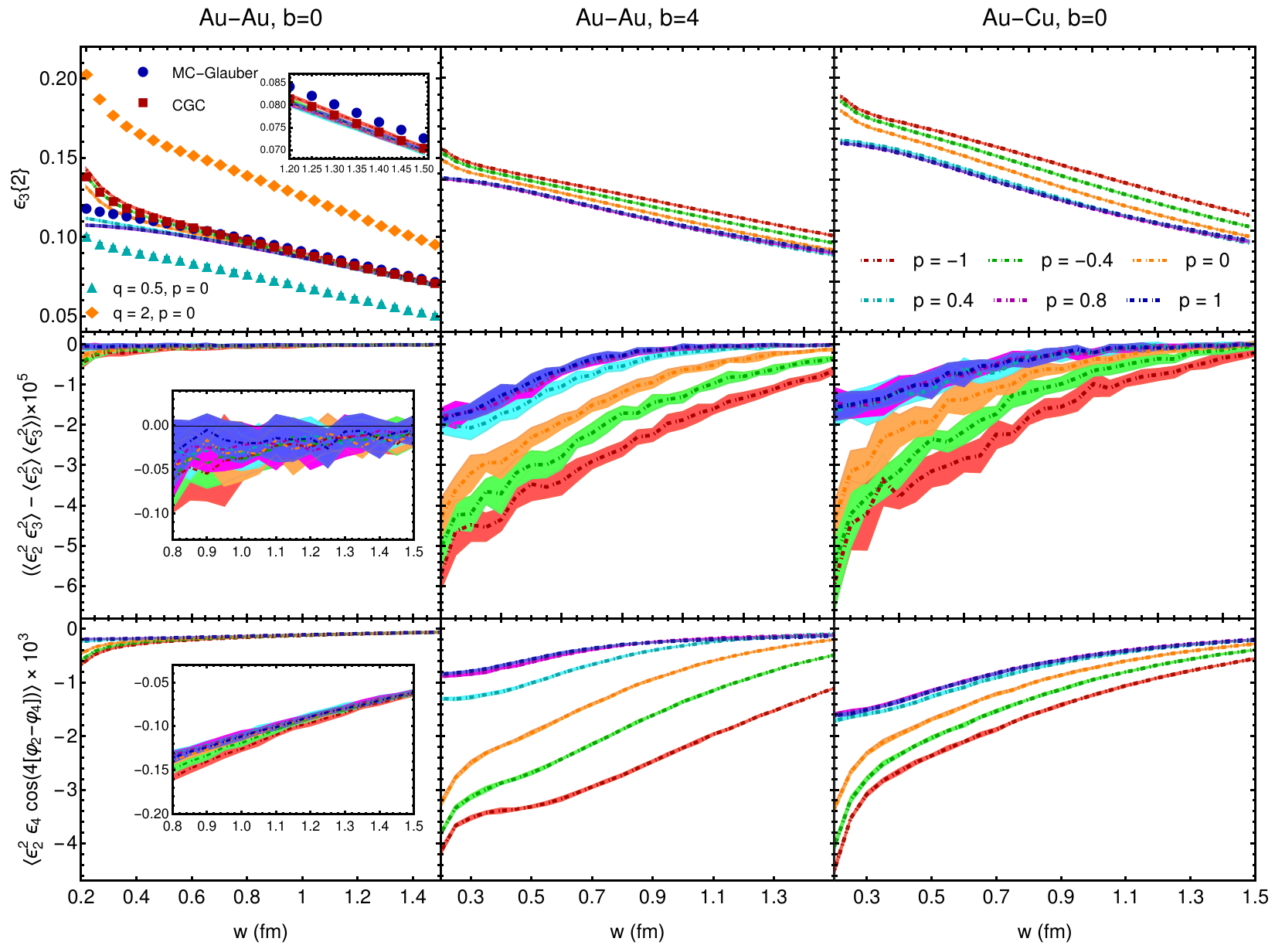}
		\caption{More moments similar to those shown in Fig.~\ref{ConvergencTestTrento}. For symmetric cumulant, convergence seems to occur even in non-central and asymmetric collisions.}
		\label{ConvergencTestTrentoII}
	\end{center}
\end{figure*}

We examine three additional initial state moments (cumulants): $\epsilon_3\{2\}$,  $\la \epsilon_2^2\epsilon_3^2\ra - \la \epsilon_2^2\ra \la\epsilon_3^2\ra$, and $\la \epsilon_2^2 \epsilon_4 \cos \left(4\varphi_2-4\varphi_4\right)\ra$. These moments exhibit the same convergence behavior as discussed in Section \ref{scaleInvSec} and are depicted in Figure \ref{ConvergencTestTrentoII}. The symmetric cumulant $\la \epsilon_2^2\epsilon_3^2\ra - \la \epsilon_2^2\ra \la\epsilon_3^2\ra$ and the symmetric plane correlation $\la \epsilon_2^2 \epsilon_4 \cos \left(4\varphi_2-4\varphi_4\right)\ra$ show significantly larger values at non-central and asymmetric collisions. In UCSC, convergence occurs at the same order of $w$ as the other moments discussed in Section \ref{scaleInvSec}.

\section{Effect of nucleon size and impact parameter on \trento{} reduced thickness function}\label{SingleShotTR}

\begin{figure}
	\includegraphics[width=0.5\textwidth]{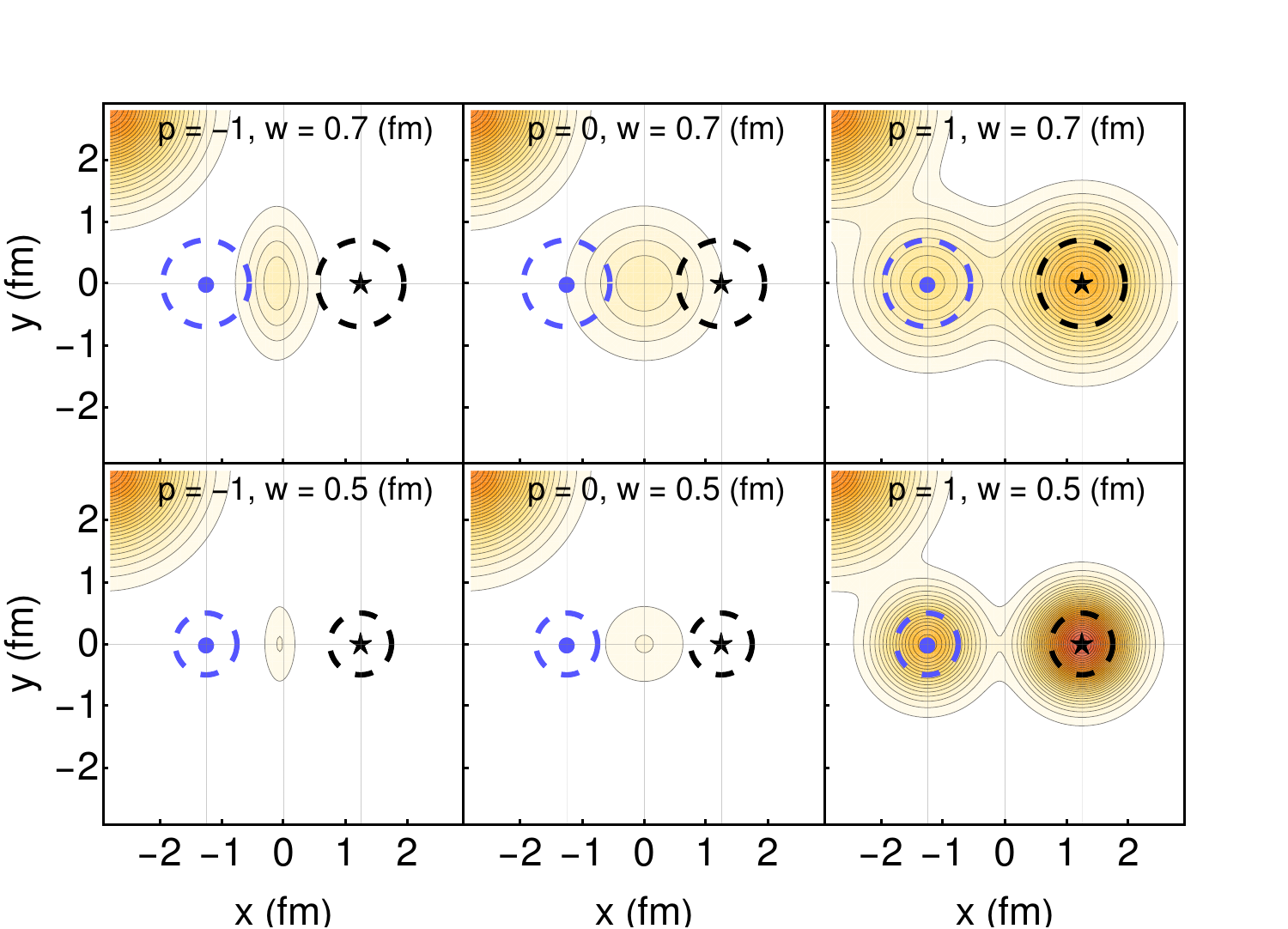} 
	\caption{The isolated participants in \trento{} reduced thickness function for $p=-1,0,1$. }\label{fig1}
\end{figure}

\textit{Isolated participants.} We start with analyzing an idealized scenario. In Fig.~\ref{fig1}, we depict one isolated participant from each target and projectile. Dashed blue and black circles show the location of these participants in the transverse space, and the radius of the circles indicates the width $w$. Here, we assume two nucleons are separated by a distance $d$ along the $x$ axis. Assuming these nucleons are  isolated from the rest of the target and projectile thickness functions (visualized by the shaded region on the upper left part of each panel), we can decompose $T_{A(B)}(\bx)$ as
\begin{equation}
	T_{A(B)}(\bx) = \overline{T}_{A(B)}(\bx)+T_{A(B)}^{\text{isol}}(\bx),
\end{equation}
such that
\begin{equation}
	\overline{T}_{A(B)}(\bx)\,T_{A(B)}^{\text{isol}}(\bx) \approx 0,
\end{equation} 
for all points in the transverse plane. This assumption leads to a decomposition in the reduced thickness function as well, 
\begin{equation}
	T_R(\bx)\approx \overline{T}_R(\bx)+T_R^{\text{isol}}(\bx).
\end{equation}
Focusing on the isolated target and projectile participants, located at $x_a= -x_b = d/2$, $y_a = y_b =0$, one finds the reduced thickness function for $p=1$,  
\begin{subequations}
	\begin{align}
		&T^{\text{isol}}_R(1;\bx)=\nonumber\\
		&\frac{\mathcal{N}}{2}\left(\gamma_t \frac{e^{-\frac{(x-d/2)^2+y^2}{2w^2}}}{2\pi w^2}+\gamma_p \frac{e^{-\frac{(x+d/2)^2+y^2}{2w^2}}}{2\pi w^2}\right),
	\end{align}
\end{subequations}
and for $p=0$,
\begin{subequations}
	\begin{align}
		&T^{\text{isol}}_R(0;\bx)=\left(\mathcal{N}\sqrt{\gamma_t\gamma_p}e^{-\frac{d^2}{8w^2}} \right) \frac{e^{-\frac{x^2+y^2}{2w^2}}}{2\pi w^2}.\label{isolP0}
	\end{align}
\end{subequations}

In Fig.~\ref{fig1}, for two distant nucleons in $p=1$, $T^{\text{isol}}_R(1;\bx)$ is the summation of two Gaussian elongated along the connecting line between the center of two nucleons (Fig.~\ref{fig1} right panel). The total entropy in this case, is a constant given by
\begin{equation}
dS^{\text{\,isol}}/d\eta = \mathcal{N}\left(\frac{\gamma_a+\gamma_b}{2}\right).\label{reducedP1}
\end{equation}
For $p=0$, the distribution is always centered at the middle of two nucleons with a round Gaussian distribution with width $w$ even in case $\gamma_t \neq \gamma_p$ as it is shown in Fig.~\ref{fig1} middle panel. The total entropy reads as
\begin{equation}
dS^{\text{\,isol}}/d\eta=\mathcal{N}\sqrt{\gamma_a \gamma_b}\, \exp\left[-\frac{d^2}{8 w^2}\right].\label{reducedP0}
\end{equation}
The exponential term leads to the fact that the more distant the two sources are, the less contribution the isolated part delivers to the total entropy. A similar analysis can be done for $p=-1$. The distribution is elongated perpendicular to the connecting line as it is seen from Fig.~\ref{fig1} left panel. For small values of $w$, the chance of finding isolated participants increases.

\textit{Moments of $T_R(1;\bx)$ for arbitrary value of $w$. } In more realistic scenarios, finding isolated sources similar to the previous part is unlikely unless we have $w \ll \bar{d}$ where $\bar{d}^2$ stands for the inverse source density in the transverse space,
\begin{equation}
	\bar{d}^2 \sim \pi R_0^2 / N_{A(B)}.
\end{equation}
Assume a typical event with $N_A$ and $N_B$ number of sources at $\{\bx_{a,i}\}$ and $\{\bx_{b,i}\}$ in the transverse space. For $p = 1$, at each point, $\bx_i \in \{ \bx_{a,i}\} \cup\{\bx_{b,i} \}$,  a Gaussian distribution is located. Hence, 
\begin{equation}\label{p1ReducedThick}
T_R(1;\bx)=\frac{1}{S(1)}\frac{1}{2}\sum_{i=1}^{N_A+N_B}\frac{1}{2\pi w^2}\exp\left[-\frac{|\bx-\bx_i|^2}{2w^2}\right],
\end{equation}
where we ignore the effect of normalization fluctuation $\gamma_i$ from now on. Here, the total entropy is given by
\begin{equation}
	dS/d\eta = \frac{1}{2}\sum_{i=1}^{N_A+N_B} 1 = \frac{N_A+N_B}{2}.
\end{equation}

The characteristic function can be calculated from Eq.~\eqref{p1ReducedThick}, 
\begin{equation}
	\begin{split}
		\{ e^{i\boldsymbol{k}\cdot\bx} \} &=\frac{1}{2}\sum_{i=1}^{N_A+N_B}\frac{1}{2\pi w^2}  \exp\left[i\boldsymbol{k}\cdot\bx-\frac{|\bx-\bx_i|^2}{2w^2}\right]  \\
		&=e^{-\frac{k^2 w^2 }{2}}\frac{1}{2}\sum_{i=1}^{N_A+N_B} \, e^{i\boldsymbol{k}\cdot\bx_i} \\
		&=e^{-\frac{k^2 w^2 }{2}} \left. \{ e^{i\boldsymbol{k}\cdot\bx} \} \right|_{w\to0}.
	\end{split}
\end{equation}
From above, we find that
\begin{subequations}
	\begin{align}
		&\{x^2 + y^2\} = 2w^2 + \left.\{x^2 + y^2\}\right|_{w=0}, \\
		&\{y^2 - x^2\} = \left.\{x^2 + y^2\}\right|_{w=0},\\
		&\{2 x y\} = \left.\{2 x y\}\right|_{w=0}.
	\end{align}
\end{subequations}
This means that for $p = 1$, we can assume the nucleons are point-like sources, and the effect of nucleon width can be included trivially. The average of the point-like source second-order moments are given by
\begin{subequations}
	\begin{align}
		&\left.\{x^2 + y^2\}\right|_{w=0}= \\
		&\hspace*{0.8cm}\frac{N_A}{N_A+N_B}  \{x_{a}^2 + y_{a}^2\}_s  + \frac{N_B}{N_A+N_B}  \{x_{b}^2+y_b^2\}_s , \nonumber\\
		&\left.\{y^2 -x^2\}\right|_{w=0} = \\
		&\hspace*{0.8cm}\frac{N_A}{N_A+N_B} \{ y_{a}^2 -x_{a}^2 \}_s + \frac{N_B}{N_A+N_B} \{y_{b}^2-x_b^2\}_s,	\nonumber\\
		&\left.\{2 x y\}\right|_{w=0} = \\
		&\hspace*{0.8cm}\frac{N_A}{N_A+N_B}  \{2 x_{a}y_{a}\}_s + \frac{N_B}{N_A+N_B}  \{2x_{b} y_b\}_s,\nonumber		
	\end{align}
\end{subequations}
where we have used the following notation:
\begin{equation}\label{p1AverageA}
	\{ \cdots \}_s  =\frac{1}{N_{A(B)}} \sum_{i=1}^{N_{A(B)}} \cdots.
\end{equation}

\begin{table*}[t]
	\centering
	\renewcommand{\arraystretch}{2}
	\begin{tabular}{| c | c | c | c |}
		\hline
		   \multicolumn{2}{|c|}{ } &  $w \ll \bar{d}$ & $w \gg \bar{d}$ \\
		\hline
		\hline
		 $p=1$  &$S$  & $(N_A+N_B)/2$  &  $(N_A+N_B)/2$  \\
		 & $ \{x^2 + y^2\} $  & $\frac{N_A}{N_A+N_B}  \{x_{a}^2 + y_{a}^2\}_s  + \frac{N_B}{N_A+N_B}  \{x_{b}^2+y_b^2\}_s+2w^2$ & $\frac{N_A}{N_A+N_B}  \{x_{a}^2 + y_{a}^2\}_s  + \frac{N_B}{N_A+N_B}  \{x_{b}^2+y_b^2\}_s+2w^2$ \\
		 &$ \{y^2 - x^2\} $  & $\frac{N_A}{N_A+N_B} \{ y_{a}^2 -x_{a}^2 \}_s + \frac{N_B}{N_A+N_B} \{y_{b}^2-x_b^2\}_s$ & $ \frac{N_A}{N_A+N_B} \{ y_{a}^2 -x_{a}^2 \}_s + \frac{N_B}{N_A+N_B} \{y_{b}^2-x_b^2\}_s $ \\		
		 &$ \{2 x y\} $  & $ \frac{N_A}{N_A+N_B}  \{2 x_{a}y_{a}\}_s + \frac{N_B}{N_A+N_B}  \{2x_{b} y_b\}_s $ & $ \frac{N_A}{N_A+N_B}  \{2 x_{a}y_{a}\}_s + \frac{N_B}{N_A+N_B}  \{2x_{b} y_b\}_s $ \\
		\hline
		 $p=0$&$S$  & $  N_A\,N_B \{ \exp\left[-d_{ij}^2/8w^2\right] \}_{ij,s} $  &  $ \sqrt{N_A\,N_B}$ \\		
		 & $ \{x^2 + y^2\} $  &  $ \left \{(x_{ij}^2+y_{ij}^2) e^{-\frac{d_{ij}^2}{8w^2}} \right \}_{ij,s} / \left  \{e^{-\frac{d_{ij}^2}{8w^2}} \right  \}_{ij,s}  + 2w^2$ & $ \frac{1}{2}\Big( \{x_{a}^2 + y_{a}^2\}_s +  \{x_{b}^2 + y_{b}^2\}_s  +2   \{x_{a}\}_s   \{x_{b}\}_s  +2   \{y_{a}\}_s   \{y_{b}\}_s   \Big)  + 2w^2  $\\
		 & $ \{y^2 -x^2 \} $  &  $   \left \{(y_{ij}^2 - x_{ij}^2 ) e^{-\frac{d_{ij}^2}{8w^2}} \right \}_{ij,s}  /  \left  \{e^{-\frac{d_{ij}^2}{8w^2}} \right  \}_{ij,s}  $ & $  \frac{1}{2} \Big( \{y_{a}^2- x_{a}^2\}_s +  \{y_{b}^2 - x_{b}^2\}_s    +2  \{y_{a}\}_s   \{y_{b}\}_s  -2  \{x_{a}\}_s   \{x_{b}\}_s    \Big) $ \\		
		 & $ \{2 x y\} $   & $ \left \{( 2 x_{ij}  y_{ij}  ) e^{-\frac{d_{ij}^2}{8w^2}} \right \}_{ij,s}   / \left  \{e^{-\frac{d_{ij}^2}{8w^2}} \right  \}_{ij,s} $ & $ \{x_{a} y_{a}\}_s    +   \{x_{b} y_{b}\}_s  
		 + \{x_{a}\}_s  \{y_{b}\}_s  +\{y_{a}\}_s   \{x_{b}\}_s$ \\		
		\hline
	\end{tabular}
	\caption{Total entropy and second order moments of reduced thickness function for $p=0$ and $p=1$ in large and small $w$.  }
	\label{tab:1}
\end{table*}

\textit{Limit of  $T_R(0;\bx)$ for small and large $w$. }
 For $p=0$, we start with rewriting the reduced thickness function as follows,
\begin{equation}\label{fullP0}
\begin{split}
&T_R(0;\bx)=\\
&\frac{1}{S}\left[\sum_{i=1}^{N_A} \sum_{j=1}^{N_B}\left( \frac{\lambda_{ij}}{2\pi w^2} \exp\left[-\frac{|\bx-\bx_{ij}|^2}{2w^2}\right]\right)^2\right]^{1/2},
\end{split}
\end{equation}
where $\lambda_{ij}(w) = \exp\left[-\frac{d_{ij}^2}{8w^2}\right]$, and 
\begin{equation}\label{p0CenterDef}
\bx_{ij}= \frac{\bx_{a,i}+\bx_{b,j}}{2},\qquad  d_{ij} = |\bx_{a,i}-\bx_{b,j}|.
\end{equation}
This way of writing is inspired by Eq.~\eqref{isolP0}, which indicates that at $p=0$, there is effectively a source at the center of a target and projectile sources weighted exponentially by their distance. However, due to the power of $1/2$, we cannot find a full analytical result for the characteristic function. For that reason, we find an analytical estimation only in two limits, small and large $w$.

In the $w\ll \bar{d}$ limit, as we argued for a single pair of isolated participants, we employ the fact that
\begin{equation}\label{decompoistion}
(\alpha^2_1(\bx)+\alpha^2_2(\bx)+\cdots+ \alpha^2_n(\bx))^{1/2} \approx \alpha_1(\bx)+\cdots+ \alpha_n(\bx)
\end{equation} 
if $\alpha_i(\bx) \alpha_j(\bx) \approx 0$ for all $i\neq j$ pairs at any given $\bx$.
The sources are well-separated in the limit $w \ll \bar{d}$ and fulfill this condition approximately. Using it, one can  simplify Eq.~\eqref{fullP0} as
\begin{equation}\label{p0Approx}
\begin{split}
&\left.T_R(0;\bx)\right|_{w\ll \bar{d}} = \\
&\hspace*{1cm}\frac{1}{S}  \sum_{i=1}^{N_A} \sum_{j=1}^{N_B} \frac{\lambda_{ij}(w)}{2\pi w^2} \exp\left[-\frac{|\bx-\bx_{ij}|^2}{2w^2}\right].
\end{split}
\end{equation}
The thickness function in this limit is reduced to that in case $p=1$, but this time, we have $N_A N_B$ number of sources centered in the middle of each pair of target and projectile participants shown by $\bx_{ij}$. The contribution of sources is weighted based on the distance $d_{ij}$ of the participant sources. The total entropy in $w \ll \bar{d}$ is given by
\begin{equation}\label{smallWTotalEnt}
\begin{split}
		dS/d\eta &=  \sum_{i=1}^{N_A} \sum_{j=1}^{N_B}  \exp\left[-\frac{d_{ij}^2}{8w^2}\right] \\
		& = N_A N_B \{ \exp\left[-d_{ij}^2/8w^2\right] \}_{ij,s}.
\end{split}
\end{equation}
and second order
 moments are given by
\begin{subequations}
	\begin{align}
		&\{x^2 + y^2\} = \frac{ \left \{(x_{ij}^2+y_{ij}^2) e^{-\frac{d_{ij}^2}{8w^2}} \right \}_{ij,s}}{\left  \{e^{-\frac{d_{ij}^2}{8w^2}} \right  \}_{ij,s} } + 2w^2, \\
		&\{y^2 - x^2\} =   \frac{ \left \{(y_{ij}^2 - x_{ij}^2 ) e^{-\frac{d_{ij}^2}{8w^2}} \right \}_{ij,s}}{\left  \{e^{-\frac{d_{ij}^2}{8w^2}} \right  \}_{ij,s} },  \\
		&\{ 2 x y  \} =  \left \{( 2 x_{ij}  y_{ij}  ) e^{-\frac{d_{ij}^2}{8w^2}} \right \}_{ij,s}   / \left  \{e^{-\frac{d_{ij}^2}{8w^2}} \right  \}_{ij,s}, 
		\end{align}
\end{subequations}
where
\begin{equation}\label{p0AverageA}
		\{ \cdots \}_{ij,s}  =\frac{1}{N_A N_B} \sum_{i=1}^{N_A} \sum_{j=1}^{N_B} \cdots.
\end{equation}

We can estimate the asymptotic behavior of the total entropy in the $w\gg \bar{d}$ limit. Considering $\alpha_i(\bx) =\frac{\lambda_{ij}}{2\pi w^2} e^{-\frac{|\bx-\bx_{ij}|^2}{2w^2}}$ in Eq.~\eqref{fullP0}, one assumes the total distribution inside the square root can be approximated with the square of a single radially Gaussian distribution,
\begin{equation}\label{LargeWApprox}
	\alpha^2_1(\bx)+\alpha^2_2(\bx)+\cdots+ \alpha^2_n(\bx) \approx \left[\frac{\Lambda}{ |\Sigma^2|}\exp\left[-\frac{|\bx|^2}{2 \Sigma^2(\varphi)}\right]\right]^2,
\end{equation}
where $\varphi=\arctan(y/x)$, and we have assumed the average location of sources is at the origin. Here, the normalization $|\Sigma^2|$ is obtained via
\begin{equation}
	|\Sigma^2| := \int_{0}^{2\pi}  d\varphi \,\Sigma^2(\varphi).
\end{equation}
In the present study, only the ellipticity of the initial state is studied.  In this case, we approximate the $\varphi$ dependent width $\Sigma^2(\varphi)$ as follows,
\begin{equation}
	\frac{1}{\Sigma^2(\varphi)}=\boldsymbol{\varphi}^T \cdot \boldsymbol{\Sigma}^{-1} \cdot \boldsymbol{\varphi},
\end{equation}
where
\begin{equation}
	\boldsymbol{\varphi} = \begin{pmatrix}
		\cos\varphi \\
		\sin\varphi
	\end{pmatrix},\qquad 
	\boldsymbol{\Sigma}= \begin{pmatrix}
		\Sigma_{xx}^2 & \Sigma_{xy}^2 \\
		\Sigma_{xy}^2 & \Sigma_{yy}^2
	\end{pmatrix}.
\end{equation}
This leads to the square of a 2D-Gaussian for the right-hand side of Eq.~\eqref{LargeWApprox}. By using the above estimation, we eventually find
\begin{equation}\label{p0LargeWA}
	\begin{split}
		&T_R(0;\bx) \approx \\
		& \frac{1}{S}\left( \frac{\Lambda}{2\pi \sqrt{\det \boldsymbol{\Sigma}}}\right)  \exp\left[-\frac{1}{2}\left(\bx^T \cdot \boldsymbol{\Sigma}^{-1} \cdot \bx\right)\right],
	\end{split}
\end{equation}
where
\begin{subequations}
	\begin{align}
		\Lambda^2&=\sum_{ij} \lambda_{ij}^2 ,\\
		\Sigma_{xx}^2&=w^2+\frac{2\sum_{ij} x_{ij}^2\lambda_{ij}^2}{\sum_{ij} \lambda_{ij}^2},\\
		\Sigma_{yy}^2&=w^2+\frac{2\sum_{ij} y_{ij}^2\lambda_{ij}^2}{\sum_{ij} \lambda_{ij}^2},\\
		\Sigma_{xy}^2&=\frac{2\sum_{ij} x_{ij} y_{ij}\lambda_{ij}^2}{\sum_{ij} \lambda_{ij}^2}.
	\end{align}
\end{subequations}
In the above, we have used $\sum_{ij} := \sum_{i=1}^{N_A} \sum_{j=1}^{N_B}$ as a shorthand notation. 

Noting the fact that $\lambda_{ij} \to 1 $ for larege $w$, the total entropy is given by
\begin{equation}
	dS/d\eta \approx \left[ \sum_{i=1}^{N_A} \sum_{j=1}^{N_B} 1  \right]^{1/2} = \sqrt{N_A N_B}.
\end{equation}
From Eqs.~\eqref{p1AverageA}, \eqref{p0CenterDef}, and \eqref{p0AverageA}, one finds,
\begin{equation}
	\begin{split}
		\left  \{x_{ij}^2 \right \}_{ij,s}  &= \frac{1}{4} \frac{1}{N_AN_B} \sum_{i=1}^{N_A} \sum_{j=1}^{N_B} (x_{i,a}+x_{j,b})^2\\
		&= \frac{1}{4} \frac{1}{N_AN_B} \sum_{i=1}^{N_A} \sum_{j=1}^{N_B}\left(x_{i,a}^2+x_{j,b}^2+2 x_{i,a}\,x_{j,b}\right)\\
		&= \frac{1}{4} \left( \{x_{a}^2\}_s +   \{x_{b}^2\}_s +2  \{x_{a}\}_s \{x_{b}\}_s  \right).
	\end{split}
\end{equation}
Similarly,
\begin{subequations}
	\begin{align}
		&\left  \{y_{ij}^2 \right \}_{ij,s}  = \frac{1}{4} \left( \{y_{a}^2\}_s +   \{y_{b}^2\}_s \right. \\
		&\hspace*{2.7cm}\left. +2  \{y_{a}\}_s \{y_{b}\}_s  \right),\nonumber\\
		&\left  \{ x_{ij}   y_{ij}  \right \}_{ij,s}  = \frac{1}{4} \left( \{x_{a} y_{a} \}_s +   \{ x_{b} y_{b}\}_s \right.\\
		&\hspace*{2.7cm}\left. +\{x_{a}\}_s \{y_{b}\}_s +\{x_{b}\}_s \{y_{a}\}_s \right).\nonumber
	\end{align}
\end{subequations}
Using above, second order moments of $T_R(0;\bx)$ in large $w$  limit are given as
\begin{subequations}
	\begin{align}
		&\{x^2 + y^2\} = \frac{1}{2}\Big( \{x_{a}^2 + y_{a}^2\}_s +  \{x_{b}^2 + y_{b}^2\}_s  \\
		&\hspace*{1.5cm} +2   \{x_{a}\}_s   \{x_{b}\}_s  +2   \{y_{a}\}_s   \{y_{b}\}_s   \Big)  + 2w^2, \nonumber\\
		&\{y^2 - x^2\} =  \frac{1}{2} \Big( \{y_{a}^2- x_{a}^2\}_s +  \{y_{b}^2 - x_{b}^2\}_s   \\
		&\hspace*{1.5cm} +2  \{y_{a}\}_s   \{y_{b}\}_s  -2  \{x_{a}\}_s   \{x_{b}\}_s    \Big),  \nonumber\\
		&\{ 2 x y  \} =   \{x_{a} y_{a}\}_s    +   \{x_{b} y_{b}\}_s  
		+ \{x_{a}\}_s  \{y_{b}\}_s \\
		&\hspace*{0.9cm} +\{y_{a}\}_s   \{x_{b}\}_s.\nonumber 
	\end{align}
\end{subequations}

We have summarized the result in Table.~\ref{tab:1}. Assuming $ \{x_a +x_b\}_s =  \{y_a +y_b\}_s = 0$, we have $\langle   \{x_a\}_s   \rangle \approx -\langle    \{x_b\}_s \rangle = \tilde{b}/2$ and $\langle   \{y_a\}_s   \rangle \approx -\langle  \{y_b\}_s  \rangle = 0$ for non-central collisions where $\tilde{b}$ should be proportional to the impact parameter of the event, $b$. For the central collisions of same nuclei, we have $\tilde{b} = 0$ and $N_A = N_B$, meaning the total entropy and second order moments are the same for $p=0$ and $p=1$.

\section{Event-by-event fluctuation and cluster-expansion}\label{app:EbyEAndClusterExpandion}

In this section, we discuss the details of calculations based on the cluster expansion.

\subsection{Decomposition of constituents contribution and source position}\label{DecompositionSubsection}

We start with the collision of two nuclei, each with mass number A. Assuming nucleons as structureless spherical objects, the reduced thickness function is given by
\begin{equation}\label{UCReducedThickness}
	T_R(\bx) \approx  \frac{\mathcal{N}}{2} \sum_{i=1}^{N_\pa} \gamma_i  \int dz\,\rho_{\text{nucleon}}(\vec{x}-\vec{x}_i),
\end{equation} 
where we have used Eq. (1) with $p=1$, which is similar to any initial state with the scale-invariance property at large nucleon size.
Here, the location of the nucleon is represented by $\vec{x}_i$, and the number of participants is denoted by $N_\pa$. Following Refs.\cite{Moreland:2014oya,Bozek:2013uha}, the parameter $\gamma_i$ is randomly selected from a gamma distribution that has a mean value of unity and a variance of $\sigma_{\text{fluc}}^2$. The shape of $T_R(\bx)$ can be analyzed by calculating its characteristic function,
\begin{equation}\label{characteristicFuc}
	e^{W_R(\boldsymbol{k})} = \frac{\int d\bx e^{i\boldsymbol{k}\cdot\bx} T_R(\bx)}{\int d\bx  T_R(\bx)}.
\end{equation}
The expansion coefficient of  $e^{W_R(\boldsymbol{k})}$  ($W_R(\boldsymbol{k})$) in terms of $\boldsymbol{k}$ leads to the moments (cumulants) of $T_R(\bx)$~\cite{Teaney:2010vd}. We label the averages with superscript ${(c)}$ if we refer to cumulants. 
Substituting $T_R$ from Eq.~\eqref{UCReducedThickness} into the above, we find that the characteristic function can be written as $e^{W_{\text{nucleon}}(\boldsymbol{k})}  e^{W_{\text{source}}(\boldsymbol{k})}$. For Gaussian nucleon with width $w$, one finds $W_{\text{nucleon}}(\boldsymbol{k}) = -|\boldsymbol{k}|^2 w^2 /2$. Therefore, the cumulants of reduced thickness function depends on the nucleons width trivially only in $\{r^2\}$. The non-trivial information, therefore, encoded in the position of the sources,
\begin{equation}
	e^{W_{\text{source}}(\boldsymbol{k})} = \frac{\sum_{i=1}^{N_\pa} \gamma_{i} e^{i\boldsymbol{k}\cdot\bx_i}}{\sum_{i=1}^{N_\pa} \gamma_{i} }.\label{sourcePositionCharacter}
\end{equation}

\subsection{Source fluctuation distribution and moments average}\label{sourceFlucObsSubSection}

\textit{Decomposing the sources into target and projectile participants.} The initial state event-by-event fluctuation comes from the fluctuation of the sources (nucleons, constituent). Suppose we have a distribution of the location of sources given by $P_s(\gamma_1,\vx_1,\ldots,\gamma_{N_\tot},\vx_{N_\tot})$, where $N_\tot = n_c N_\pa$, and assume large $\sigma_{\text{inel}}^\text{NN}$ such that all nucleons participate in the collision.  In this case, no matter how complex the distribution is, the scale-invariance in UCSC allows us to break it down into two separate distributions of target and projectile.
\begin{equation}\label{decompose}
	\begin{aligned}
		P_s(\gamma,\vx) =  P_{A}(\gamma,\vx) P_{B}(\gamma,\vx),
	\end{aligned}
\end{equation}
where $P_{A(B)}$ is the distribution of participant sources from the target (projectile). The correction to this decomposition is discussed in appendix.~\ref{appendix:cross-section}. The parameters $\gamma$ and $\vx$ are used as collective coordinates that run through the number of sources in the target (projectile) in $P_{A(B)}$, or their combination in $P_s$. The average over position of sources is given as follows (see Eq.~\eqref{sourcePositionCharacter}),
\begin{equation}\label{totalObs}
	\{f\}_s =  \frac{\sum_{i=1}^{N} \gamma_i f(\bx_i)}{\sum_{i=1}^{N} \gamma_i},
\end{equation}
where $f(\bx_i)$ is a given function, defining our moment.  The effect of nucleon width  appears only in the cumulants $\{r^2\}^{(c)}$.  In the present study, we are particularly interested in averages $\{r^2\}^{(c)}$ and $\{ r^2 e^{i2\varphi}  \}^{(c)}$ where due to the rotational invariance of WS at the leading order, these cumulants are equivalent with the moments $\{r^2\}$ and $\{ r^2 e^{i2\varphi}  \}$.

Assuming that the $\gamma$ variables fluctuate independently of the source location, we can calculate the average of the moments using
\begin{equation}
	\la \{f\}_s \ra = \int d^{N }\gamma \,d^{2N }x\, d^{N}z \, 	\{f\}_s \,P_s(\gamma,\vx).
\end{equation}
In particular, one finds the following relation between moments and the nucleons position averages:
\begin{subequations}\label{SourceToObs}
	\begin{align}
		&\la \{r^2\} \ra = \la \{r^2\}_s \ra + 2w^2,\label{SourceToObsA} \\
		&\la \{r^2\}^2 \ra = \left\la \left(\{r^2\}_s+2w^2\right)^2 \right\ra,\label{SourceToObsB} \\
		&\la \left|\{ r^2 e^{i2\varphi}  \} \right|^2  \ra =  \la \left| \{ r^2 e^{i2\varphi}  \}_s \right|^2 \ra. \label{SourceToObsC}
	\end{align}
\end{subequations}
For nucleons with substructure, $w$ should be replaced by $v$.
For symmetric collisions $A = B $, one can assume $\sum_{i=1}^{N_A} \gamma_i \approx \sum_{i=1+N_A}^{N_A+N_B} \gamma_i$, and decompose Eq.~\eqref{totalObs} into two parts, $\{f\}_s  \approx  \frac{1}{2} \left[\{f\}_{s,A}+\{f\}_{s,B}\right]$. Since $P_A$ and $P_B$ are same distributions in Eq.~\eqref{decompose}, one immediately deduce that 
\begin{subequations}\label{ObservableDecomposition}
	\begin{align}
		&\la \{f\}_s\ra = \la \{f\}_{s,A} \ra,\label{ObservableDecompositionA}\\	
		&\la \{f_1\}_{s} \{f_2\}_{s} \ra = \frac{1}{2}\left[ \la \{f_1\}_{s,A} \{f_2\}_{s,A} \ra + \la \{f_1\}_{s,A} \ra \la \{f_2\}_{s,A} \ra \right], \label{ObservableDecompositionB}
	\end{align}
\end{subequations}
The average on the left-hand side is taken over $P_s$, while on the right-hand side, it is taken over $P_A$, given $\OO_A$. Therefore, we only need to calculate observables for one nucleus participant. The correction due to the finite nucleon-nucleon inelastic cross-section will be discussed in appendix~\ref{appendix:cross-section}.

\textit{The effect of the constituent weight fluctuation on observables.} As mentioned, the constituent weight fluctuation is followed by the gamma distribution in Eq.~\eqref{GammaDist}. Given that fluctuations are assumed independent, referring to Eq.~\eqref{totalObs}, the the  averages such as $\la \{f\}_s \ra$ or $\la \{f_1\}_{s}  \{f_2\}_{s} \ra$ can be written as the average of numerator divided by denominator.  Using $p_\fluc(\gamma)$ to find the denominator average, one estimates that
\begin{subequations}\label{obserAndFluc}
	\begin{align}
		\la \{f\}_s  \ra &\approx   \left \la  \frac{1}{N}\sum_{i=1}^{N} \gamma_i f(\bx_i)  \right \ra, \label{observableAv}\\
		\la \{f_1\}_s  \{f_2\}_s  \ra 
		&\approx \left \la \frac{1}{N^2}\sum_{i,j=1}^{N} \gamma_i \gamma_j f_1(\bx_i) f_2(\bx_j) \right \ra.\label{observableAvb}
	\end{align}
\end{subequations}
There is an extra term proportional to $\sigma_{\text{fluc}}^2/N^3$ in Eq.~\eqref{observableAvb}. This is ignored in this study as it is a subleading contribution in the $1/N$ expansion.

\subsection{Initial state fluctuation from cluster expansion method}\label{app:CESection}

For independent nucleons without structure, the distribution of A sources is given by:
\begin{equation}\label{structureLess}
	P_{A}(\gamma,\vx)=\prod_{i=1}^{A} p_\fluc(\gamma_i) p_{A}(\vx_i),
\end{equation}
which allows us to decompose the average on the right-hand side of equation \eqref{observableAv} and perform the integration over a single source distribution. This leads to Eq.~\eqref{SingleObsAvreave}. 
In the case of calculating second-order moments average, $\la \{f_1\}_{s,A}  \{f_2\}_{s,A}  \ra$, we find Eq.~\eqref{TwoObsAvreave}
from equation \eqref{observableAvb} where we have used $\int d\gamma\,\gamma^2 \; p_\fluc(\gamma)=1+\sigma_\fluc^2$. To include the effect of short-range correlation, we employ the statistical mechanics analogy as discussed in Sec.~\ref{UCSCClusterExpansion} where the distribution is defined as $P_A(\vx) = F_A(\vx) / \mathcal{Z}_A$.

Although we usually assume a fixed number of sources, when studying fluctuation in cluster expansion, closed-form results can be obtained in grand-canonical systems. It should be noted that the calculated observables from canonical and grand-canonical systems may differ with terms in the order of $O(1/A)$. With this caveat in mind, we can define grand-canonical partition functions as follows:
\begin{equation}\label{GrandPartFubc}
	\mathscr{Q}(z,V,\lambda_a) = 
	\sum_{n=0}^\infty z^n \Q_n(V,\lambda_a),
\end{equation}
where the canonical partition function is given as $\Q_A(V,\lambda_a) = \mathcal{Z}_A(V,\lambda_a) / h^{3A}A!$. The dimensionful parameter $h$ is included to make the partition functions dimensionless. The Mayer function $f_{ij}$ is defined as $c_2(\vx_i,\vx_j)=1+f_{ij}$, which leads to 
\begin{equation}\label{canonicalCluster}
	\Q_A(V,\lambda_a) = \frac{1}{h^{3A} A!} \int \D^{3A}x \prod_{1\leq i < j \leq A}(1+f_{ij}),
\end{equation}
where the measure is defined as
\begin{equation}\label{definitionOfMeasure}
	\D^3x(\lambda_a) = d^{3}x\; p_A(\vx)\;e^{\frac{1}{A}\sum_a\lambda_a f_a(\vx_i)}.
\end{equation}
In the limit of no correlation, we have $f_{ij}=0$, and
\begin{equation}\label{CanonDecompos}
	\Q_A(V,\lambda_a) = \Q_1^A(V,\lambda_a)/A!= \left[\int \D^3x(\lambda_a) \right]^A/h^{3A} A!.
\end{equation}
The averages $\la \{f_1\} \ra $ and $\la \{f_1\} \{f_2\} \ra $ corresponds to $f_1(\bx)$ and $f_2(\bx)$ in Eq.~\eqref{SingleObsAvreave} and \eqref{TwoObsAvreave} can be obtained by appropriately differentiating $\Q_A(V,\lambda_a)$ with respect to $\lambda_a$ and setting $\lambda_a$ to zero in the end.

To calculate the canonical partition function in the presence of correlation, one needs to evaluate integrals of the form $\int \D^3 x_1\cdots\D^3 x_N f_{\alpha_1} f_{\alpha_2} \cdots f_{\alpha_k}$, where $\alpha_i$ represents a pair of distinct numbers between 1 and $A$. Depending on the values of the pairs $\alpha_i$, the integral may or may not be decomposed into lower-dimensional integrals or clusters. In the presence of correlations, finding the grand-canonical partition function is an easier task, and it is given by:
\begin{equation}\label{grandCanoCluster}
	\log \mathscr{Q}(z,\lambda_a)= \sum_{\ell=1} b_\ell(\lambda_a)z^\ell.
\end{equation}
The first term in the above expansion corresponds to the grand-canonical partition function between sources with no correlations. The coefficients $b_\ell(\lambda_a)$ are Mayer cluster integrals. The 1-cluster and 2-cluster integrals read as:
\begin{subequations}
	\begin{align}
		b_1(\lambda_a) &=  \int  \D^3x(\lambda_a),\\
		b_2(\lambda_a) &= \frac{1}{2} \int  \D^3x_1(\lambda_a)  \D^3x_2(\lambda_a)\;f_{12}.
	\end{align}
\end{subequations}

Now, we are able to connect the cluster integrals to the desired moments.
The average value of $A$ can be obtained as follows,
\begin{equation}\label{N_z_expansion}
	\la A \ra = \left.  z \frac{\partial \log\mathscr{Q}}{\partial z}  \right|_{\lambda_a\to 0} = \sum_{\ell=1}^{\infty}\ell b_\ell(0) z^\ell.
\end{equation}
In the absence of correlation, only 1-cluster integral is non-vanishing, meaning $\la A \ra$ is equal to $b_1(0) z$. From the definition in Eq.~\eqref{definitionOfMeasure}, we know that $b_1(0) = 1$. Therefore, we obtain $z = \la A \ra$. Also, assuming $A \equiv \la A \ra$, we can conclude from Eqs.~\eqref{GrandPartFubc} and \eqref{CanonDecompos} that grand-canonical partition function is written in terms of canonical partition function as $\mathscr{Q} = e^{A \Q_1}$. To account for the effect of correlation, we need to find the correction to $z=\la A\ra$. We do this by substituting $z=\sum_{\ell=1}^\infty a_\ell \la A \ra^\ell$ into Eq.~\eqref{N_z_expansion} and finding $a_\ell$ order by order. Afterward, we substitute the expansion of $z$ into Eq.~\eqref{grandCanoCluster} to obtain the partition function, which is written as
\begin{equation}
	\begin{split}
		\log \mathscr{Q}&(\la A \ra,\lambda_a) = \\ 
		b_1(\lambda_a)&\;\la A \ra + \left[b_2(\lambda_a)- 2b_2(0)b_1(\lambda_a) \right] \;\la A \ra^2 +\cdots.
	\end{split}
\end{equation}
From Eq.~\eqref{GrandPartFubc}, we note that the dominant term in $\sum_{n=0}^\infty z^n \mathcal{Z}_n(\lambda_a) / n!$ is around $n \approx z$ for large $z \approx \la A \ra$. Therefore, we can approximately assume $\mathscr{Q}(\la A \ra,\lambda_a) \propto \Q_A(\lambda_a)$.  The moments average can be obtained by appropriately differentiating the grand  partition function with respect to $\lambda_a$ as discussed in the main text.

The introduced method is generic, and a realistic one-body density and two-body correlation can be used to perform the cluster integrals. For particular choice of one-body density and two-body correlation, we can perform the integrals analytically. In particular, we assume $p_A(\vx)$ keeps the WS distribution structure.  Considering that the value of $R_0$ and $a_0$ in WS is fixed from experimental measurements, we start with unknown parameters $\tilde{R}_0$ and $\tilde{a}_0$, and then adjust it so that $\{r^2\}$ leads to a correct nucleus radius. This is a reasonable approximation for small values of $d_\mi$ (see subsection.~\ref{radiusRenorm}). To impose two-body correlation, we assume a hard central core for each nucleon,
\begin{equation}\label{stepFuncC2}
	c_2(\vx_1,\vx_2) = 1 + f_{12}(\vx_1,\vx_2) = \begin{cases}
		0 & |\vx_1-\vx_2| \leq d_\mi \\
		1      & |\vx_1-\vx_2| > d_\mi
	\end{cases}.
\end{equation}
The details of the moments calculation can be found in appendix~\ref{appendix:ClusterExpansioMoment}. The results are presented in Eqs.~\eqref{StepFubcCorrA} and  \eqref{dprimePhi}.

\onecolumngrid

\section{Analytical moments from cluster expansion}\label{appendix:ClusterExpansioMoment}

The initial state moments can be written in terms of the following integrals:
\begin{subequations}
	\begin{align}
		I_{m}&=\int  d^{3}x\; |\bx|^{2m} \; p_A(\vx), \\ 
		J^{(2)}_{m,n,\ell}&=\int  d^{3}x_1d^{3}x_2\; (x_1^{2}+x_2^{2})^m (y_1^{2}+y_2^{2})^n (x_1y_1+x_2y_2)^\ell p_A(\vx_1)p_A(\vx_2)f_{12}(\vx_1,\vx_2). \label{Jmoments} \\
		J^{(4)}_{m,n,\ell}&=\int  d^{3}x_1d^{3}x_2\; (x_1^{4}+x_2^{4})^m (y_1^{4}+y_2^{4})^n (x_1^2y_1^2+x_2^2y_2^2)^\ell p_A(\vx_1)p_A(\vx_2)f_{12}(\vx_1,\vx_2). \label{Jmoments}
	\end{align}
\end{subequations}
The interested moment averages is obtained from $\partial_{\lambda_1}^{m_1} \partial_{\lambda_2}^{m_2} \mathscr{Q}(\la A \ra,\lambda_a) / \mathscr{Q}(\la A \ra,\lambda_a)$ leading to the following result:
\begin{subequations}\label{corrInTermsMoment}
	\begin{align}
		\left \la \{r^2\}_{s,A} \right\ra &= I_{1} + \frac{A}{2} \left(J^{(2)}_{1,0,0}+J^{(2)}_{0,1,0} -2 J^{(2)}_{0,0,0}I_{1}\right) \label{r2_sA_corr},\\
		\left\la\left|\{r^2 e^{2i\varphi}\}_{s,A}\right|^2\right\ra &=	\frac{I_{2}}{A}  +\frac{1}{2} \left(J^{(2)}_{2,0,0}+J^{(2)}_{0,2,0} -2 J^{(2)}_{1,1,0}+4J^{(2)}_{0,0,2}-2J^{(2)}_{0,0,0}I_{2} \right) \\
		&+\left[\frac{I_{2}}{A}  + \left(J^{(4)}_{1,0,0} + J^{(4)}_{0,1,0}+2J^{(4)}_{0,0,1}-2J^{(2)}_{0,0,0}I_{2} \right)\right]\sigma_{\text{fluc}}^2 +O(J^2), \nonumber \\
		\left \la \{r^2\}^2_{s,A} \right\ra - \left \la \{r^2\}_{s,A} \right\ra^2 & = \frac{I_{2}}{A}  +\frac{1}{2} \left(J^{(2)}_{2,0,0}+J^{(2)}_{0,2,0} +2 J^{(2)}_{1,1,0}-2J^{(2)}_{0,0,0}I_{2} \right) \\
		&+\left[\frac{I_{2}}{A}  + \left(J^{(4)}_{1,0,0} + J^{(4)}_{0,1,0}+2J^{(4)}_{0,0,1}-2J^{(2)}_{0,0,0}I_{2} \right)\right]\sigma_{\text{fluc}}^2 +O(J^2). \nonumber
	\end{align}
\end{subequations}
Interestingly, we have the following relation,
\begin{equation}
	\left(\left \la \{r^2\}^2_{s,A} \right\ra - \left \la \{r^2\}_{s,A} \right\ra^2\right) - \left\la\left|\{r^2 e^{2i\varphi}\}_{s,A}\right|^2\right\ra = 2( J^{(2)}_{1,1,0}-J^{(2)}_{0,0,2})+O(J^2, 1/A),
\end{equation}
which merely depends on the two-body density. Ignoring the correlation, one can use canonical partition function to calculate this difference which leads to an important contribution from terms proportional to $1/A$. 

The one-body density is given by
\begin{equation}\label{pAWS}
	p_{A}(\vx)\approx  \frac{1}{\rho_0\,} \frac{\rho_\text{WS}(\vec{x})}{-8 a_0^3 \text{Li}_{3}(e^{R_0/a_0})}.
\end{equation}
The moments of WS distribution can be obtained analytically as follows:
\begin{equation}
	\begin{split}
		I_{m}(R_0,a_0)&= \frac{4^m a_0^{2m}\,\Gamma(m+2)\Gamma(m+1)\;\text{Li}_{3+2m}\left(-e^{R_0/a_0}\right)}{\text{Li}_{3}\left(-e^{R_0/a_0}\right)}\\
		&= \frac{3}{4}\left(\frac{\Gamma(1/2) \Gamma(m+1)}{\Gamma(m+5/2)}\right) R_0^{2m}\left[1+\frac{2m(2m+5)\pi^2}{6}\left(\frac{a_0}{R_0}\right)^2+\cdots\right].
	\end{split}
\end{equation}
 In particular, following moments are relevant for us in this study, 
\begin{subequations}\label{Imoments}
	\begin{align}
		I_1(R_0,a_0) &= \frac{2R_0^2}{5}\left[1 + \frac{7\pi^2}{3}\left(\frac{a_0}{R_0}\right)^2+\cdots\right],\label{ImomentsA}\\
		I_2(R_0,a_0) &= \frac{8 R_0^4}{35}\left[1+6\pi^2 \left(\frac{a_0}{R_0}\right)^2   +\frac{31}{3}\pi^4 \left(\frac{a_0}{R_0}\right)^4+\cdots\right].\nonumber\label{ImomentsB}
	\end{align}
\end{subequations}

In the following, we suggest an analytical estimation for $J_{m,n,\ell}$ integral when a minimum distance between nucleons is imposed. To calculate $J_{m,n,\ell}$ analytically, we still need to approximate WS distribution. It is important to note that any approximation to WS will result in a slight error to our estimation since the integral $J_{m,n,\ell}$ is the correction to the observable. We replace WS with a step-function $\Theta(\tilde{R}_0^2-|\vec{x}|^2)$ and calculate the integral in Eq.~\eqref{Jmoments} analytically by using the integral representation of the step-function.

In cases where there is no correlation between sources, the value of $f_{ij}$ becomes zero. In such cases, the absence of correlation causes the moments $J_{m,n,\ell}$ to be equal to zero as well. To keep the leading contribution in calculating $J_{m,n,\ell}$, we ignore the effect of skin thickness, and therefore WS is approximated by a step-function, $\rho_{\text{WS}}(|\vx|)\propto\theta(R_0^2-|\vx_1|^2)$. To calculate a generic integral 
\begin{equation}
	J(R_0,d_\mi) = \frac{1}{(4/3\pi R_0^3)^2}\int  d^{3}x_1d^{3}x_2\; f(\bx_1,\bx_2)  \; \theta(R_0^2-|\vx_1|^2)\theta(R_0^2-|\vx_2|^2)f_{12}(\vx_1,\vx_2),
\end{equation}
we employ the integral representation of the step-function 
\begin{equation}
	\Theta(x) = \lim_{\substack{\epsilon\to 0^+ \\ \Lambda\to \infty}} \frac{1}{2\pi i } \int_{-\Lambda}^{\Lambda} \frac{e^{i\tau x}}{\tau - i\epsilon} d\tau.
\end{equation}
Performing the change of variable $\vx_1 = (\vx'_2-\vx'_1) / 2$ and $ \vx_2 = (\vx'_2+\vx'_1) / 2$, the integral is written as
\begin{equation}
	J(R_0,d_\mi) =  \lim_{\substack{\epsilon\to 0^+ \\ \Lambda\to \infty}}  \lim_{\substack{\epsilon'\to 0^+ \\ \Lambda'\to \infty}} \tilde{J}(R_0,d_\mi,\Lambda,\Lambda',\epsilon,\epsilon'),
\end{equation}
where
\begin{equation}
	\tilde{J}=\frac{1}{8}\frac{1}{(2\pi i)^2}\int d\tau d\tau'\, d^3x_1' d^3x_2'\; \frac{1}{\tau - i\epsilon} \frac{1}{\tau' - i\epsilon'}\; f(\bx'_1,\bx'_2)\; e^{iR_0^2  (\tau+\tau') + \frac{i (\tau-\tau') }{2} r_1 r_2 \cos\Omega_{12}-\frac{i(\tau+\tau')}{4}(r_1^2+r_2^2) } f_{12}(r_1). 
\end{equation}
Here, we have used the $|\vx'_i|^2 = r_i^2$, and
\begin{equation}
	\begin{split}
		|\vx_1|^2 &= \frac{1}{4} \left[r_1^2+r_2^2-2r_1 r_2 \cos\Omega_{12}\right], \qquad 
		|\vx_2|^2  = \frac{1}{4} \left[r_1^2+r_2^2+2r_1 r_2 \cos\Omega_{12}\right].
	\end{split}
\end{equation}
In present study, the function $f(\bx'_1,\bx'_2)$ has the following form:  
\begin{equation}
	f(\bx'_1,\bx'_2) = r_1^{q_1}r_2^{q_2} g_1(\Omega_1) g_2(\Omega_2),
\end{equation}
where $\Omega_i$ represents the collective azimuthal and polar angles. 
Using the following integral, 
\begin{equation}
	\int_0^d dx\, x^\alpha e^{-a x^2-b x}= \frac{d^{\alpha+1}}{\alpha+1}+\cdots,
\end{equation}
one can perform the radial integrals along $r_1$ and $r_2$ to find 
\begin{equation}
	\begin{split}
		\tilde{J} =  -\frac{1}{16}\frac{1}{(2\pi i)^2} \left(\int d\Omega_1 g_1(\Omega_1)\right)\left(\frac{d_\text{min}^{q_1+3}}{q_1+3}\right)\left(\int d\Omega_2 g_2(\Omega_2)\right)\Gamma\left(\frac{q_2+3}{2}\right) \left(\frac{4}{i}\right)^{\frac{q_2+3}{2}}\mathcal{I}_{\frac{q_2+3}{2}},
	\end{split}
\end{equation}
where
\begin{equation}
	\mathcal{I}_{n} = \int  d\tau'\,  \frac{e^{i R_0^2 \tau'}}{\tau' - i\epsilon'}  \mathcal{J}_n(\tau'),\qquad 	\mathcal{J}_n(\tau')=\lim_{\eta\to0}\int d\tau \,  \frac{e^{i R_0^2 \tau}}{\tau - i\epsilon}   \frac{1}{ \left(\tau +\tau'-i\eta \right)^{n}},\qquad \epsilon,\epsilon',\eta > 0.
\end{equation}
The parameter $n$  is non-integer in our case. Here, we assume it is an integer and find a general formula for the integral and then analytically continue the result into fraction values of $n$.

To calculate the integral $\mathcal{J}_n(\tau')$, we should close the contour from above. In this case, the integrand has pols at $\tau = i \epsilon $ and $  \tau = -\tau'+i\eta$ in the upper half plane, leading to
\begin{equation}
	\begin{split}
		\lim_{\substack{\epsilon\to 0^+ \\ \Lambda\to \infty }} \mathcal{J}_n(\tau')& = \frac{2\pi i}{\tau'^n}-2\pi i \frac{e^{-i R_0^2 \tau'} f_{n-1}(\tau')}{\tau'^n},
	\end{split}
\end{equation}
where $f_{n}(\tau')$ is a polynomial of order $n$, $f_n(\tau') = \sum_{\ell=1}^{n} \frac{(i R_0^2 \tau')^\ell}{(\ell-1)!}$. Therefore, the integral $\mathcal{I}_{n}$ is written as
\begin{equation}\label{C19}
	\begin{split}
		\lim_{\substack{\epsilon'\to 0^+ \\ \Lambda'\to \infty }}\mathcal{I}_{n}(1,1,\eta) &=  \int  d\tau'\,  \frac{e^{i R_0^2 \tau'}}{\tau' - i\epsilon'}\left(  \frac{2\pi i}{\tau'^n}-2\pi i \frac{e^{-i R_0^2 \tau'} f_{n-1}(\tau')}{\tau'^n}\right).
	\end{split}
\end{equation}
The first part of the above integral is obtained as follows: 
\begin{equation}
	\lim_{\substack{\epsilon',\epsilon''\to 0^+ }}\int  d\tau'\,  \frac{e^{i R_0^2 \tau'}}{\tau' - i\epsilon'}\left(  \frac{2\pi i}{(\tau'-i\epsilon'')^n}\right) =  -\frac{4 i^n \pi^2 R_0^{2n}}{\Gamma(n+1)}.
\end{equation}
For the second part of the integral in Eq.~\eqref{C19}, we have
\begin{equation}
	\int_{-\Lambda'}^{\Lambda'}  d\tau'\,  \frac{1}{\tau' - i\epsilon'}  \frac{ f_{n-1}(\tau')}{(\tau'-i\epsilon'')^n} = -\frac{1}{\Lambda'} \left(-2\frac{(i R_0^2 )^{n-1}}{(n-2)!}\right),
\end{equation}
which vanishes in the limit $\Lambda'$ to infinity.
Therefore
\begin{equation}
	\lim_{\substack{\epsilon'\to 0^+ \\ \Lambda'\to \infty }}\mathcal{I}_{n} = -\frac{4 i^n \pi^2 R_0^{2n}}{\Gamma(n+1)}.
\end{equation}
In the following, we calculate the moments required in Eq.~\eqref{corrInTermsMoment}.

\textit{Moment $J^{(2)}_{0,0,0}$}: In this case, the function $f(\bx'_1,\bx'_2) = 1$ and consequently $q_1 = q_2 =0$, and  $g_1(\Omega_1) = g_2(\Omega_2) = 1$, leading to
\begin{equation}\label{Jmoment1}
	J^{(2)}_{0,0,0} = -\left(\frac{d_\text{min}}{R_0}\right)^3.
\end{equation}

\textit{Moment $J^{(2)}_{1,0,0} + J^{(2)}_{0,1,0}$}:  To calculate this moment combination, one needs to choose the function $f(\bx'_1,\bx'_2)$ as
\begin{equation}
	f(\bx'_1,\bx'_2) = x_1^2+x_2^2+y_1^2+y_2^2 = \frac{1}{4} \left(r_1^2(1-\cos 2\theta_1)+ r_2^2(1-\cos 2\theta_2)\right).
\end{equation}
Therefore, the integral has two parts. For the first part, we have
\[ q_1 =2, \quad g_1(\Omega_1) = \frac{1}{4} (1-\cos 2\theta_1), \qquad q_2 = 0,\quad g_2(\Omega_2) = 1, \]
corresponding to a term proportional to  $d_\mi^5$, which is subleading in our calculation. For the second part, one  chooses
\[ q_1 =0 \quad g_1(\Omega_1) = 1, \qquad q_2 = 2,\quad g_2(\Omega_2) = \frac{1}{4} (1-\cos 2\theta_2), \]
leading to term $d_\mi^3$. As a result, we finally find,
\begin{equation}\label{Jmoment2}
	J^{(2)}_{1,0,0} + J^{(2)}_{0,1,0} =-2 \left(\frac{d_\mi}{R_0}\right)^3 \left(\frac{2 R_0^2}{5}\right).
\end{equation}

\textit{Moment $J^{(2)}_{2,0,0}+J^{(2)}_{0,2,0}-2 J^{(2)}_{1,1,0}+4J^{(2)}_{0,0,2}$}: The function $f(\bx'_1,\bx'_2) $  need to be chosen as the following:
\begin{equation}
	f(\bx'_1,\bx'_2) =\left(\left(x_2+y_1\right)^2+\left(x_1-y_2\right)^2\right)\left(\left(x_2-y_1\right)^2+\left(x_1+y_2\right)^2\right)
\end{equation}
In $r_i$, $\Omega_i$ coordinate, this function is written as
\begin{equation}
	r_1^4 \left(\frac{\sin^4\theta_1}{4}\right)+\frac{1}{2}r_1^2 r_2^2\left(\cos\left(2\phi_1-2\phi_2\right)\sin^2\theta_1\sin^2\theta_2\right) + r_2^4 \left(\frac{\sin^4\theta_2}{4}\right).
\end{equation}
The last term leads to the leading term $d_\mi^3$ which corresponds to
\[ q_1=0,\quad g_1(\Omega_1) = 1,\qquad q_2 = 4, \quad g_2(\Omega_2) = \frac{\sin^4\theta_2}{4}.  \]
As a result, the contribution proportional to $d_\mi^3$ in this case is as follows:
\begin{equation}\label{Jmoment3}
	J^{(2)}_{2,0,0}+J^{(2)}_{0,2,0}-2 J^{(2)}_{1,1,0}+4J^{(2)}_{0,0,2} = -4\left(\frac{8 R_0^4}{35}\right)\left(\frac{d_\mi}{R_0}\right)^3.
\end{equation}

\textit{Moment $J^{(4)}_{1,0,0} + J^{(4)}_{0,1,0}+2J^{(4)}_{0,0,1}$}: these combination of moments  corresponds to the following function:
\begin{equation}
	f(\bx'_1,\bx'_2) =\left(x_1^2+y_1^2\right)^2+\left(x_2^2+y_2^2\right)^2.
\end{equation}
Similar to the previous case, in spherical coordinate the leading contribution is coming from the term $r_2^4 \left(\frac{\sin^4\theta_2}{8}\right)$. This means that
\begin{equation}\label{Jmoment4}
J^{(4)}_{1,0,0} + J^{(4)}_{0,1,0}+2J^{(4)}_{0,0,1} = -2\left(\frac{8 R_0^4}{35}\right)\left(\frac{d_\mi}{R_0}\right)^3.
\end{equation}
The contribution from $\sigma_{\text{fluc}}^2$ at the presence of non-vanishing $d_\mi$ is given by
\begin{equation}
	J^{(4)}_{1,0,0} + J^{(4)}_{0,1,0}+2J^{(4)}_{0,0,1}-2 I_2 J^{(2)}_{0,0,0} \approx 0.
\end{equation} 
The result in Eqs.~\eqref{StepFubcCorrA} and  \eqref{dprimePhi} can be obtained by substituing the moments into Eqs~\eqref{corrInTermsMoment}.

\vspace{0cm}
\section{Two-body correlation in a 1D toy model}\label{1DToyModel}

We aim to determine the fluctuation of a pair of sources on a one-dimensional line distributed via a one-dimensional Gaussian distribution. The distribution is given by the equation $p(x) = \frac{1}{\sqrt{2\pi}\sigma} e^{-\frac{x^2}{2\sigma^2}}$. We also assume that a minimum distance is imposed among these sources. The two-body density is defined as follows:
\begin{equation}
	\rho_2(x_1,x_2) = \frac{p(x_1)p(x_2)\Theta(|x_1-x_2| - d_\mi)}{\int dx_1 dx_2 p(x_1)p(x_2)\Theta(|x_1-x_2| - d_\mi) },
\end{equation}
where $\Theta$ is the Heaviside step function. We can redefine the two-body density as $\rho_2(x_1,x_2) =  \rho_1(x_1) \rho_1(x_2) c_2(x_1,x_2)$,
where the one-body density is given by:
\begin{equation}\label{1DOneBody}
	\rho_1(x_1) = \int dx_2 \rho_2(x_1,x_2) = p(x_1)\left[1 + \left(\frac{1}{\sqrt{\pi}}- 2\sigma p(x_1)\right)\left(\frac{d_\mi}{\sigma}\right)+\cdots  \right],
\end{equation}
and the two-body correlation reads as:
\begin{equation}\label{1DTrueC2}
	c_2(x_1,x_2) =  \Theta(|x_1-x_2| - d_\mi) \left[ 1- \left(\frac{d_\mi}{\sigma}\right)\left(\frac{1}{\sqrt{\pi}}- 2\sigma p(x_1)- 2\sigma p(x_2)\right) +\cdots   \right].
\end{equation}

\begin{figure*}
	\begin{center}
		\begin{tabular}{c c c}
			\includegraphics[width=0.35\textwidth]{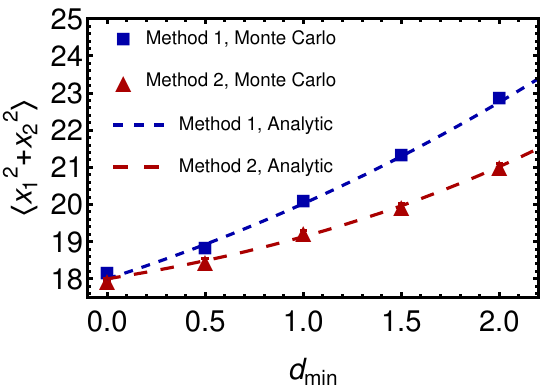} & \hspace*{0.7cm} &
			\includegraphics[width=0.375\textwidth]{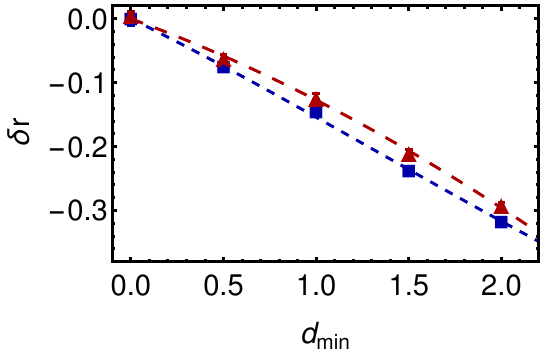}
		\end{tabular}
	\end{center}
	\captionof{figure}{ This figure illustrates the implementation of two-body correlation on a 1D toy model. The blue dots represent the results obtained using a toy Monte Carlo based on the direct method, while the red dots correspond to the iterative method with the step-function as a two-body correlation. The dashed curves depict the analytical calculation.}
	\label{1DCorrelationFig}
\end{figure*}

In the following, we test three scenarios of generating two sources at positions $x_1$ and $x_2$. We analytically calculate $\la x_1^2 +x_2^2 \ra$ and $\la (x_1^2 +x_2^2)^2 \ra$ and compare our calculation with toy Monte Carlo simulations.

\textit{Method 1: Direct method. }We first generate two random numbers and calculate their distance. If the distance is smaller than $d_\mi$, we discard the pair; otherwise, we accept it. This method matches with the following calculation,
\begin{subequations}\label{method1R2}
	\begin{align}
		\la x_1^2 + x_2^2\ra &= \int dx_1 dx_2 (x_1^2+x_2^2) \rho_2(x_1,x_2) = 2\sigma^2 + \frac{\sigma d_\mi}{\sqrt{\pi}}+ \cdots,\\
		\la (x_1^2 + x_2^2)^2\ra &= \int dx_1 dx_2 (x_1^2+x_2^2)^2 \rho_2(x_1,x_2) = 8\sigma^4 + \frac{5\sigma^3 d_\mi}{\sqrt{\pi}}+ \cdots.
	\end{align}
\end{subequations} 

\textit{Method 2: Iterative method with $p(x)$ as one-body density and step-function as two-body correlation. } First, we generate a random source. Then, we generate a second source and calculate its distance from the first one. If the distance is less than the value of $d_\mi$, we discard the second source and generate a new one. If the distance is greater than or equal to $d_\mi$, we accept the second source. The analytical calculation for this method is as follows: The second source's position depends on the first source's position. Therefore, we first calculate $\la x_2^2 \ra(x_1) \propto \int dx_2 x_2^2 p(x_2) \Theta(|x_1-x_2| - d_\mi)$. Then, we calculate $\la x_1^2 + \la x_2^2 \ra(x_1) \ra = \int dx_1 (x_1^2 + \la x_2^2 \ra(x_1)) p(x_1)$ using the result from the previous step. We follow the same process for $(x_1^2+x_2^2)^2$,
\begin{subequations}
	\begin{align}
		\la x_1^2 + x_2^2\ra &=  2\sigma^2 + \frac{\sigma d_\mi}{2\sqrt{\pi}}+ \cdots,\\
		\la (x_1^2 + x_2^2)^2\ra &=  8\sigma^4 + \frac{7\sigma^3 d_\mi}{4\sqrt{\pi}}+ \cdots.
	\end{align}			
\end{subequations}

\textit{Method 3: Iterative method  with $\rho_1(x)$ as one-body density and and $c_2(x_1,x_2)$ as two-body correlation. } This method is similar to the previous one while we replace $p(x)$ with $\rho_1(a)$ (Eq.~\eqref{1DOneBody}) and step-function with $c_2(x_1,x_2)$ (Eq.~\eqref{1DTrueC2}). The result is given by
\begin{subequations}\label{method3DeltaR}
	\begin{align}
		\la x_1^2 + x_2^2\ra &=  2\sigma^2 + \frac{\sigma d_\mi}{\sqrt{\pi}}+ \cdots,\\
		\la (x_1^2 + x_2^2)^2\ra &= 8\sigma^4 + \frac{5\sigma^3 d_\mi}{\sqrt{\pi}}+ \cdots.
	\end{align}			
\end{subequations}

The above analysis shows that method 2 produces inconsistent outcomes compared to methods 1 and 3. It can be concluded that using an iterative method with true $\rho_1(x)$ and $c_2(x_1,x_2)$ leads to a result similar to that obtained by starting directly from the two-body density. The difference between methods can be explained as follows: In the direct method (method 1), the most likely position for uncorrelated sources is around the origin. However, when correlation is present, finding sources very close to the origin becomes unlikely, and the probability of finding their position is symmetric for both sources. In method 2, the first source is generated independently and is usually located around the center. On the other hand, the second source cannot be too close to the center and needs to be placed further away from the origin. In method 3, the one-body density modifies, resulting in a distribution that differs from the Gaussian distribution. This distribution is already more populated symmetrically in points on both sides of the origin, making it similar to method 1.

In Fig.~\ref{1DCorrelationFig}, we have compared the toy Monte Carlo of methods 1 and 2 with our analytical calculation, where we define $\delta r$ as follows:

\begin{equation}
	\delta r = \frac{\la (x_1^2+x_2^2)^2 \ra - 2\la (x_1^2+x_2^2) \ra^2}{\la (x_1^2+x_2^2) \ra^2}.
\end{equation}
In equations \eqref{method1R2} to \eqref{method3DeltaR}, we presented the leading terms, while in the figure, we showed the full calculation. The figure makes it clear that there is a more drastic effect on $\langle x_1^2 + x_2^2 \rangle$, which corresponds to the size of the one-body density, compared to $\delta r$, which captures the two-body correlation.

\section{Effect of substructure, finite cross-section, and deformation}\label{appendix:extraParameters}

\subsection{Estimation of the nucleon substructure contribution}\label{appendix:substructure}

It is possible to assume substructure constituents for the nucleons. The reduced thickness function can then be written as follows:
\begin{equation}\label{UCReducedThicknessSub}
	T_R(\bx) \approx  \frac{\mathcal{N}}{2} \sum_{i=1}^{N_\pa} \sum_{j=1}^{n_c}\gamma_{i,j}  \int dz\,\rho_{\text{c}}(\vec{x}-\vec{x}_i-\vec{\xi}_{i,j}),
\end{equation} 
where $\vec{\xi}_{i,j}$ indicates the location of $j$th constituent of the $i$th nucleon, and $n_c$ is the number of constituents. Here, we assume the constituents are distributed inside the nucleons via a normal distribution with width $\sigma_w$, and constituents have a round Gaussian profile with width $v$.
We have similar decomposition as that in appendix~\ref{DecompositionSubsection}, $e^{W_R}=e^{W_{c}}  e^{W_{\text{source}}}$.

 Considering that the substructure of nucleons has a small contribution, we will exclude the two-body correlation from our calculations. Additionally, we will assume that the density of the nucleus is Gaussian in shape with a width of $\sigma_G \approx R_0/\sqrt{5}$. It is important to note that the contribution of the nucleon substructure is local, and the overall shape of the nucleus should not have a significant impact. These assumptions will allow us to use  Eqs~\eqref{SingleObsAvreave} and \eqref{TwoObsAvreave}.

Suppose a nucleon with a density distribution of $\rho_\text{nucleon}(\vx)$ and width of $\sigma_w$ is located at position $\vx_i$. Hence, the constituents are distributed based on this distribution at locations $\vec{\xi}_{i,1},\ldots, \vec{\xi}_{i,n_c}$. The fluctuation of these constituents is given by $\rho_\text{nucleon}(\vec{\xi}_{i,1}-\vx_i)\cdots \rho_\text{nucleon}(\vec{\xi}_{i,{n_c}}-\vx_i)$. However, we impose a constraint that the average location of $n_c$ constituents must coincide with the location of the nucleon at $\vx_i$. This constraint is given by
\[   \rho_\text{nucleon}(\vec{\xi}_{i,1}-\vx_i)\cdots \rho_\text{nucleon}(\vec{\xi}_{i,{n_c}}-\vx_i) \; \delta([\vx]_i-\vx_i), \] 
where $[\vx]_i = \frac{1}{n_c} \sum_{j=1}^{n_c} \vec{\xi}_{i,j}$ is the location of the nucleon's constituent center. Hence, we obtain the following expression for the source fluctuation distribution, 
\begin{equation}\label{substructureDist}
	\begin{split}
		P_{A}=
		\left[\prod_{i=1}^{A} g([\vx]_i)\prod_{j=1}^{n_c} \rho_\text{nucleon}(\vec{\xi}_{i,j}-[\vx]_i)p_\fluc(\gamma_{i,j}) \right],
	\end{split}
\end{equation}
which is the generalized version of Eq.~\eqref{structureLess} at the presence of nucleon substructure. Here, $g(\vx)$ is a Gaussian distribution with width $\sigma_G$. Evidently, the Gaussian distribution for nucleus density is a poor approximation. For that reason, we only single out the contribution of the substructure to the observables and incorporate it with those obtained in the previous parts based on WS distribution.

To isolate the effect of the substructure, it is useful to define the following measures:
\begin{equation}
	\begin{split}
		\D\xi_i &= \left( d^3\xi_{i,1}\cdots d^3\xi_{i,n_c} g([\vx]_i)\prod_{j=1}^{n_c} \rho_\text{nucleon}(\vec{\xi}_{i,j}-[\vx]_i)\right),\\
		\D\gamma &= d\gamma_{i,1}p_\fluc(\gamma_{i,1})\cdots d\gamma_{i,n_c} p_\fluc(\gamma_{i,n_c}).
	\end{split}
\end{equation}
We also define the following function:
\begin{equation}
	F(\gamma_{i,j},\vec{\xi}_{i,j})=\frac{1}{n_c} \sum_{j=1}^{n_c} \gamma_{ij}f(\vec{\xi}_{i,j}).
\end{equation}
Using these definitions, the moments read as
\begin{equation}
	\OO =\frac{1}{N_A}\sum_{i=1}^{N_A} F_{i}(\gamma_{i,j},\vec{\xi}_{i,j}),
\end{equation}
and one can straightforwardly find
\begin{subequations}
	\begin{align}
		\la \OO \ra &= \frac{\int \D\gamma_1\D\xi_1 F(\gamma_{1,j},\vec{\xi}_{1,j}) }{\int \D\xi}, \\
		\la \OO_1  \OO_2 \ra &= \frac{1}{N_A}\frac{\int \D\gamma_1\D\xi_1 F_1(\gamma_{1,j},\vec{\xi}_{1,j}) F_2(\gamma_{1,j},\vec{\xi}_{1,j})}{\int \D\xi}+\frac{N_A-1}{N_A}\left(\frac{\int \D\gamma_1\D\xi_1 F_1(\gamma_{1,j},\vec{\xi}_{1,j}) }{\int \D\xi}\right)\left(\frac{\int \D\gamma_1\D\xi_1 F_2(\gamma_{1,j},\vec{\xi}_{1,j}) }{\int \D\xi}\right).\label{substrucVarianceObs}
	\end{align}
\end{subequations}
Therefore, we need to calculate the integrals over substructures. These integrals are $n_c$ dimensions and cannot be decomposed into smaller dimensions due to the constraint$[\vx]_i = \frac{1}{n_c} \sum_{j=1}^{n_c} \vec{\xi}_{i,j}$. We still, however, relate the integration over $F(\gamma_{1,j},\vec{\xi}_{1,j})$ to the integration over $f(\vec{\xi}_{i,j})$,
\begin{equation}
	\int \D\gamma_1\D\xi_1 F(\gamma_{1,j},\vec{\xi}_{1,j}) = \int \D\gamma_1\D\xi_1 \frac{1}{n_c} \sum_{j=1}^{n_c} \gamma_{1,j}f(\vec{\xi}_{1,j}) = \int \D\xi_1 f(\vec{\xi}_{1,1}),
\end{equation}
where we used the fact that partons are indistinguishable in the second equality.
Similarly, 
\begin{equation}
	\begin{split}
		\int \D\gamma_1\D\xi_1 F_1(\gamma_{1,j},\vec{\xi}_{1,j}) F_2(\gamma_{1,j},\vec{\xi}_{1,j}) &= \frac{1}{n_c^2}\int \D\gamma_1\D\xi_1  \left[\sum_{j=1}^{n_c} \gamma_{1,j}^2f_1(\vec{\xi}_{1,j}) f_2(\vec{\xi}_{1,j}) + \sum_{j\neq k}^{n_c} \gamma_{1,j}\gamma_{1,k}f_1(\vec{\xi}_{1,j}) f_2(\vec{\xi}_{1,k})\right] \\
		&=\frac{1}{n_c} \left(1+\sigma_\fluc^2\right)\int  \D\xi_1\; f_1(\vec{\xi}_{1,1}) f_2(\vec{\xi}_{1,1})+\frac{n_c-1}{n_c}\int  \D\xi_1\; f_1(\vec{\xi}_{1,1}) f_2(\vec{\xi}_{1,2}).
	\end{split}
\end{equation}
Now, we calculate the integration over $f(\vec{\xi}_{1,2})$. We are particularly interested in the following choices,
\begin{equation}
	f_1(\vec{\xi}) = |\boldsymbol{\xi}|^2,\qquad f_\pm(\vec{\xi}) = |\boldsymbol{\xi}|^2 e^{\pm 2i\varphi},
\end{equation}
where $\vec{\xi}=(\boldsymbol{\xi},\xi_z)$, and $\boldsymbol{\xi}=(\xi,\eta)$. The first function corresponds to observable $ \{r^2\}_{s,A}$, 
\begin{equation}
	\begin{split}
		\left\la \{r^2\}_{s,A} \right\ra &= \frac{\int \D\xi \left|\boldsymbol{\xi}\right|^2}{\int \D\xi}= \frac{2\int \D\xi \; \xi^2}{\int \D\xi},
	\end{split}
\end{equation}
and the functions $f_\pm(\vec{\xi})$ correspond to observable $\left|\{r^2 e^{2i\varphi}\}_{s,A}\right|^2$,
\begin{equation}
	\begin{split}
		\left\la \left|\{r^2 e^{2i\varphi}\}_{s,A}\right|^2 \right\ra &=\frac{\left(1+\sigma_\fluc^2\right)}{N_A n_c} \frac{\int \D\xi \left|\boldsymbol{\xi}\right|^4}{\int \D\xi}+\frac{n_c-1}{N_A n_c} \frac{\int \D\xi_1 \left|\boldsymbol{\xi}_{1,1}\right|^2\left|\boldsymbol{\xi}_{1,2}\right|^2 e^{2i(\varphi_{1,1}-\varphi_{1,2})}}{\int \D\xi},\\
		&=\frac{2\left(1+\sigma_\fluc^2\right)}{N_A n_c} \left[\frac{\int \D\xi \; \xi^4+\int \D\xi \; \xi^2\eta^2}{\int \D\xi}\right]\\
		&+\frac{2(n_c-1)}{N_A n_c} \left[\frac{\int \D\xi \; \xi_1^2\xi_2^2-\int \D\xi \; \xi_1^2\eta_2^2 +2 \int \D\xi \; \xi_1\xi_2\eta_1\eta_2 }{\int \D\xi}\right],
	\end{split}
\end{equation}
where we have used the symmetry in $\boldsymbol{\xi}=(\xi,\eta)$ space to simplify equations. 
The required integrals are listed in the following,
\begin{equation}
	\begin{split}
		&\frac{\int \D\xi \, \xi^2}{\int \D\xi} = \sigma_R^2, \qquad \frac{\int \D\xi \, \xi^4}{\int \D\xi} = 3\sigma_R^4,\qquad \frac{\int \D\xi \, \xi^2 \eta^2}{\int \D\xi} = \frac{\int \D\xi \, \xi_1^2 \eta_2^2}{\int \D\xi} = \sigma_R^4,\\
		&\frac{\int \D\xi \, \xi_1^2 \xi_2^2}{\int \D\xi} = 3\sigma_R^4-4\sigma_R^2\sigma_w^2+2\sigma_w^4,\qquad \frac{\int \D\xi \, \xi_1 \xi_2\eta_1\eta_2 \eta_2}{\int \D\xi} =\left(\sigma_R^2-\sigma_w^2\right)^2,
	\end{split}
\end{equation} 
where
\begin{equation}
	\sigma_R^2=\sigma_G^2 \left[1+\frac{n_c-1}{n_c}\left(\frac{\sigma_w}{\sigma_G}\right)^2\right].
\end{equation}
Based on the above calculation, one finds,
\begin{subequations}\label{substructureResults}
	\begin{align}
		&\frac{1}{2\sigma_G^2}\left.\left\la \{r^2\}_{s,A} \right\ra \right|_\text{substr} 
		= 5\left(\frac{n_c-1}{n_c}\right)\left(\frac{\sigma_w}{R_0}\right)^2,\label{substructureResultsA}\\
		&\frac{A}{8\sigma_G^4}\left.\left\la\left|\{r^2 e^{2i\varphi}\}_{s,A}\right|^2\right\ra \right|_\text{substr} =  \frac{\sigma_\fluc^2}{n_c}\left(1+10\frac{n_c-1}{n_c}\left(\frac{\sigma_w}{R_0}\right)^2\right)+\cdots.\label{substructureResultsB}
	\end{align}
\end{subequations}
The right-hand side is the correction that needs to be incorporated with the cluster expansion calculation.

We will now discuss the different limits of Eqs.~\eqref{substructureResults}. First, we will ignore the effect of skin thickness and short-range correlation in Eq.~\eqref{StepFubcCorrA}. Next, we will incorporate Eq.~\eqref{substructureResultsA} with Eq.~\eqref{StepFubcCorrA} and implement them into Eqs.~\eqref{SourceToObsA} and \eqref{ObservableDecompositionA} where we replaced $w$ with $v$ as we are specifically discussing nucleons with substructure,
\begin{equation}
	\left\la \{r^2\} \right\ra = \frac{2R_0^2}{5}+2\frac{n_c-1}{n_c}\sigma_w^2+2v^2.
\end{equation}
This equation suggests defining a new \textit{effective} width for the nucleons as follows: $	w^2 = \frac{n_c-1}{n_c}\sigma_w^2+v^2$.
In \trento{} model, $w$ is the input parameter not $\sigma_w$ which can be written in terms of other parameters via~\cite{CiteDrive2022}
\begin{equation}\label{nucleonFlucWidth}
	\sigma_w = \sqrt{\frac{w^2-v^2}{1-1/n_c}}.
\end{equation}
In the case where there is no substructure ($n_c=1$), the expected result is $w=v$, which leads to the correction term in Eq.~\eqref{substructureResultsB} vanishing. According to subsection~\ref{DecompositionSubsection}, only constituents that are round in shape contribute to $\{r^2\}$. When $n_c$ approaches infinity, the fluctuation contributions from nucleon constituents vanish, and the nucleons behave like smooth round objects again. Therefore, the correction term in Eq.~\eqref{substructureResultsB} also vanishes in this limit. The constituent weight fluctuation contributions also vanish in this limit because they are only applied to nucleon constituents. As $\sigma_{\text{fluc}}\to0$, the substructure contribution appears at the order of $(\sigma_w/R_0)^4$ in Eq.~\eqref{substructureResultsB}.

\subsection{Effect of finite cross-section }\label{appendix:cross-section}

Previously, we considered that all nucleons in nuclei were involved in the collision, which means $N_\pa = 2A$. However, this assumption does not take into account the mutual correlation between the sources in the target and the sources in the projectile. When the inelastic cross-section is finite, the correlation between the nucleons in the target and the projectile is induced, which means that the decompression in Eq.~\eqref{decompose} requires a correction, and Eq.~\eqref{ObservableDecomposition} must be modified. 

\trento{} model calculates the probability of colliding sources forming the target and the projectile. This probability is given by a distribution $P_\text{coll}(\boldsymbol{d})$, which depends on the size of each nucleon and the inelastic cross-section between two nucleons. In the black-disk approximation, $P_\text{coll}(\boldsymbol{d})$ is assumed to be a step-function, and collisions occur when the distance between two nucleons is less than $\sqrt{\sigma_{\text{fluc}}^\text{NN}/\pi}$. After the position of the sources is sampled, the distance between pairs from the target and the projectile is calculated, and they are accepted as participants based on $P_\text{coll}(\boldsymbol{d})$.

It is a challenging task to include the correlation induced by collisions directly because a nucleon in the target may be correlated with more than one nucleon in the projectile. To estimate this correlation analytically, we adopt the optical Glauber model approach (see Ref.~\cite{Miller:2007ri}). Instead of assuming nucleons that collide with some other nucleons, we begin with $2A$ sources and then deduct the effect of those which did not collide. The probability of a nucleon from nucleus $A$ passing through a nucleus with $B$ nucleons and not experiencing any collision is given by:
\begin{equation}
	\begin{split}
		T_A^\text{opt}(\bx)&\left(1-\frac{1}{B}T_B^\text{opt}(\bx) \sigma_{\text{inel}}^\text{NN}\right)^B \\
		&\approx T_A^\text{opt}(\bx)e^{-T_A^\text{opt}(\bx) \sigma_{\text{inel}}^\text{NN}},
	\end{split}
\end{equation}
Here, $T_{A}^\text{opt}(\bx)$ is defined as $T_{A}^\text{opt}(\bx) = A \int dz \rho_\text{WS}(\vec{x})$.
We have used the approximation that $T_A(\bx)\approx T_B(\bx)$ and the large $A$ and $B$ limit. The number of sources in a nucleus that do not participate in the collision can be obtained via the following relation,
\begin{equation}
	\overline{N} =  \int d\bx \; T_A^\text{opt}(\bx)e^{-T_A^\text{opt}(\bx) \sigma_{\text{inel}}^\text{NN}}.
\end{equation}

To estimate the impact of finite cross-section, we use an approximation for the WS distribution. The effect of the cross-section is expected to be small, so any error associated with the approximated WS will be negligible. To do this, we substitute the WS distribution with the following function:
\begin{equation}\label{rhoGauss}
	\rho_G(|\vx|) = \frac{\rho_0}{1+\exp\left[ (|\vx|^2-R_0^2)/(2 a_0 R_0) \right]}. 
\end{equation}
For small $a_0$, the above distribution is similar to the WS distribution. We will also keep the leading term in $a_0$, where the distribution becomes $\rho_0$ for $|\vx|^2<R_0^2$, and $\rho_0e^{-(|\vx|^2-R_0^2)/(2 a_0 R_0)}$ for $|\vx|^2 > R_0^2$. The approximation of distribution in Eq.~\eqref{rhoGauss} in the $a_0\to 0$ limit yields the following approximation for the thickness function, 
\begin{equation}
	T_A^\text{opt}(\bx) \approx  \frac{3A}{4\pi R_0^3} \left[2\Theta(R_0-|\bx|)\sqrt{R_0^2-|\bx|^2} + \Theta(|\bx|-R_0) \sqrt{2\pi  a_0 R_0}\, e^{-\frac{|\bx|^2-R_0^2}{2a_0R_0}}\right].
\end{equation} 
We employ the above to calculate the following moment,
\begin{equation}
	\begin{split}
		&\int d\bx\; |\bx|^{2m}\, T_A^\text{opt}(\bx) e^{-T_A^\text{opt}(\bx) \sigma^{\text{NN}}_\text{inel}} = \mathcal{I}_{1,m} + \mathcal{I}_{2,m},
	\end{split}
\end{equation}
where ($s_\perp = \pi R_0^2/A$)
\begin{subequations}
	\begin{align}
		\mathcal{I}_{1,m} & = \frac{ 2\pi A}{4\pi R_0^3/3}\int_0^{R_0} rdr \, r^{2m}\, 2\sqrt{R_0^2-r^2}\, \exp\left[-\frac{3}{2 }\left(\frac{\sigma^{\text{NN}}_\text{inel}}{s_\perp}\right)\sqrt{1-\left(\frac{r}{R_0}\right)^2}\right], \\
		\mathcal{I}_{2,m} & = \frac{A (2\pi)^{3/2}  \sqrt{  a_0 R_0}}{4\pi R_0^3 / 3} \,\int_{R_0}^\infty r dr \, r^{2m}  \, \exp\left[-\frac{1}{2}\left(\frac{R_0}{a_0}\right)\left[\left(\frac{r}{R_0}\right)^2-1\right]\right.\\
		&\hspace*{6cm}\left.-\frac{3}{2}\left(\frac{\sigma^{\text{NN}}_\text{inel}}{s_\perp}\right)\sqrt{\frac{\pi}{2}}\left(\frac{a_0}{R_0}\right)^{1/2} e^{-\frac{1}{2}\left(\frac{R_0}{a_0}\right)\left[\left(\frac{r}{R_0}\right)^2-1\right]} \right].\nonumber
	\end{align}
\end{subequations}

To perform the first integral, we change the variable as $\tilde{r} = \sqrt{1-\left(\frac{r}{R_0}\right)^2}$. The result reads as
\begin{equation}
	\begin{split}
		\mathcal{I}_{1,m} &= 3A R_0^{2m}\int_0^{1} \tilde{r}^2d\tilde{r} \, \left(1-\tilde{r}^2\right)^{m} \, \exp\left[-\frac{3}{2 }\left(\frac{\sigma^{\text{NN}}_\text{inel}}{s_\perp}\right)\tilde{r}\right]\\
		&= A \,R^{2m} \,\frac{16}{9}   \left(\frac{s_\perp}{\sigma^{\text{NN}}_\text{inel}}\right)^3 + \cdots.
	\end{split}
\end{equation}

To calculate the second integral, we change the variable as $\tilde{r} = \sqrt{\left(\frac{r}{R_0}\right)^2-1}$ which leads to
\begin{equation}
	\begin{split}
		\mathcal{I}_{2,m} &= 3A  \sqrt{\frac{\pi}{2}}\sqrt{\frac{a_0}{R_0}} \;R_0^{2m}\int_0^\infty \tilde{r} d\tilde{r} \left(1+\tilde{r}^2\right)^m \,\exp\left[-\frac{1}{2}\left(\frac{R_0}{a_0}\right)\tilde{r}^2-\frac{3}{2}\left(\frac{\sigma^{\text{NN}}_\text{inel}}{s_\perp}\right)\sqrt{\frac{\pi}{2}}\left(\frac{a_0}{R_0}\right)^{1/2} e^{-\frac{1}{2}\left(\frac{R_0}{a_0}\right)\tilde{r}^2} \right]\\
		&= 2A\, R^{2m}\,\left(\frac{a_0}{R_0}\right)\left(\frac{s_\perp}{\sigma^{\text{NN}}_\text{inel}}\right) +\cdots.
	\end{split}
\end{equation}
The combination of these two integrals leads to, 
\begin{equation}\label{IcorrExpansion}
	\begin{split}
		&\int d\bx\, |\bx|^{2m} \, T_A^\text{opt}(\bx)e^{-T_A^\text{opt}(\bx) \sigma_{\text{fluc}}^\text{NN}}=  A R_0^{2m}\left[2\left(\frac{a_0}{R_0}\right)\left(\frac{\pi R_0^2/A}{\sigma^{\text{NN}}_\text{inel}}\right) + \frac{16}{9}   \left(\frac{\pi R_0^2/A}{\sigma^{\text{NN}}_\text{inel}}\right)^3 + \cdots\right].
	\end{split}
\end{equation} 
In the above, the first term represents the contribution from the correction of skin thickness, while the second term represents the contribution when the WS distribution is approximated with a step-function. Usually, the first term has a greater contribution than the second term, except when $(a_0/R_0)^2 \sim (\pi R_0^2/A) / \sigma^{\text{NN}}_\text{inel}$. Therefore, we only consider the first term. This can be explained as follows: the primary modification to the source distribution should come from the tail of the WS distribution. The probability of finding sources in the skin part of the distribution is lower compared to the other parts. Hence, the sources from this region are more likely to miss the collision. Using Eq.~\eqref{IcorrExpansion}, we can estimate $\overline{N}$,
\begin{equation}\label{NpartCrossSection}
	\begin{split}
		\overline{N} &=  2A\left(\frac{a_0}{R_0}\right)\left(\frac{\pi R_0^2/A}{\sigma^{\text{NN}}_\text{inel}}\right)+\cdots,
	\end{split}
\end{equation}
and consequently, the number of participants,
\begin{equation}
	N_\pa = 2(A-\overline{N}).
\end{equation}
In the presence of interaction, we can write the moments as
\begin{equation}
	\OO_s = \frac{2A}{N_\pa}\OO_s^{(0)}-\frac{2\overline{N}}{N_\pa}\overline{\OO}_s.
\end{equation}
where
\begin{subequations}
	\begin{align}
		&\OO_s = \frac{1}{N_\pa}\sum_{i=1}^{N_\pa} f(\bx_i), \qquad \OO_s^{(0)} = \frac{1}{2A}\sum_{i=1}^{2A} f(\bx_i),\qquad \overline{\OO}_s = \frac{1}{2\overline{N}}\sum_{i=1}^{2\overline{N}} f(\bx_i).
	\end{align}
\end{subequations}

In the previous subsections, we have computed the averages of $\la \OO_s^{(0)} \ra$ and $\la \OO_{s,1}^{(0)} \OO_{s,2}^{(0)} \ra$. In this subsection, we assume that there is a small correlation between the total sources and those that do not participate in the collision, meaning that $\la \OO_{s,i}^{(0)} \overline{\OO}_{s,j} \ra$ is approximately equal to $\la \OO_{s,i}^{(0)} \ra \la \overline{\OO}_{s,j} \ra$. Therefore, we only need to calculate $\la \overline{\OO}_s \ra$ and $\la \overline{\OO}_{s,1} \overline{\OO}_{s,2}  \ra$ where the average
\begin{equation}
	\la \overline{\OO}_s \ra= \frac{\int d\bx\, \overline{\OO}_s \, T_A^\text{opt}(\bx)e^{-T_A^\text{opt}(\bx) \sigma_{\text{fluc}}^\text{NN}}}{\int d\bx\,  T_A^\text{opt}(\bx)e^{-T_A^\text{opt}(\bx) \sigma_{\text{fluc}}^\text{NN}}}
\end{equation}
can be derived from the moments in Eq.~\eqref{IcorrExpansion}. Since the $m$ dependence at the leading order in these moments only appears in  $R_0^{2m}$, one finds
\begin{subequations}
	\begin{align}
		\la \overline{ \{r^2\}  }_s \ra &= R_0^2\left[1+\cdots\right],\\
		\la \overline{ \left| \{r^2 e^{2i\varphi} \} \right|^2 }_s   \ra &= \frac{1}{2\overline{N}} R_0^4\left[1+\cdots\right].
	\end{align}
\end{subequations}
To calculate the correction to Eq.~\eqref{ObservableDecomposition}, we need to find $\la \OO_s \ra$ and $\la \OO_{s,1} \OO_{s,2} \ra$ in terms of $\la \OO_s^{(0)} \ra$, $\la \OO_{s,1}^{(0)} \OO_{s,2}^{(0)} \ra$, $\la \overline{\OO}_s \ra$, and $\la \overline{\OO}_{s,1} \overline{\OO}_{s,2} \ra$,
\begin{subequations}\label{observableFullMinusNonInt}
	\begin{align}
		\la \OO_s \ra &= 
		\frac{2A}{N_\pa}\la \OO_s^{(0)} \ra -\frac{2\overline{N}}{N_\pa}\la \overline{\OO}_s \ra,\\
		\la \OO_{s,1} \OO_{s,2} \ra &= \left(\frac{2A}{N_\pa}\right)^2\la \OO_{s,1}^{(0)} \OO_{s,2}^{(0)} \ra +\left(\frac{2\overline{N}}{N_\pa}\right)^2\la \overline{\OO}_{s,1} \overline{\OO}_{s,2}  \ra -\left(\frac{2A}{N_\pa}\right)\left(\frac{2\overline{N}}{N_\pa}\right)\left( \la \OO_{s,1}^{(0)} \overline{\OO}_{s,2} \ra + \la \OO_{s,2}^{(0)} \overline{\OO}_{s,1} \ra\right).\nn
	\end{align}
\end{subequations}
  In the following, one finds the correction to the moment averages we are interested in the present work,
\begin{subequations}\label{cross_section_effect}
	\begin{align}
		&\frac{5}{2R_0^2}\left.\left\la \{r^2\}_{s} \right\ra \right|_\text{crs-sct} 
		= -3\left(\frac{a_0}{R_0}\right)\left(\frac{\pi R_0^2/A}{\sigma^{\text{NN}}_\text{inel}}\right),\\
		&\frac{35A}{4 R_0^4}\left.\left\la\left|\{r^2 e^{2i\varphi}\}_{s}\right|^2\right\ra \right|_\text{crs-sct} = \frac{51}{4}\left(\frac{a_0}{R_0}\right)\left(\frac{\pi R_0^2/A}{\sigma^{\text{NN}}_\text{inel}}\right).
	\end{align}
\end{subequations}
In the limit $\sigma^{\text{NN}}_\text{inel}$ is much larger than the inverse density of sources in the transverse area, $\pi R_0^2/A$, we expect all nucleons to participate in the collision and the above corrections approach to zero.

\subsection{Effect of nucleus deformation}\label{subsec:deform}

The one-body density of a deformed nucleus, denoted as $p_A(\vx)$, is not rotationally invariant. For nuclei with quadrupole deformation, an elliptical shape is expected in the initial entropy density in the transverse space, even in the infinite number of nucleons at zero impact parameter. In the collision of such nuclei, we expect the maximum ellipticity value in the body-body collision and the minimum in the tip-tip collision. For randomly rotating nuclei and an infinite number of nucleons, the ellipticity of the overlapping region fluctuates event-by-event between these two values. 

The scale-invariance approximation in UCSC events is valid when two deformed nuclei have maximum overlap, i.e., the target and projectile have the same orientation in space, or they must be mirror reflections compared to the transverse plane. However, selecting such events is not an easy task. At finite values of the cross-section, it is always possible that a nucleon from the projectile passes through the target without any interaction. On the other hand, at very large values of the cross-section, all the nucleons participate in the collision, irrespective of the orientation of the target and projectile. As a result, events with the same number of participants (and the same multiplicity) might correspond to event topologies when the geometrical overlapping region of the target and projectile is not the same. Ignoring such complications in selecting the most accurate, maximally overlapping events and following Ref.~\cite{Jia:2021tzt}, we assume that the target and projectile rotate randomly and study the mean deformation and its fluctuation from the independent random orientation of colliding nuclei. 

The rotated one-body density is given by $p_A( \mathcal{R} \cdot \vx)$, where $\mathcal{R}(\alpha_e,\beta_e,\gamma_e)$ is the rotation matrix written in terms of the Euler anglers. The moment of the overlapping region in a single event is obtained as follows,
\begin{equation}
	\begin{split}
		& \OO(\mathcal{R}_1, \mathcal{R}_1) 
		=  \frac{1}{2}\int d^3 x \left(p_A( \mathcal{R}_1 \cdot \vx) + p_A( \mathcal{R}_2 \cdot \vx)\right) f(\bx),
	\end{split}
\end{equation}
and the many event average of the above moment is given by
\begin{equation}
	\begin{split}
		\la \OO_1 \OO_2 \cdots \ra = \int \, \OO_1 \OO_2 \cdots  \, d\Omega_1 d\Omega_2,
	\end{split}
\end{equation}
where $d\Omega = \sin\beta_e d\alpha_e d\beta_e d\gamma_e$. One can estimate the effect of random nuclei orientation for a given deformed one-body density by performing the integral numerically.

\begin{figure*}
	\centering
	\begin{tabular}{c}
		\includegraphics[width=0.78\textwidth]{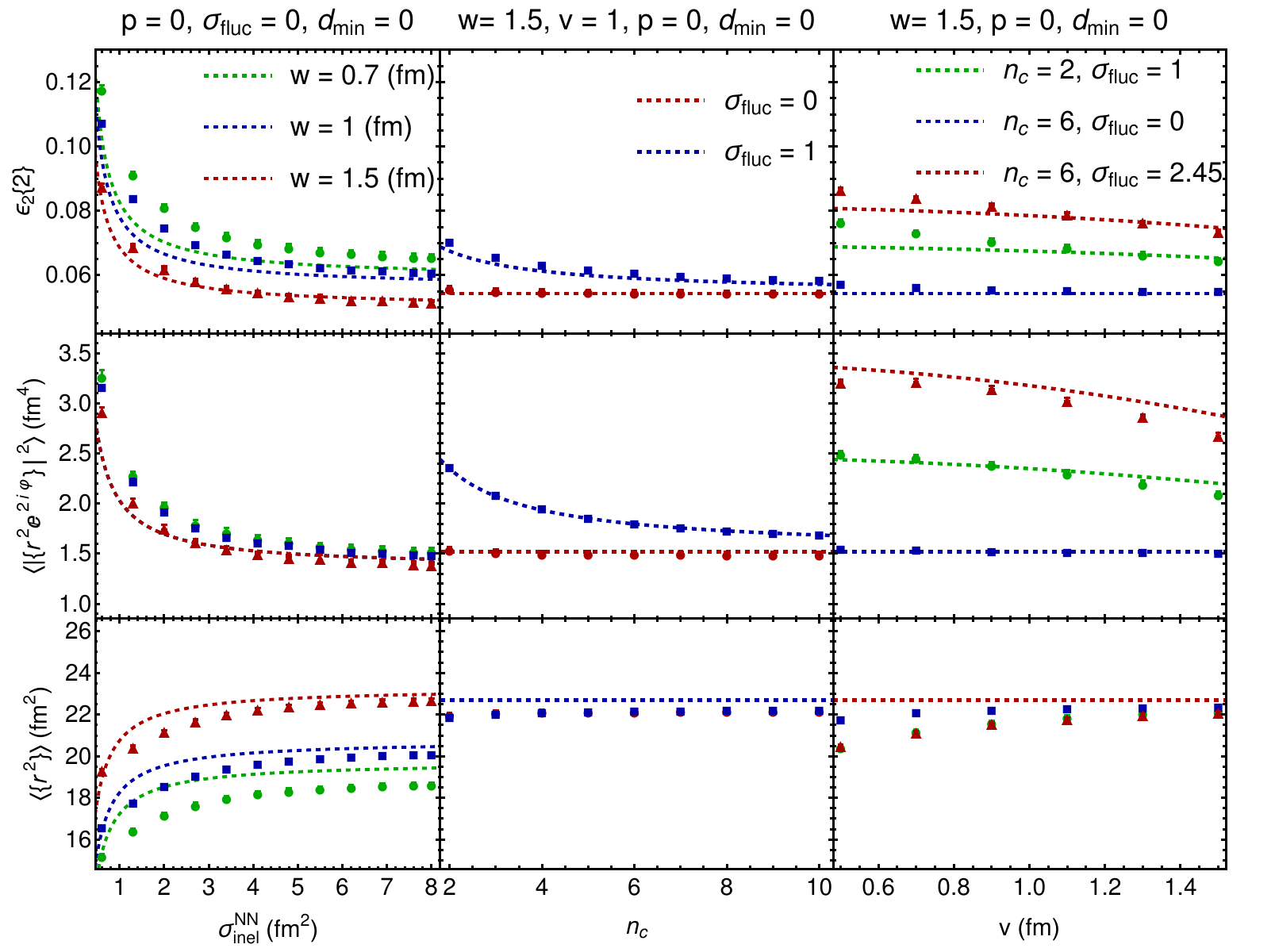} 
	\end{tabular}
	\caption{ Comparing the analytical estimation (curves) for $\la \{r^2\}\ra $, $\left\la\left|\{r^2 e^{2i\varphi}\}\right|^2\right\ra$, and $\epsilon_2\{2\}$ with \trento{} model outcome (dots) for Au--Au collisions  with $\sigma_{\textbf{inel}}^{NN}=4.2\,$[fm$^2$].}
	\label{trentoValidationII}
\end{figure*}

An analytical estimation for the mentioned integrals has been done in Red.~\cite{Jia:2021tzt} where the WS is replaced by a step-function with a surface at $R(\theta,\varphi)$. The integrals can be done analytically for a WS distribution when its radius is given by 
\begin{equation}
	\begin{split}
		R(\theta,\varphi) = R_0 &\left(1 + \beta_2 \sum_{m=-2}^2 \mathcal{D}^{(2)}_{0,m}\,Y_2^m(\theta,\varphi)   + \beta_3 \sum_m \mathcal{D}^{(3)}_{0,m}\,Y_3^m(\theta,\varphi)+\cdots\right).
	\end{split}
\end{equation}
The Wigner D-matrix $\mathcal{D}^{(\ell)}_{0,m}(\alpha_e,\beta_e,\gamma_e)$ rotates the nucleus into an arbitrary orientation in space. For randomly rotating nuclei with an infinite number of point-like sources, we obtain
\begin{subequations}\label{deform_effect}
	\begin{align}
		&\left.\left\la\{r^2 \}_{s}^2\right\ra \right|_{A\to\infty}= I_1^2(R_0,a_0) + B_1(\beta_2,\beta_3), \\
		&\left.\left\la\left|\{r^2 e^{2i\varphi}\}_{s}\right|^2\right\ra \right|_{A\to\infty}= B_2(\beta_2),
	\end{align}
\end{subequations}
where $I_1$ is given in Eq.~\eqref{ImomentsA} and 
\begin{subequations}
	\begin{align}
		&B_1(\beta_2,\beta_3)  = \frac{4 R_0^2}{25} \left[ \frac{29}{8\pi}\beta_2^2 + \frac{7}{2\pi}\beta_3^2  + \cdots\right], \\
		&B_2(\beta_2)  = \frac{4 R_0^2}{25} \left[\frac{3}{4\pi} \beta_2^2   + \cdots\right].
	\end{align}
\end{subequations}
The effect of skin thickness appears in higher order terms in $B_i$ and has a negligible contribution.

\subsection{Validation via \trento{} model}

The validation of parameters $a_0$, $d_\mi$, and $\sigma_{\text{fluc}}$ is presented in section.~\ref{EllipticityFluctuationSec}. Here, we validate $n_c$, $v$, and $\sigma_{\textbf{inel}}^{\text{NN}}$ dependence of our calculation with comparing it with \trento{} simulations.

We can observe from the Fig.~\ref{trentoValidationII} that the moment $ \la \{r^2\} \ra$ is not significantly affected by the parameters $n_c$, and $v$, which is consistent with our calculations. Similar to $\sigma_{\text{fluc}}$, these parameters mostly modify the fluctuation, while the main contribution to the value of $ \la \{r^2\} \ra$ is the size of the nucleus.  Decreasing $\sigma_{\text{inel}}^\text{NN}$ leads to a decrease in the total number of participants, starting from those that are more populated at the tail of the distribution. This results in a decrease in $ \la \{r^2\} \ra$. 

The figure shows  increasing $n_c$ leads to a decrease in the fluctuation related to the nucleon substructure, which results in a smaller value for this moment. When $\sigma_\fluc = 0$, the $n_c$ dependence of this moment begins from higher order terms in $(w^2-v^2)/R_0^2$, appears as a small dependence on $n_c$ in the simulation. As the value of constituent width $v$ approaches $w$, the effect of the substructure fluctuation is decreased, leading to a decreasing trend in the $v$ dependence.
By decreasing the cross-section $\sigma_{\text{inel}}^\text{NN}$, the number of participants reduces, leading to more fluctuations and larger $\left\la\left|\{r^2 e^{2i\varphi}\}\right|^2\right\ra$. The dependence of $\epsilon_2\{2\}$ on the parameters is dominated by $\left\la\left|\{r^2 e^{2i\varphi}\}\right|^2\right\ra$. Our calculation shows a smaller difference compared to the \trento{} outcome since some errors are canceled out in calculating the ratio. However, inaccuracies due to small $w$ still affect $\epsilon_2\{2\}$ because $\left\la\left|\{r^2 e^{2i\varphi}\}\right|^2\right\ra$ has a smaller dependence on $w$ compared to $\left\la\{r^2\}\right\ra$. This dependence is reduced when we calculate the isobar ratio.

\twocolumngrid
\bibliography{references.bib}%

\begin{thebibliography}{39}%
\makeatletter
\providecommand \@ifxundefined [1]{%
 \@ifx{#1\undefined}
}%
\providecommand \@ifnum [1]{%
 \ifnum #1\expandafter \@firstoftwo
 \else \expandafter \@secondoftwo
 \fi
}%
\providecommand \@ifx [1]{%
 \ifx #1\expandafter \@firstoftwo
 \else \expandafter \@secondoftwo
 \fi
}%
\providecommand \natexlab [1]{#1}%
\providecommand \enquote  [1]{``#1''}%
\providecommand \bibnamefont  [1]{#1}%
\providecommand \bibfnamefont [1]{#1}%
\providecommand \citenamefont [1]{#1}%
\providecommand \href@noop [0]{\@secondoftwo}%
\providecommand \href [0]{\begingroup \@sanitize@url \@href}%
\providecommand \@href[1]{\@@startlink{#1}\@@href}%
\providecommand \@@href[1]{\endgroup#1\@@endlink}%
\providecommand \@sanitize@url [0]{\catcode `\\12\catcode `\$12\catcode
  `\&12\catcode `\#12\catcode `\^12\catcode `\_12\catcode `\%12\relax}%
\providecommand \@@startlink[1]{}%
\providecommand \@@endlink[0]{}%
\providecommand \url  [0]{\begingroup\@sanitize@url \@url }%
\providecommand \@url [1]{\endgroup\@href {#1}{\urlprefix }}%
\providecommand \urlprefix  [0]{URL }%
\providecommand \Eprint [0]{\href }%
\providecommand \doibase [0]{https://doi.org/}%
\providecommand \selectlanguage [0]{\@gobble}%
\providecommand \bibinfo  [0]{\@secondoftwo}%
\providecommand \bibfield  [0]{\@secondoftwo}%
\providecommand \translation [1]{[#1]}%
\providecommand \BibitemOpen [0]{}%
\providecommand \bibitemStop [0]{}%
\providecommand \bibitemNoStop [0]{.\EOS\space}%
\providecommand \EOS [0]{\spacefactor3000\relax}%
\providecommand \BibitemShut  [1]{\csname bibitem#1\endcsname}%
\let\auto@bib@innerbib\@empty
\bibitem [{\citenamefont {Hergert}(2020)}]{Hergert:2020bxy}%
  \BibitemOpen
  \bibfield  {author} {\bibinfo {author} {\bibfnamefont {H.}~\bibnamefont
  {Hergert}},\ }\bibfield  {title} {\bibinfo {title} {{A Guided Tour of $ab$
  $initio$ Nuclear Many-Body Theory}},\ }\href
  {https://doi.org/10.3389/fphy.2020.00379} {\bibfield  {journal} {\bibinfo
  {journal} {Front. in Phys.}\ }\textbf {\bibinfo {volume} {8}},\ \bibinfo
  {pages} {379} (\bibinfo {year} {2020})},\ \Eprint
  {https://arxiv.org/abs/2008.05061} {arXiv:2008.05061 [nucl-th]} \BibitemShut
  {NoStop}%
\bibitem [{\citenamefont {Brown}(2000)}]{Brown:2000pd}%
  \BibitemOpen
  \bibfield  {author} {\bibinfo {author} {\bibfnamefont {B.~A.}\ \bibnamefont
  {Brown}},\ }\bibfield  {title} {\bibinfo {title} {{Neutron radii in nuclei
  and the neutron equation of state}},\ }\href
  {https://doi.org/10.1103/PhysRevLett.85.5296} {\bibfield  {journal} {\bibinfo
   {journal} {Phys. Rev. Lett.}\ }\textbf {\bibinfo {volume} {85}},\ \bibinfo
  {pages} {5296} (\bibinfo {year} {2000})}\BibitemShut {NoStop}%
\bibitem [{\citenamefont {Horowitz}\ and\ \citenamefont
  {Piekarewicz}(2001)}]{Horowitz:2000xj}%
  \BibitemOpen
  \bibfield  {author} {\bibinfo {author} {\bibfnamefont {C.~J.}\ \bibnamefont
  {Horowitz}}\ and\ \bibinfo {author} {\bibfnamefont {J.}~\bibnamefont
  {Piekarewicz}},\ }\bibfield  {title} {\bibinfo {title} {{Neutron star
  structure and the neutron radius of Pb-208}},\ }\href
  {https://doi.org/10.1103/PhysRevLett.86.5647} {\bibfield  {journal} {\bibinfo
   {journal} {Phys. Rev. Lett.}\ }\textbf {\bibinfo {volume} {86}},\ \bibinfo
  {pages} {5647} (\bibinfo {year} {2001})},\ \Eprint
  {https://arxiv.org/abs/astro-ph/0010227} {arXiv:astro-ph/0010227}
  \BibitemShut {NoStop}%
\bibitem [{\citenamefont {Giacalone}\ \emph {et~al.}(2021)\citenamefont
  {Giacalone}, \citenamefont {Jia},\ and\ \citenamefont
  {Zhang}}]{Giacalone:2021udy}%
  \BibitemOpen
  \bibfield  {author} {\bibinfo {author} {\bibfnamefont {G.}~\bibnamefont
  {Giacalone}}, \bibinfo {author} {\bibfnamefont {J.}~\bibnamefont {Jia}},\
  and\ \bibinfo {author} {\bibfnamefont {C.}~\bibnamefont {Zhang}},\ }\bibfield
   {title} {\bibinfo {title} {{Impact of Nuclear Deformation on Relativistic
  Heavy-Ion Collisions: Assessing Consistency in Nuclear Physics across Energy
  Scales}},\ }\href {https://doi.org/10.1103/PhysRevLett.127.242301} {\bibfield
   {journal} {\bibinfo  {journal} {Phys. Rev. Lett.}\ }\textbf {\bibinfo
  {volume} {127}},\ \bibinfo {pages} {242301} (\bibinfo {year} {2021})},\
  \Eprint {https://arxiv.org/abs/2105.01638} {arXiv:2105.01638 [nucl-th]}
  \BibitemShut {NoStop}%
\bibitem [{\citenamefont {Giacalone}\ \emph {et~al.}(2023)\citenamefont
  {Giacalone}, \citenamefont {Nijs},\ and\ \citenamefont {van~der
  Schee}}]{Giacalone:2023cet}%
  \BibitemOpen
  \bibfield  {author} {\bibinfo {author} {\bibfnamefont {G.}~\bibnamefont
  {Giacalone}}, \bibinfo {author} {\bibfnamefont {G.}~\bibnamefont {Nijs}},\
  and\ \bibinfo {author} {\bibfnamefont {W.}~\bibnamefont {van~der Schee}},\
  }\bibfield  {title} {\bibinfo {title} {{Determination of the Neutron Skin of
  Pb208 from Ultrarelativistic Nuclear Collisions}},\ }\href
  {https://doi.org/10.1103/PhysRevLett.131.202302} {\bibfield  {journal}
  {\bibinfo  {journal} {Phys. Rev. Lett.}\ }\textbf {\bibinfo {volume} {131}},\
  \bibinfo {pages} {202302} (\bibinfo {year} {2023})},\ \Eprint
  {https://arxiv.org/abs/2305.00015} {arXiv:2305.00015 [nucl-th]} \BibitemShut
  {NoStop}%
\bibitem [{\citenamefont {Jia}(2022)}]{Jia:2021tzt}%
  \BibitemOpen
  \bibfield  {author} {\bibinfo {author} {\bibfnamefont {J.}~\bibnamefont
  {Jia}},\ }\bibfield  {title} {\bibinfo {title} {{Shape of atomic nuclei in
  heavy ion collisions}},\ }\href {https://doi.org/10.1103/PhysRevC.105.014905}
  {\bibfield  {journal} {\bibinfo  {journal} {Phys. Rev. C}\ }\textbf {\bibinfo
  {volume} {105}},\ \bibinfo {pages} {014905} (\bibinfo {year} {2022})},\
  \Eprint {https://arxiv.org/abs/2106.08768} {arXiv:2106.08768 [nucl-th]}
  \BibitemShut {NoStop}%
\bibitem [{\citenamefont {Zhang}\ and\ \citenamefont
  {Jia}(2022)}]{Zhang:2021kxj}%
  \BibitemOpen
  \bibfield  {author} {\bibinfo {author} {\bibfnamefont {C.}~\bibnamefont
  {Zhang}}\ and\ \bibinfo {author} {\bibfnamefont {J.}~\bibnamefont {Jia}},\
  }\bibfield  {title} {\bibinfo {title} {{Evidence of Quadrupole and Octupole
  Deformations in Zr96+Zr96 and Ru96+Ru96 Collisions at Ultrarelativistic
  Energies}},\ }\href {https://doi.org/10.1103/PhysRevLett.128.022301}
  {\bibfield  {journal} {\bibinfo  {journal} {Phys. Rev. Lett.}\ }\textbf
  {\bibinfo {volume} {128}},\ \bibinfo {pages} {022301} (\bibinfo {year}
  {2022})},\ \Eprint {https://arxiv.org/abs/2109.01631} {arXiv:2109.01631
  [nucl-th]} \BibitemShut {NoStop}%
\bibitem [{\citenamefont {Zhang}\ \emph {et~al.}(2022)\citenamefont {Zhang},
  \citenamefont {Bhatta},\ and\ \citenamefont {Jia}}]{Zhang:2022fou}%
  \BibitemOpen
  \bibfield  {author} {\bibinfo {author} {\bibfnamefont {C.}~\bibnamefont
  {Zhang}}, \bibinfo {author} {\bibfnamefont {S.}~\bibnamefont {Bhatta}},\ and\
  \bibinfo {author} {\bibfnamefont {J.}~\bibnamefont {Jia}},\ }\bibfield
  {title} {\bibinfo {title} {{Ratios of collective flow observables in
  high-energy isobar collisions are insensitive to final-state interactions}},\
  }\href {https://doi.org/10.1103/PhysRevC.106.L031901} {\bibfield  {journal}
  {\bibinfo  {journal} {Phys. Rev. C}\ }\textbf {\bibinfo {volume} {106}},\
  \bibinfo {pages} {L031901} (\bibinfo {year} {2022})},\ \Eprint
  {https://arxiv.org/abs/2206.01943} {arXiv:2206.01943 [nucl-th]} \BibitemShut
  {NoStop}%
\bibitem [{\citenamefont {Jia}\ \emph {et~al.}(2023)\citenamefont {Jia},
  \citenamefont {Giacalone},\ and\ \citenamefont {Zhang}}]{Jia:2022qgl}%
  \BibitemOpen
  \bibfield  {author} {\bibinfo {author} {\bibfnamefont {J.}~\bibnamefont
  {Jia}}, \bibinfo {author} {\bibfnamefont {G.}~\bibnamefont {Giacalone}},\
  and\ \bibinfo {author} {\bibfnamefont {C.}~\bibnamefont {Zhang}},\ }\bibfield
   {title} {\bibinfo {title} {{Separating the Impact of Nuclear Skin and
  Nuclear Deformation in High-Energy Isobar Collisions}},\ }\href
  {https://doi.org/10.1103/PhysRevLett.131.022301} {\bibfield  {journal}
  {\bibinfo  {journal} {Phys. Rev. Lett.}\ }\textbf {\bibinfo {volume} {131}},\
  \bibinfo {pages} {022301} (\bibinfo {year} {2023})},\ \Eprint
  {https://arxiv.org/abs/2206.10449} {arXiv:2206.10449 [nucl-th]} \BibitemShut
  {NoStop}%
\bibitem [{\citenamefont {Miller}\ \emph {et~al.}(2007)\citenamefont {Miller},
  \citenamefont {Reygers}, \citenamefont {Sanders},\ and\ \citenamefont
  {Steinberg}}]{Miller:2007ri}%
  \BibitemOpen
  \bibfield  {author} {\bibinfo {author} {\bibfnamefont {M.~L.}\ \bibnamefont
  {Miller}}, \bibinfo {author} {\bibfnamefont {K.}~\bibnamefont {Reygers}},
  \bibinfo {author} {\bibfnamefont {S.~J.}\ \bibnamefont {Sanders}},\ and\
  \bibinfo {author} {\bibfnamefont {P.}~\bibnamefont {Steinberg}},\ }\bibfield
  {title} {\bibinfo {title} {{Glauber modeling in high energy nuclear
  collisions}},\ }\href {https://doi.org/10.1146/annurev.nucl.57.090506.123020}
  {\bibfield  {journal} {\bibinfo  {journal} {Ann. Rev. Nucl. Part. Sci.}\
  }\textbf {\bibinfo {volume} {57}},\ \bibinfo {pages} {205} (\bibinfo {year}
  {2007})},\ \Eprint {https://arxiv.org/abs/nucl-ex/0701025}
  {arXiv:nucl-ex/0701025} \BibitemShut {NoStop}%
\bibitem [{\citenamefont {Alver}\ \emph {et~al.}(2008)\citenamefont {Alver},
  \citenamefont {Baker}, \citenamefont {Loizides},\ and\ \citenamefont
  {Steinberg}}]{Alver:2008aq}%
  \BibitemOpen
  \bibfield  {author} {\bibinfo {author} {\bibfnamefont {B.}~\bibnamefont
  {Alver}}, \bibinfo {author} {\bibfnamefont {M.}~\bibnamefont {Baker}},
  \bibinfo {author} {\bibfnamefont {C.}~\bibnamefont {Loizides}},\ and\
  \bibinfo {author} {\bibfnamefont {P.}~\bibnamefont {Steinberg}},\ }\bibfield
  {title} {\bibinfo {title} {{The PHOBOS Glauber Monte Carlo}},\ }\href@noop {}
  {\  (\bibinfo {year} {2008})},\ \Eprint {https://arxiv.org/abs/0805.4411}
  {arXiv:0805.4411 [nucl-ex]} \BibitemShut {NoStop}%
\bibitem [{\citenamefont {Kharzeev}\ \emph {et~al.}(2005)\citenamefont
  {Kharzeev}, \citenamefont {Levin},\ and\ \citenamefont
  {Nardi}}]{Kharzeev:2001yq}%
  \BibitemOpen
  \bibfield  {author} {\bibinfo {author} {\bibfnamefont {D.}~\bibnamefont
  {Kharzeev}}, \bibinfo {author} {\bibfnamefont {E.}~\bibnamefont {Levin}},\
  and\ \bibinfo {author} {\bibfnamefont {M.}~\bibnamefont {Nardi}},\ }\bibfield
   {title} {\bibinfo {title} {{The Onset of classical QCD dynamics in
  relativistic heavy ion collisions}},\ }\href
  {https://doi.org/10.1103/PhysRevC.71.054903} {\bibfield  {journal} {\bibinfo
  {journal} {Phys. Rev. C}\ }\textbf {\bibinfo {volume} {71}},\ \bibinfo
  {pages} {054903} (\bibinfo {year} {2005})},\ \Eprint
  {https://arxiv.org/abs/hep-ph/0111315} {arXiv:hep-ph/0111315} \BibitemShut
  {NoStop}%
\bibitem [{\citenamefont {Drescher}\ \emph {et~al.}(2006)\citenamefont
  {Drescher}, \citenamefont {Dumitru}, \citenamefont {Hayashigaki},\ and\
  \citenamefont {Nara}}]{Drescher:2006pi}%
  \BibitemOpen
  \bibfield  {author} {\bibinfo {author} {\bibfnamefont {H.-J.}\ \bibnamefont
  {Drescher}}, \bibinfo {author} {\bibfnamefont {A.}~\bibnamefont {Dumitru}},
  \bibinfo {author} {\bibfnamefont {A.}~\bibnamefont {Hayashigaki}},\ and\
  \bibinfo {author} {\bibfnamefont {Y.}~\bibnamefont {Nara}},\ }\bibfield
  {title} {\bibinfo {title} {{The Eccentricity in heavy-ion collisions from
  color glass condensate initial conditions}},\ }\href
  {https://doi.org/10.1103/PhysRevC.74.044905} {\bibfield  {journal} {\bibinfo
  {journal} {Phys. Rev. C}\ }\textbf {\bibinfo {volume} {74}},\ \bibinfo
  {pages} {044905} (\bibinfo {year} {2006})},\ \Eprint
  {https://arxiv.org/abs/nucl-th/0605012} {arXiv:nucl-th/0605012} \BibitemShut
  {NoStop}%
\bibitem [{\citenamefont {Moreland}\ \emph {et~al.}(2015)\citenamefont
  {Moreland}, \citenamefont {Bernhard},\ and\ \citenamefont
  {Bass}}]{Moreland:2014oya}%
  \BibitemOpen
  \bibfield  {author} {\bibinfo {author} {\bibfnamefont {J.~S.}\ \bibnamefont
  {Moreland}}, \bibinfo {author} {\bibfnamefont {J.~E.}\ \bibnamefont
  {Bernhard}},\ and\ \bibinfo {author} {\bibfnamefont {S.~A.}\ \bibnamefont
  {Bass}},\ }\bibfield  {title} {\bibinfo {title} {{Alternative ansatz to
  wounded nucleon and binary collision scaling in high-energy nuclear
  collisions}},\ }\href {https://doi.org/10.1103/PhysRevC.92.011901} {\bibfield
   {journal} {\bibinfo  {journal} {Phys. Rev. C}\ }\textbf {\bibinfo {volume}
  {92}},\ \bibinfo {pages} {011901} (\bibinfo {year} {2015})},\ \Eprint
  {https://arxiv.org/abs/1412.4708} {arXiv:1412.4708 [nucl-th]} \BibitemShut
  {NoStop}%
\bibitem [{\citenamefont {Moreland}\ \emph {et~al.}(2020)\citenamefont
  {Moreland}, \citenamefont {Bernhard},\ and\ \citenamefont
  {Bass}}]{Moreland:2018gsh}%
  \BibitemOpen
  \bibfield  {author} {\bibinfo {author} {\bibfnamefont {J.~S.}\ \bibnamefont
  {Moreland}}, \bibinfo {author} {\bibfnamefont {J.~E.}\ \bibnamefont
  {Bernhard}},\ and\ \bibinfo {author} {\bibfnamefont {S.~A.}\ \bibnamefont
  {Bass}},\ }\bibfield  {title} {\bibinfo {title} {{Bayesian calibration of a
  hybrid nuclear collision model using p-Pb and Pb-Pb data at energies
  available at the CERN Large Hadron Collider}},\ }\href
  {https://doi.org/10.1103/PhysRevC.101.024911} {\bibfield  {journal} {\bibinfo
   {journal} {Phys. Rev. C}\ }\textbf {\bibinfo {volume} {101}},\ \bibinfo
  {pages} {024911} (\bibinfo {year} {2020})},\ \Eprint
  {https://arxiv.org/abs/1808.02106} {arXiv:1808.02106 [nucl-th]} \BibitemShut
  {NoStop}%
\bibitem [{\citenamefont {Schenke}\ \emph
  {et~al.}(2012{\natexlab{a}})\citenamefont {Schenke}, \citenamefont
  {Tribedy},\ and\ \citenamefont {Venugopalan}}]{Schenke:2012wb}%
  \BibitemOpen
  \bibfield  {author} {\bibinfo {author} {\bibfnamefont {B.}~\bibnamefont
  {Schenke}}, \bibinfo {author} {\bibfnamefont {P.}~\bibnamefont {Tribedy}},\
  and\ \bibinfo {author} {\bibfnamefont {R.}~\bibnamefont {Venugopalan}},\
  }\bibfield  {title} {\bibinfo {title} {{Fluctuating Glasma initial conditions
  and flow in heavy ion collisions}},\ }\href
  {https://doi.org/10.1103/PhysRevLett.108.252301} {\bibfield  {journal}
  {\bibinfo  {journal} {Phys. Rev. Lett.}\ }\textbf {\bibinfo {volume} {108}},\
  \bibinfo {pages} {252301} (\bibinfo {year} {2012}{\natexlab{a}})},\ \Eprint
  {https://arxiv.org/abs/1202.6646} {arXiv:1202.6646 [nucl-th]} \BibitemShut
  {NoStop}%
\bibitem [{\citenamefont {Schenke}\ \emph
  {et~al.}(2012{\natexlab{b}})\citenamefont {Schenke}, \citenamefont
  {Tribedy},\ and\ \citenamefont {Venugopalan}}]{Schenke:2012hg}%
  \BibitemOpen
  \bibfield  {author} {\bibinfo {author} {\bibfnamefont {B.}~\bibnamefont
  {Schenke}}, \bibinfo {author} {\bibfnamefont {P.}~\bibnamefont {Tribedy}},\
  and\ \bibinfo {author} {\bibfnamefont {R.}~\bibnamefont {Venugopalan}},\
  }\bibfield  {title} {\bibinfo {title} {{Event-by-event gluon multiplicity,
  energy density, and eccentricities in ultrarelativistic heavy-ion
  collisions}},\ }\href {https://doi.org/10.1103/PhysRevC.86.034908} {\bibfield
   {journal} {\bibinfo  {journal} {Phys. Rev. C}\ }\textbf {\bibinfo {volume}
  {86}},\ \bibinfo {pages} {034908} (\bibinfo {year} {2012}{\natexlab{b}})},\
  \Eprint {https://arxiv.org/abs/1206.6805} {arXiv:1206.6805 [hep-ph]}
  \BibitemShut {NoStop}%
\bibitem [{\citenamefont {Pandit}(2013)}]{Pandit:2013uiv}%
  \BibitemOpen
  \bibfield  {author} {\bibinfo {author} {\bibfnamefont {Y.}~\bibnamefont
  {Pandit}} (\bibinfo {collaboration} {STAR}),\ }\bibfield  {title} {\bibinfo
  {title} {{Azimuthal Anisotropy in U+U Collisions at $\sqrt{s_{NN}} = 193 $
  GeV with STAR Detector at RHIC}},\ }\href
  {https://doi.org/10.1088/1742-6596/458/1/012003} {\bibfield  {journal}
  {\bibinfo  {journal} {J. Phys. Conf. Ser.}\ }\textbf {\bibinfo {volume}
  {458}},\ \bibinfo {pages} {012003} (\bibinfo {year} {2013})},\ \Eprint
  {https://arxiv.org/abs/1305.0173} {arXiv:1305.0173 [nucl-ex]} \BibitemShut
  {NoStop}%
\bibitem [{\citenamefont {Wang}\ and\ \citenamefont
  {Sorensen}(2014)}]{Wang:2014qxa}%
  \BibitemOpen
  \bibfield  {author} {\bibinfo {author} {\bibfnamefont {H.}~\bibnamefont
  {Wang}}\ and\ \bibinfo {author} {\bibfnamefont {P.}~\bibnamefont {Sorensen}}
  (\bibinfo {collaboration} {STAR}),\ }\bibfield  {title} {\bibinfo {title}
  {{Azimuthal anisotropy in U+U collisions at STAR}},\ }\href
  {https://doi.org/10.1016/j.nuclphysa.2014.09.111} {\bibfield  {journal}
  {\bibinfo  {journal} {Nucl. Phys. A}\ }\textbf {\bibinfo {volume} {932}},\
  \bibinfo {pages} {169} (\bibinfo {year} {2014})},\ \Eprint
  {https://arxiv.org/abs/1406.7522} {arXiv:1406.7522 [nucl-ex]} \BibitemShut
  {NoStop}%
\bibitem [{\citenamefont {Bozek}\ and\ \citenamefont
  {Broniowski}(2013)}]{Bozek:2013uha}%
  \BibitemOpen
  \bibfield  {author} {\bibinfo {author} {\bibfnamefont {P.}~\bibnamefont
  {Bozek}}\ and\ \bibinfo {author} {\bibfnamefont {W.}~\bibnamefont
  {Broniowski}},\ }\bibfield  {title} {\bibinfo {title} {{Collective dynamics
  in high-energy proton-nucleus collisions}},\ }\href
  {https://doi.org/10.1103/PhysRevC.88.014903} {\bibfield  {journal} {\bibinfo
  {journal} {Phys. Rev. C}\ }\textbf {\bibinfo {volume} {88}},\ \bibinfo
  {pages} {014903} (\bibinfo {year} {2013})},\ \Eprint
  {https://arxiv.org/abs/1304.3044} {arXiv:1304.3044 [nucl-th]} \BibitemShut
  {NoStop}%
\bibitem [{\citenamefont {Everett}\ \emph {et~al.}(2021)\citenamefont {Everett}
  \emph {et~al.}}]{JETSCAPE:2020mzn}%
  \BibitemOpen
  \bibfield  {author} {\bibinfo {author} {\bibfnamefont {D.}~\bibnamefont
  {Everett}} \emph {et~al.} (\bibinfo {collaboration} {JETSCAPE}),\ }\bibfield
  {title} {\bibinfo {title} {{Multisystem Bayesian constraints on the transport
  coefficients of QCD matter}},\ }\href
  {https://doi.org/10.1103/PhysRevC.103.054904} {\bibfield  {journal} {\bibinfo
   {journal} {Phys. Rev. C}\ }\textbf {\bibinfo {volume} {103}},\ \bibinfo
  {pages} {054904} (\bibinfo {year} {2021})},\ \Eprint
  {https://arxiv.org/abs/2011.01430} {arXiv:2011.01430 [hep-ph]} \BibitemShut
  {NoStop}%
\bibitem [{\citenamefont {Nijs}\ and\ \citenamefont {van~der
  Schee}(2023{\natexlab{a}})}]{Nijs:2023yab}%
  \BibitemOpen
  \bibfield  {author} {\bibinfo {author} {\bibfnamefont {G.}~\bibnamefont
  {Nijs}}\ and\ \bibinfo {author} {\bibfnamefont {W.}~\bibnamefont {van~der
  Schee}},\ }\bibfield  {title} {\bibinfo {title} {{A generalized
  hydrodynamizing initial stage for Heavy Ion Collisions}},\ }\href@noop {} {\
  (\bibinfo {year} {2023}{\natexlab{a}})},\ \Eprint
  {https://arxiv.org/abs/2304.06191} {arXiv:2304.06191 [nucl-th]} \BibitemShut
  {NoStop}%
\bibitem [{\citenamefont {Ollitrault}(1992)}]{Ollitrault:1992bk}%
  \BibitemOpen
  \bibfield  {author} {\bibinfo {author} {\bibfnamefont {J.-Y.}\ \bibnamefont
  {Ollitrault}},\ }\bibfield  {title} {\bibinfo {title} {{Anisotropy as a
  signature of transverse collective flow}},\ }\href
  {https://doi.org/10.1103/PhysRevD.46.229} {\bibfield  {journal} {\bibinfo
  {journal} {Phys. Rev. D}\ }\textbf {\bibinfo {volume} {46}},\ \bibinfo
  {pages} {229} (\bibinfo {year} {1992})}\BibitemShut {NoStop}%
\bibitem [{\citenamefont {Barrette}\ \emph {et~al.}(1997)\citenamefont
  {Barrette} \emph {et~al.}}]{E877:1996czs}%
  \BibitemOpen
  \bibfield  {author} {\bibinfo {author} {\bibfnamefont {J.}~\bibnamefont
  {Barrette}} \emph {et~al.} (\bibinfo {collaboration} {E877}),\ }\bibfield
  {title} {\bibinfo {title} {{Energy and charged particle flow in a
  10.8-A/GeV/c Au + Au collisions}},\ }\href
  {https://doi.org/10.1103/PhysRevC.55.1420} {\bibfield  {journal} {\bibinfo
  {journal} {Phys. Rev. C}\ }\textbf {\bibinfo {volume} {55}},\ \bibinfo
  {pages} {1420} (\bibinfo {year} {1997})},\ \bibinfo {note} {[Erratum:
  Phys.Rev.C 56, 2336--2336 (1997)]},\ \Eprint
  {https://arxiv.org/abs/nucl-ex/9610006} {arXiv:nucl-ex/9610006} \BibitemShut
  {NoStop}%
\bibitem [{\citenamefont {Niemi}\ \emph {et~al.}(2016)\citenamefont {Niemi},
  \citenamefont {Eskola},\ and\ \citenamefont {Paatelainen}}]{Niemi:2015qia}%
  \BibitemOpen
  \bibfield  {author} {\bibinfo {author} {\bibfnamefont {H.}~\bibnamefont
  {Niemi}}, \bibinfo {author} {\bibfnamefont {K.~J.}\ \bibnamefont {Eskola}},\
  and\ \bibinfo {author} {\bibfnamefont {R.}~\bibnamefont {Paatelainen}},\
  }\bibfield  {title} {\bibinfo {title} {{Event-by-event fluctuations in a
  perturbative QCD + saturation + hydrodynamics model: Determining QCD matter
  shear viscosity in ultrarelativistic heavy-ion collisions}},\ }\href
  {https://doi.org/10.1103/PhysRevC.93.024907} {\bibfield  {journal} {\bibinfo
  {journal} {Phys. Rev. C}\ }\textbf {\bibinfo {volume} {93}},\ \bibinfo
  {pages} {024907} (\bibinfo {year} {2016})},\ \Eprint
  {https://arxiv.org/abs/1505.02677} {arXiv:1505.02677 [hep-ph]} \BibitemShut
  {NoStop}%
\bibitem [{\citenamefont {Abdallah}\ \emph {et~al.}(2022)\citenamefont
  {Abdallah} \emph {et~al.}}]{STAR:2021mii}%
  \BibitemOpen
  \bibfield  {author} {\bibinfo {author} {\bibfnamefont {M.}~\bibnamefont
  {Abdallah}} \emph {et~al.} (\bibinfo {collaboration} {STAR}),\ }\bibfield
  {title} {\bibinfo {title} {{Search for the chiral magnetic effect with isobar
  collisions at $\sqrt {s_{NN}}$=200 GeV by the STAR Collaboration at the BNL
  Relativistic Heavy Ion Collider}},\ }\href
  {https://doi.org/10.1103/PhysRevC.105.014901} {\bibfield  {journal} {\bibinfo
   {journal} {Phys. Rev. C}\ }\textbf {\bibinfo {volume} {105}},\ \bibinfo
  {pages} {014901} (\bibinfo {year} {2022})},\ \Eprint
  {https://arxiv.org/abs/2109.00131} {arXiv:2109.00131 [nucl-ex]} \BibitemShut
  {NoStop}%
\bibitem [{\citenamefont {Aamodt}\ \emph {et~al.}(2011)\citenamefont {Aamodt}
  \emph {et~al.}}]{ALICE:2010mlf}%
  \BibitemOpen
  \bibfield  {author} {\bibinfo {author} {\bibfnamefont {K.}~\bibnamefont
  {Aamodt}} \emph {et~al.} (\bibinfo {collaboration} {ALICE}),\ }\bibfield
  {title} {\bibinfo {title} {{Centrality dependence of the charged-particle
  multiplicity density at mid-rapidity in Pb-Pb collisions at
  $\sqrt{s_{NN}}=2.76$ TeV}},\ }\href
  {https://doi.org/10.1103/PhysRevLett.106.032301} {\bibfield  {journal}
  {\bibinfo  {journal} {Phys. Rev. Lett.}\ }\textbf {\bibinfo {volume} {106}},\
  \bibinfo {pages} {032301} (\bibinfo {year} {2011})},\ \Eprint
  {https://arxiv.org/abs/1012.1657} {arXiv:1012.1657 [nucl-ex]} \BibitemShut
  {NoStop}%
\bibitem [{\citenamefont {Abelev}\ \emph {et~al.}(2013)\citenamefont {Abelev}
  \emph {et~al.}}]{ALICE:2013hur}%
  \BibitemOpen
  \bibfield  {author} {\bibinfo {author} {\bibfnamefont {B.}~\bibnamefont
  {Abelev}} \emph {et~al.} (\bibinfo {collaboration} {ALICE}),\ }\bibfield
  {title} {\bibinfo {title} {{Centrality determination of Pb-Pb collisions at
  $\sqrt{s_{NN}}$ = 2.76 TeV with ALICE}},\ }\href
  {https://doi.org/10.1103/PhysRevC.88.044909} {\bibfield  {journal} {\bibinfo
  {journal} {Phys. Rev. C}\ }\textbf {\bibinfo {volume} {88}},\ \bibinfo
  {pages} {044909} (\bibinfo {year} {2013})},\ \Eprint
  {https://arxiv.org/abs/1301.4361} {arXiv:1301.4361 [nucl-ex]} \BibitemShut
  {NoStop}%
\bibitem [{\citenamefont {Giacalone}\ \emph {et~al.}(2019)\citenamefont
  {Giacalone}, \citenamefont {Mazeliauskas},\ and\ \citenamefont
  {Schlichting}}]{Giacalone:2019ldn}%
  \BibitemOpen
  \bibfield  {author} {\bibinfo {author} {\bibfnamefont {G.}~\bibnamefont
  {Giacalone}}, \bibinfo {author} {\bibfnamefont {A.}~\bibnamefont
  {Mazeliauskas}},\ and\ \bibinfo {author} {\bibfnamefont {S.}~\bibnamefont
  {Schlichting}},\ }\bibfield  {title} {\bibinfo {title} {{Hydrodynamic
  attractors, initial state energy and particle production in relativistic
  nuclear collisions}},\ }\href
  {https://doi.org/10.1103/PhysRevLett.123.262301} {\bibfield  {journal}
  {\bibinfo  {journal} {Phys. Rev. Lett.}\ }\textbf {\bibinfo {volume} {123}},\
  \bibinfo {pages} {262301} (\bibinfo {year} {2019})},\ \Eprint
  {https://arxiv.org/abs/1908.02866} {arXiv:1908.02866 [hep-ph]} \BibitemShut
  {NoStop}%
\bibitem [{\citenamefont {Borghini}\ \emph {et~al.}(2023)\citenamefont
  {Borghini}, \citenamefont {Borrell}, \citenamefont {Feld}, \citenamefont
  {Roch}, \citenamefont {Schlichting},\ and\ \citenamefont
  {Werthmann}}]{Borghini:2022iym}%
  \BibitemOpen
  \bibfield  {author} {\bibinfo {author} {\bibfnamefont {N.}~\bibnamefont
  {Borghini}}, \bibinfo {author} {\bibfnamefont {M.}~\bibnamefont {Borrell}},
  \bibinfo {author} {\bibfnamefont {N.}~\bibnamefont {Feld}}, \bibinfo {author}
  {\bibfnamefont {H.}~\bibnamefont {Roch}}, \bibinfo {author} {\bibfnamefont
  {S.}~\bibnamefont {Schlichting}},\ and\ \bibinfo {author} {\bibfnamefont
  {C.}~\bibnamefont {Werthmann}},\ }\bibfield  {title} {\bibinfo {title}
  {{Statistical analysis of initial-state and final-state response in heavy-ion
  collisions}},\ }\href {https://doi.org/10.1103/PhysRevC.107.034905}
  {\bibfield  {journal} {\bibinfo  {journal} {Phys. Rev. C}\ }\textbf {\bibinfo
  {volume} {107}},\ \bibinfo {pages} {034905} (\bibinfo {year} {2023})},\
  \Eprint {https://arxiv.org/abs/2209.01176} {arXiv:2209.01176 [hep-ph]}
  \BibitemShut {NoStop}%
\bibitem [{\citenamefont {Huang}(1987)}]{huang1987statistical}%
  \BibitemOpen
  \bibfield  {author} {\bibinfo {author} {\bibfnamefont {K.}~\bibnamefont
  {Huang}},\ }\href {https://books.google.de/books?id=M8PvAAAAMAAJ} {\emph
  {\bibinfo {title} {Statistical Mechanics}}}\ (\bibinfo  {publisher} {Wiley},\
  \bibinfo {year} {1987})\BibitemShut {NoStop}%
\bibitem [{\citenamefont {Luzum}\ \emph {et~al.}(2023)\citenamefont {Luzum},
  \citenamefont {Hippert},\ and\ \citenamefont {Ollitrault}}]{Luzum:2023gwy}%
  \BibitemOpen
  \bibfield  {author} {\bibinfo {author} {\bibfnamefont {M.}~\bibnamefont
  {Luzum}}, \bibinfo {author} {\bibfnamefont {M.}~\bibnamefont {Hippert}},\
  and\ \bibinfo {author} {\bibfnamefont {J.-Y.}\ \bibnamefont {Ollitrault}},\
  }\bibfield  {title} {\bibinfo {title} {{Methods for systematic study of
  nuclear structure in high-energy collisions}},\ }\href
  {https://doi.org/10.1140/epja/s10050-023-01021-8} {\bibfield  {journal}
  {\bibinfo  {journal} {Eur. Phys. J. A}\ }\textbf {\bibinfo {volume} {59}},\
  \bibinfo {pages} {110} (\bibinfo {year} {2023})},\ \Eprint
  {https://arxiv.org/abs/2302.14026} {arXiv:2302.14026 [nucl-th]} \BibitemShut
  {NoStop}%
\bibitem [{\citenamefont {Alvioli}\ \emph {et~al.}(2009)\citenamefont
  {Alvioli}, \citenamefont {Drescher},\ and\ \citenamefont
  {Strikman}}]{Alvioli:2009ab}%
  \BibitemOpen
  \bibfield  {author} {\bibinfo {author} {\bibfnamefont {M.}~\bibnamefont
  {Alvioli}}, \bibinfo {author} {\bibfnamefont {H.~J.}\ \bibnamefont
  {Drescher}},\ and\ \bibinfo {author} {\bibfnamefont {M.}~\bibnamefont
  {Strikman}},\ }\bibfield  {title} {\bibinfo {title} {{A Monte Carlo generator
  of nucleon configurations in complex nuclei including Nucleon-Nucleon
  correlations}},\ }\href {https://doi.org/10.1016/j.physletb.2009.08.067}
  {\bibfield  {journal} {\bibinfo  {journal} {Phys. Lett. B}\ }\textbf
  {\bibinfo {volume} {680}},\ \bibinfo {pages} {225} (\bibinfo {year}
  {2009})},\ \Eprint {https://arxiv.org/abs/0905.2670} {arXiv:0905.2670
  [nucl-th]} \BibitemShut {NoStop}%
\bibitem [{\citenamefont {Hammelmann}\ \emph {et~al.}(2020)\citenamefont
  {Hammelmann}, \citenamefont {Soto-Ontoso}, \citenamefont {Alvioli},
  \citenamefont {Elfner},\ and\ \citenamefont {Strikman}}]{Hammelmann:2019vwd}%
  \BibitemOpen
  \bibfield  {author} {\bibinfo {author} {\bibfnamefont {J.}~\bibnamefont
  {Hammelmann}}, \bibinfo {author} {\bibfnamefont {A.}~\bibnamefont
  {Soto-Ontoso}}, \bibinfo {author} {\bibfnamefont {M.}~\bibnamefont
  {Alvioli}}, \bibinfo {author} {\bibfnamefont {H.}~\bibnamefont {Elfner}},\
  and\ \bibinfo {author} {\bibfnamefont {M.}~\bibnamefont {Strikman}} (\bibinfo
  {collaboration} {SMASH}),\ }\bibfield  {title} {\bibinfo {title} {{Influence
  of the neutron-skin effect on nuclear isobar collisions at energies available
  at the BNL Relativistic Heavy Ion Collider}},\ }\href
  {https://doi.org/10.1103/PhysRevC.101.061901} {\bibfield  {journal} {\bibinfo
   {journal} {Phys. Rev. C}\ }\textbf {\bibinfo {volume} {101}},\ \bibinfo
  {pages} {061901} (\bibinfo {year} {2020})},\ \Eprint
  {https://arxiv.org/abs/1908.10231} {arXiv:1908.10231 [nucl-th]} \BibitemShut
  {NoStop}%
\bibitem [{\citenamefont {Yan}\ and\ \citenamefont
  {Ollitrault}(2014)}]{Yan:2013laa}%
  \BibitemOpen
  \bibfield  {author} {\bibinfo {author} {\bibfnamefont {L.}~\bibnamefont
  {Yan}}\ and\ \bibinfo {author} {\bibfnamefont {J.-Y.}\ \bibnamefont
  {Ollitrault}},\ }\bibfield  {title} {\bibinfo {title} {{Universal
  fluctuation-driven eccentricities in proton-proton, proton-nucleus and
  nucleus-nucleus collisions}},\ }\href
  {https://doi.org/10.1103/PhysRevLett.112.082301} {\bibfield  {journal}
  {\bibinfo  {journal} {Phys. Rev. Lett.}\ }\textbf {\bibinfo {volume} {112}},\
  \bibinfo {pages} {082301} (\bibinfo {year} {2014})},\ \Eprint
  {https://arxiv.org/abs/1312.6555} {arXiv:1312.6555 [nucl-th]} \BibitemShut
  {NoStop}%
\bibitem [{\citenamefont {Pritychenko}\ \emph {et~al.}(2016)\citenamefont
  {Pritychenko}, \citenamefont {Birch}, \citenamefont {Singh},\ and\
  \citenamefont {Horoi}}]{Pritychenko:2013gwa}%
  \BibitemOpen
  \bibfield  {author} {\bibinfo {author} {\bibfnamefont {B.}~\bibnamefont
  {Pritychenko}}, \bibinfo {author} {\bibfnamefont {M.}~\bibnamefont {Birch}},
  \bibinfo {author} {\bibfnamefont {B.}~\bibnamefont {Singh}},\ and\ \bibinfo
  {author} {\bibfnamefont {M.}~\bibnamefont {Horoi}},\ }\bibfield  {title}
  {\bibinfo {title} {{Tables of E2 Transition Probabilities from the first
  $2^{+}$ States in Even-Even Nuclei}},\ }\href
  {https://doi.org/10.1016/j.adt.2015.10.001} {\bibfield  {journal} {\bibinfo
  {journal} {Atom. Data Nucl. Data Tabl.}\ }\textbf {\bibinfo {volume} {107}},\
  \bibinfo {pages} {1} (\bibinfo {year} {2016})},\ \bibinfo {note} {[Erratum:
  Atom.Data Nucl.Data Tabl. 114, 371--374 (2017)]},\ \Eprint
  {https://arxiv.org/abs/1312.5975} {arXiv:1312.5975 [nucl-th]} \BibitemShut
  {NoStop}%
\bibitem [{\citenamefont {Nijs}\ and\ \citenamefont {van~der
  Schee}(2023{\natexlab{b}})}]{Nijs:2021kvn}%
  \BibitemOpen
  \bibfield  {author} {\bibinfo {author} {\bibfnamefont {G.}~\bibnamefont
  {Nijs}}\ and\ \bibinfo {author} {\bibfnamefont {W.}~\bibnamefont {van~der
  Schee}},\ }\bibfield  {title} {\bibinfo {title} {{Inferring nuclear structure
  from heavy isobar collisions using Trajectum}},\ }\href
  {https://doi.org/10.21468/SciPostPhys.15.2.041} {\bibfield  {journal}
  {\bibinfo  {journal} {SciPost Phys.}\ }\textbf {\bibinfo {volume} {15}},\
  \bibinfo {pages} {041} (\bibinfo {year} {2023}{\natexlab{b}})},\ \Eprint
  {https://arxiv.org/abs/2112.13771} {arXiv:2112.13771 [nucl-th]} \BibitemShut
  {NoStop}%
\bibitem [{\citenamefont {Teaney}\ and\ \citenamefont
  {Yan}(2011)}]{Teaney:2010vd}%
  \BibitemOpen
  \bibfield  {author} {\bibinfo {author} {\bibfnamefont {D.}~\bibnamefont
  {Teaney}}\ and\ \bibinfo {author} {\bibfnamefont {L.}~\bibnamefont {Yan}},\
  }\bibfield  {title} {\bibinfo {title} {{Triangularity and Dipole Asymmetry in
  Heavy Ion Collisions}},\ }\href {https://doi.org/10.1103/PhysRevC.83.064904}
  {\bibfield  {journal} {\bibinfo  {journal} {Phys. Rev. C}\ }\textbf {\bibinfo
  {volume} {83}},\ \bibinfo {pages} {064904} (\bibinfo {year} {2011})},\
  \Eprint {https://arxiv.org/abs/1010.1876} {arXiv:1010.1876 [nucl-th]}
  \BibitemShut {NoStop}%
\bibitem [{Cit()}]{CiteDrive2022}%
  \BibitemOpen
  \href {http://qcd.phy.duke.edu/trento/usage.html} {\bibinfo {title}
  {http://qcd.phy.duke.edu/trento/usage.html}}\BibitemShut {NoStop}%
\end{thebibliography}%

\end{document}